\documentclass[aps,prd,longbibliography,nofootinbib]{revtex4-2}
\usepackage{graphicx}% Include figure files
\usepackage{dcolumn}% Align table columns on decimal point
\usepackage{amsmath,amssymb,epsfig}
\usepackage{paralist}
\usepackage{comment}
\usepackage{multirow}
\usepackage{color,soul}
\usepackage{suffix}
\usepackage{mathtools}

\usepackage{siunitx}
\DeclareSIUnit\parsec{pc}

\usepackage{tikz}
\usetikzlibrary{arrows.meta,positioning}

\usepackage{array}
\usepackage{hyperref}
\hypersetup{colorlinks=true,allcolors=blue}

\usepackage{physics}
\usepackage{microtype}
\usepackage{enumitem}
\usepackage{booktabs}
\usepackage{tabularx}
\usepackage{longtable}
\usepackage{slashed}

\newcommand{\Order}{\mathcal{O}}
\newcommand{\Lag}{\mathcal{L}}
\newcommand{\mpl}{M_{\mathrm{Pl}}}

\newcommand{\vEW}{v}

\newcommand{\alphaem}{\alpha}
\newcommand{\LCDM}{\texorpdfstring{$\Lambda$CDM}{LambdaCDM}}
\newcommand{\Neff}{N_{\mathrm{eff}}}
\newcommand{\sigSI}{\sigma_{\mathrm{SI}}}

\newcommand{\keti}{k_{\mathrm{eff}}}
\newcommand{\Tobs}{T_{\mathrm{obs}}}
\newcommand{\SNR}{\mathrm{SNR}}
\newcommand{\eV}{\si{\electronvolt}}
\newcommand{\GeV}{\si{\giga\electronvolt}}
\newcommand{\TeV}{\si{\tera\electronvolt}}
\newcommand{\Hz}{\si{\hertz}}

\newcommand{\cm}{\si{\centi\meter}}
\newcommand{\yr}{\mathrm{yr}}
\newcommand{\kmsMpc}{\si{\kilo\meter\per\second\per\mega\parsec}}
\newcommand{\Hn}{H_0}

\newcommand{\urlfoot}[2]{\footnote{#1: \url{#2} (accessed 2025-12-22).}}

\newcommand{\tcell}[2]{\parbox[t]{#1}{\raggedright\strut #2\strut}}

\begin{document}

\title{Fundamental Physics in 2025: Status, Decisive Targets, and Path Forward}
\author{Slava G. Turyshev} 
\affiliation{ 
Jet Propulsion Laboratory, California Institute of Technology,\\
4800 Oak Grove Drive, Pasadena, CA 91109-0899, USA
}

\date{\today}

\begin{abstract}
Fundamental physics today is best defined operationally: it is the program of identifying the microscopic degrees of freedom, symmetries, and dynamical laws that (i) reproduce the Standard Model (SM) of particle physics, General Relativity (GR), and the \LCDM\ cosmological model in their regimes of validity, and (ii) explain the observed phenomena that these baseline theories do not account for (dark matter, neutrino masses, baryogenesis, dark energy), while resolving conceptual inconsistencies (quantum gravity, naturalness, the cosmological constant problem, the measurement problem in quantum theory, information in black holes) and providing predictive unification. This review first lays out the SM+GR+$\Lambda$CDM baseline, the best current evidence for its parameters, and the concrete anomalies and missing ingredients. It then surveys the most relevant theoretical directions (effective field theories; amplitude/positivity programs; lattice and many-body methods; symmetry-based model building; cosmological EFTs; quantum information approaches to QFT/gravitation) and the experimental/observational landscape, including ground and space platforms, astronomical messengers, and in-situ tests. Throughout we emphasize: (a) how each observable maps to energy scales and couplings; (b) the dominant statistical and systematic limitations;  (c) the sensitivity required for decisive progress. A staged roadmap is given only after the technical review, organized by decision points and cross-checks rather than by specific projects.

\end{abstract}

\maketitle
\tableofcontents

% ==========================================
\section{Introduction}
\label{sec:intro}

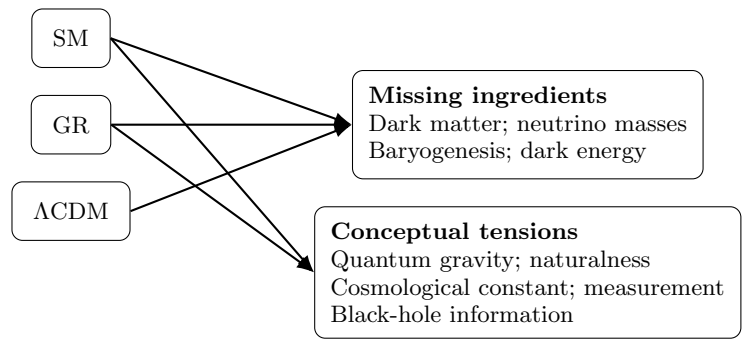
\begin{figure}[h]
\centering
\begin{tikzpicture}[
  box/.style={draw, rounded corners, align=left, inner sep=6pt},
  stack/.style={draw, rounded corners, align=center, inner sep=8pt},
  arr/.style={-Latex, thick},
  node distance=1.1em and 2.2em
]
\node[stack] (sm) {SM};
\node[stack, below=of sm] (gr) {GR};
\node[stack, below=of gr] (lcdm) {\LCDM};

\node[box, right=3.2cm of gr] (miss) {\textbf{Missing ingredients}\\
Dark matter; neutrino masses\\
Baryogenesis; dark energy};
\node[box, below=of miss] (concept) {\textbf{Conceptual tensions}\\
Quantum gravity; naturalness\\
Cosmological constant; measurement\\
Black-hole information};

\draw[arr] (sm.east) -- (miss.west);
\draw[arr] (gr.east) -- (miss.west);
\draw[arr] (lcdm.east) -- (miss.west);

\draw[arr] (gr.east) -- (concept.west);
\draw[arr] (sm.east) -- (concept.west);
\end{tikzpicture}
\caption{Baseline theory framework (SM+GR+\LCDM) and the empirically motivated ``missing ingredients`` and conceptual gaps that define the decision problems summarized in Table~\ref{tab:openproblems}.}
\label{fig:stack_overview}
\end{figure}

Fundamental physics is the effort to (i) identify the minimal degrees of freedom, symmetries, and dynamical principles that reproduce the experimentally verified baseline theory framework of the Standard Model (SM), General Relativity (GR), and the $\Lambda$CDM cosmological model, and (ii) design empirical tests that can falsify \emph{decisive} extensions of that baseline. 
In this review, \emph{fundamental physics} is understood operationally as the empirical program to (i) test the universality of the SM and GR in the regimes where they are established, and (ii) identify (or tightly constrain) the minimal extensions required by phenomena that the baseline stack does not explain. The baseline SM+GR+$\Lambda$CDM framework is quantitatively successful and highly overconstrained by redundant observables across disparate scales, yet it is empirically incomplete: neutrino masses, dark matter, and the baryon asymmetry require ingredients beyond the minimal SM, and late-time cosmic acceleration is well described phenomenologically by $\Lambda$ but not explained microscopically by SM+GR (Figure~\ref{fig:stack_overview}). Standard reviews of the baseline and its present status are given in
\cite{PDG2024RPP,Planck2018Cosmo,Will2014LRR,Turyshev2007IJMPD}.

In 2025 the baseline framework is not a qualitative picture but a quantitatively overconstrained framework: the SM is a renormalizable gauge theory with spontaneous electroweak symmetry breaking, validated from MeV scales through multi-TeV collider processes and high-precision low-energy observables \cite{PDG2024RPP}; GR is validated in weak-field regimes by solar-system timing and ranging, precision time/frequency metrology, and equivalence-principle tests, and in strong-field dynamical regimes by compact-binary gravitational-wave signals \cite{Abbott2016GW150914}; and $\Lambda$CDM fits CMB anisotropies and large-scale structure with percent-level parameter determinations in the minimal parameter set \cite{Planck2018Cosmo}. At the same time, the baseline is empirically incomplete: dark matter and neutrino masses are established by multiple independent datasets, the baryon asymmetry requires new ingredients beyond the minimal SM, and late-time acceleration is well parameterized by $\Lambda$ but not explained microscopically by SM+GR (see Table~\ref{tab:acronyms} for acronyms and abbreviations).

% ==========================================
\begin{table*}[t]
\centering
\caption{Acronyms and abbreviations}
\label{tab:acronyms}
\renewcommand{\arraystretch}{1.08}
\begin{tabular}{ll}
\hline
Acronym & Meaning \\
\hline\hline
SM & Standard Model \\
GR & General Relativity \\
\LCDM & $\Lambda$ cold dark matter cosmology \\
EFT / SMEFT & Effective field theory / Standard Model EFT \\
SME & Standard-Model Extension (Lorentz/CPT-violation framework) \\
EP / UFF & Equivalence principle / universality of free fall \\
ISL & Inverse-square law \\
LPI & Local position invariance \\
LLR / ILR & lunar laser ranging / interplanetary laser ranging \\
CLFV & Charged-lepton flavor violation \\
EDM & Electric dipole moment \\
CE$\nu$NS & Coherent elastic neutrino--nucleus scattering \\
PTA & Pulsar timing array \\
ULDM & Ultralight dark matter \\
LMT & Large momentum transfer (atom interferometry) \\
ROI & Region of interest (analysis window) \\
BBR & Blackbody radiation \\
GW & Gravitational wave \\
\hline
\end{tabular}
\end{table*}

A technically useful way to frame ``where the field stands'' is to anchor the discussion in \emph{representative decision-grade constraints} that already delimit large regions of model space.
For example: composition-dependent free fall in space is constrained at the $|\eta|\sim 10^{-15}$ level (MICROSCOPE) \cite{MICROSCOPE2022}; direct searches for spin-independent WIMP--nucleon scattering have reached $\sigma_{\rm SI}\simeq 2.2\times 10^{-48}\ \mathrm{cm^2}$ at $m_\chi\simeq 40~\mathrm{GeV}$ (LZ, $4.2$ tonne-year exposure) \cite{LZ2024}; direct kinematic neutrino-mass measurements constrain the effective electron-(anti)neutrino mass to $m_\beta<0.45~\mathrm{eV}$ (KATRIN, 90\% C.L.) \cite{KATRIN2025}; and laboratory searches for new CP violation reach $|d_e|<4.1\times 10^{-30}\ e\cdot\mathrm{cm}$ (90\% C.L.) \cite{Roussy2023eEDM}. These numbers are not merely ``best limits'': they already force any viable extension of the baseline stack to satisfy correlated constraints across particle physics, gravity, and cosmology, and they quantify the systematics regimes that now dominate progress.

\begin{figure}[h!]
\centering
\begin{tikzpicture}[
  box/.style={draw, rounded corners, align=center, inner sep=6pt},
  arr/.style={-Latex, thick},
  node distance=1.6em and 2.2em
]
\node[box] (inst) {Instrument /\\ selection model};
\node[box, right=of inst] (like) {$\mathcal{L}(\text{data}\mid \text{observable},\nu)$};
\node[box, right=of like] (eft) {EFT /\\ parameter layer};
\node[box, right=of eft] (uv) {UV model\\ classes};

\draw[arr] (inst) -- (like);
\draw[arr] (like) -- (eft);
\draw[arr] (eft) -- (uv);

\node[box, below=of like] (sys) {Correlated systematics\\ \& nuisance model $\nu$};
\draw[arr] (sys) -- (like);

\node[box, below=of eft] (xchk) {Redundancy /\\ cross-checks};
\draw[arr] (xchk) -- (eft);
\end{tikzpicture}
\caption{Schematic ``translation chain`` corresponding to Eq.~(\ref{eq:approach}): instrument and selection effects enter through an explicit likelihood with nuisance parameters, which maps to EFT coefficients and then to UV model classes.}
\label{fig:translation_chain}
\end{figure}
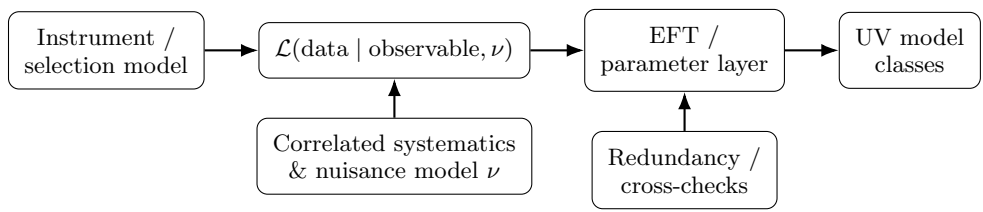

The important methodological shift relative to earlier eras is that many frontier questions are no longer limited by raw counting statistics.
They are limited by (i) correlated instrumental systematics and calibration hierarchies, (ii) degeneracies with astrophysical or nuclear modeling,
and (iii) theory uncertainties that enter the likelihood as structured nuisance parameters rather than as after-the-fact error bars.
Consequently, the review emphasizes an explicit translation chain (see Figure~\ref{fig:translation_chain})
\begin{equation}
\label{eq:approach}
\text{instrument/selection model} \;\to\; \mathcal{L}(\text{data}\mid \text{observable},\nu)\;\to\; \{\text{EFT parameters}\}\;\to\;\{\text{UV model classes}\},
\end{equation}
together with the requirement of \emph{redundancy}: at least two conceptually independent observables constraining the same underlying parameter combinations.  Effective field theory (EFT) is the common language enabling this translation across frontiers: for heavy new physics it is encoded in SMEFT Wilson coefficients \cite{BrivioTrott2019,Grzadkowski2010Warsaw}, while for gravity and cosmology it is encoded in controlled expansions around tested regimes (post-Newtonian expansions, higher-curvature EFTs, and cosmological perturbation EFTs). Related consistency programs---unitarity, analyticity, and positivity constraints on scattering amplitudes---further restrict the UV-completable region of low-energy parameter space and can be implemented as quantitative theory priors in global fits \cite{Adams2006Positivity}.

A central point of this review is that \emph{physics is not platform-defined}. Whether a result is obtained on the ground, in space, or via astronomy matters only insofar as the platform changes the accessible observables and the dominant error terms. Accordingly, a second organizing principle is that ``new approaches'' are justified only insofar as they access \emph{new observables} or change the dominant noise scalings in a way that enables qualitatively different tests \cite{Turyshev2007IJMPD}. The experimental and observational landscape naturally separates into three categories:
\begin{enumerate}[leftmargin=2.2em]
\item \emph{Ground-based controlled experiments} (colliders; intensity and precision frontiers; laboratory gravitation; clocks and atom interferometers),
where the defining advantage is a controllable environment and the ability to engineer null channels and configuration reversals.
The limiting factors are typically thermal/mechanical drifts, electromagnetic backgrounds, and theory systematics (hadronic/nuclear inputs) that must be propagated end-to-end.

\item \emph{Astronomical observations} (CMB and large-scale structure; time-domain transients; strong-field compact objects; neutrinos and gravitational waves),
where the defining advantage is access to extreme baselines, energies, and curvatures unavailable in the laboratory.
The limiting factors are source/foreground modeling, selection functions, and population inference; therefore astronomical tests become decisive primarily when multiple messengers
or multiple observables constrain the same deformation of the baseline theory framework.

\item \emph{In-situ space experiments} (active probes using controlled payloads in space: redshift tests with clocks, drag-free inertial references, precision tracking and ranging,
equivalence-principle tests, long-baseline interferometry), where the defining advantage is the ability to vary a small set of control parameters that enter sensitivity scalings: long free-fall interrogation time $T$, large baselines $L$, and large well-modeled potential differences $\Delta U$. Space does not automatically improve measurement quality; it reshapes the noise budget, so the correct figure of merit is the \emph{total} propagated uncertainty after orbit determination, time/frequency transfer, spacecraft self-gravity, thermal gradients, charging, and control-loop couplings are included.
\end{enumerate}

In this review we emphasize that multiple experimental modalities provide comparably sharp ``lever arms'' on UV parameters once translated into a common EFT/likelihood language. For high-energy colliders, the relevant scaling is energy growth of EFT-induced deviations: for a generic dimension-6 effect, a fractional deviation in a differential rate scales schematically as $\delta \sigma/\sigma \sim c \, (E^2/\Lambda^2)$ in the kinematic regime where $E$ is the characteristic momentum transfer and $c$ is an $\mathcal{O}(1)$ Wilson coefficient. With a total fractional uncertainty $\Delta$ (statistical $\propto N^{-1/2}$ plus theory and detector systematics), the implied reach is $\Lambda \gtrsim E/\sqrt{\Delta}$, and the decision-grade question is whether theory+detector systematics can be reduced to (or below) the statistical floor in the relevant EFT-sensitive tails.

For astronomical observations, the scaling is often set by cosmic variance plus calibration/foreground systematics: e.g. for CMB spectra, $\mathrm{Var}(C_\ell) \simeq \frac{2}{(2\ell+1)f_{\rm sky}}\,(C_\ell+N_\ell)^2$, while for large-scale structure and weak lensing, selection functions, baryonic feedback, and photometric calibration act as nuisance-parameter subspaces that must be closed by cross-survey, cross-probe consistency tests. For rare-event laboratories, exposure improvements are only decision-grade when backgrounds and nuclear/hadronic inputs are propagated with a complete, correlated uncertainty model, so that null results and anomalies can be compared consistently across targets and channels.

We treat coherent sensors (clocks, atom interferometers, and related devices sometimes grouped under ``quantum sensors'') in the same physics-first way: their relevance is determined by which fundamental parameters they constrain and by which noise scalings they modify. For clocks, the clean mapping is between fractional frequency comparisons and gravitational potential or fundamental-constant variations: $\Delta f/f=\Delta U/c^2$ for redshift physics, and $\delta(\nu_1/\nu_2)/(\nu_1/\nu_2)=\sum_a \Delta K_a\,\delta X_a/X_a$ for variations of dimensionless parameters $X_a$ (e.g.\ $\alpha$, $\mu$), with sensitivity set by stability, accuracy, and the time/frequency transfer link. For atom interferometers, the relevant scaling is phase accumulation $\Delta\phi \simeq k_{\rm eff} a T^2$ and hence $\delta a\propto 1/(k_{\rm eff}T^2\sqrt{N})$, so microgravity or long baselines are enabling \emph{only} when platform acceleration/rotation noise and wavefront/self-gravity systematics remain below the targeted differential-acceleration channel. Throughout, we therefore separate cases where these instruments provide primarily incremental improvements on existing null tests from cases where they enable qualitatively new observables (long-coherence searches for ultralight fields, low-frequency inertial channels, long-baseline redshift modulation).

The remainder of the paper is organized as follows.
Section~\ref{sec:FunPhys} formalizes the operational definition and introduces a quantitative baseline theory classification by probed scales and couplings. Section~\ref{sec:theory} summarizes the SM+GR+$\Lambda$CDM baseline in the regimes where it is overconstrained, emphasizing the parameter-precision status and the structure of controlled expansions. Section~\ref{sec:open-problems} reformulates the empirically grounded open problems as specific observables with indicative decision-level targets.
Section~\ref{sec:theory-frameworks} reviews theory tools that make the observable-to-parameter map quantitative (EFTs, consistency/positivity constraints, global inference with explicit nuisance models, and nonperturbative QCD/nuclear inputs).
Section~\ref{sec:exp-frontiers} surveys the experimental and observational frontiers (colliders, intensity/precision probes,
neutrinos, dark matter searches, cosmology, and gravitational waves) in a common quantitative language. Section~\ref{sec:ground-space} treats ground experiments, passive astronomical probes, and in-situ space tests as distinct categories with distinct noise scalings and systematics.
Section~\ref{sec:quantum-tech} evaluates clocks, atom interferometers, and related coherent sensors strictly by the fundamental parameters they constrain and by the size of realistically achievable improvements.  Section~\ref{sec:sensitivity} collects decision-level sensitivity targets and cross-check requirements; only after this technical review, Section~\ref{sec:roadmap} presents a staged roadmap organized by physics decision points rather than by project advocacy. Section~\ref{sec:concl} concludes with a synthesis of discovery potential and realistic prioritization principles.
Appendix~\ref{app:landscapes} discussion various experiment classes including their dominant systematics and  parameter combinations.

% =========================================
\section{Fundamental physics today}
\label{sec:FunPhys}

Fundamental physics is the study of \emph{universal} laws and degrees of freedom whose domains cut across systems and scales. In 2025, the subject is not a single subfield but an interconnected set of problems tied together by:
\begin{enumerate}[leftmargin=2.2em]
\item \textit{The baseline theory framework}: the SM (a renormalizable non-abelian gauge theory with spontaneous symmetry breaking), GR (a classical field theory of spacetime geometry), and \LCDM\ (the minimal phenomenological model of cosmology with cold dark matter and a cosmological constant).
\item \textit{Known incompletions}: neutrino masses and mixing, dark matter, baryon asymmetry, dark energy, quantum gravity, and the hierarchy/cosmological constant problems (Figure~\ref{fig:stack_overview}).
\item \textit{Methodological unity}: the use of symmetry, locality, unitarity, and EFT as organizing principles; and the use of precision metrology, rare processes, multi-messenger astronomy, and controlled many-body theory as discovery engines.
\end{enumerate}

As a result, the entire field is tied together by three requirements that are largely independent of experimental platform.
First, it targets \emph{universal} laws: symmetries, conserved (or weakly violated) quantum numbers,
and spacetime dynamics that apply across systems.
Second, it is organized by a \emph{hierarchy of scales} (energy, momentum transfer, distance, time,
curvature), which makes controlled effective descriptions possible and makes it meaningful to compare
constraints obtained in very different regimes.
Third, it requires an explicit \emph{translation layer} from measured observables to a parameterized
description of deviations (typically EFT coefficients and symmetry structures), so that disparate
measurements can be combined and UV model classes can be falsified rather than merely ``constrained.``

\subsection{A quantitative perspective}
A useful way to categorize fundamental physics is by the energy/momentum scales and curvature scales probed:
\begin{itemize}[leftmargin=2.2em]
\item \emph{Microscopic/accelerator}: \SI{1}{\GeV}--\SI{10}{\TeV} (and possibly beyond), testing new particles and interactions directly.
\item \emph{Precision/rare}: effective scales \(\Lambda\sim\SI{1}{\TeV}\) to \(10^4\ \TeV\) through dimension-6 operators in low-energy observables (EDMs, CLFV, rare meson decays).
\item \emph{Cosmic/astronomical}: horizon-scale and nonlinear structure probes, sensitive to feeble couplings and ultra-light fields (masses down to \(m\sim10^{-22}\ \eV\) and below), plus early-universe physics at energy densities up to \(V^{1/4}\sim 10^{16}\ \GeV\) if primordial tensors are detected.
\item \emph{Gravitational}: curvature, equivalence principle (EP), inverse-square law (ISL), gravitational waves (GWs), and strong-field tests (black holes, neutron stars).
\item \emph{Quantum-information/AMO}: engineered quantum systems and sensors (optical clocks, atom interferometers, quantum-limited amplifiers) that turn coherence and entanglement into probes of tiny energy shifts and phase accumulations.
\end{itemize}

\subsection{Methodological unity across scales}
\label{sec:method_unity}

Across particle, gravitational, and cosmological domains, the common technical structure is not the platform but the inference problem: a forward model for an observable, a likelihood (including correlated systematics), and a parameterization that admits controlled interpretation in terms of symmetries and effective operators. In practice this implies three requirements:  First, deviations should be expressed in languages that can be matched to UV completions (e.g.\ EFT Wilson coefficients, symmetry-breaking spurions, or parameterized modifications of generation/propagation laws). Second, the limiting uncertainties are frequently systematic rather than statistical; therefore calibration hierarchies, nuisance models, and cross-check logic are part of the \emph{physics} content, not auxiliary engineering. Third, decisive progress typically requires degeneracy breaking: two (or more) conceptually independent measurements constraining the \emph{same} parameter combination, so that agreement or tension is interpretable without relying on a single
instrument model.

% =========================================
\section{The baseline theory framework: SM + GR + \LCDM}
\label{sec:theory}

\subsection{Standard Model: structure, parameters, and precision status}
The SM is defined by gauge group \(SU(3)_c\times SU(2)_L\times U(1)_Y\) with fermion content in three generations, a scalar Higgs doublet \(H\), and renormalizable interactions:
\begin{equation}
\Lag_{\mathrm{SM}} = -\frac14 \sum_{a} G^a_{\mu\nu}G^{a\,\mu\nu}
-\frac14 \sum_{i} W^i_{\mu\nu}W^{i\,\mu\nu}
-\frac14 B_{\mu\nu}B^{\mu\nu}
+ \sum_{\psi} \bar\psi i\slashed{D}\psi
+ |D_\mu H|^2 - V(H) + \Lag_{\mathrm{Yuk}},
\end{equation}
with
\begin{equation}
V(H)= -\mu^2 H^\dagger H + \lambda (H^\dagger H)^2,
\qquad \vEW=\sqrt{\mu^2/\lambda}\simeq \SI{246}{\GeV}.
\end{equation}
At the renormalizable level (without neutrino masses) the SM has \(19\) parameters (gauge couplings, Higgs potential parameters, Yukawa eigenvalues and CKM parameters, and \(\theta_{\mathrm{QCD}}\)). Neutrino masses require extension (e.g.\ Weinberg operator at dimension 5 or additional fermions \cite{Weinberg1979}).

\paragraph{Dimension-5 neutrino mass operator (Weinberg).}
At the EFT level the leading gauge-invariant operator that generates Majorana masses is
\begin{equation}
\Lag_{5}=\frac{c_{ij}}{\Lambda}\,\big(\bar L_i^{\,c}\,\tilde H^\ast\big)\,\big(\tilde H^\dagger L_j\big)+\mathrm{h.c.},
\qquad \tilde H \equiv i\sigma_2 H^\ast ,
\label{eq:weinberg}
\end{equation}
which yields
\begin{equation}
(m_\nu)_{ij}\simeq c_{ij}\,\frac{\vEW^2}{\Lambda}.
\end{equation}
For a representative light-neutrino mass scale \(m_\nu\sim 0.05\,\eV\) and \(c_{ij}\sim \mathcal{O}(1)\), one infers
\(\Lambda\sim \vEW^2/m_\nu \sim 10^{14\text{--}15}\,\GeV\). This quantitative lever arm is why neutrino masses are often viewed as an indirect probe of very high-scale physics even when the responsible states are kinematically inaccessible.

\paragraph{Quantitative success.}
The SM has been validated across many orders of magnitude. A representative example of its predictive power is the agreement between electroweak loop-corrected observables and data. In QED, the electron anomalous magnetic moment \(a_e\) is predicted and measured at extraordinary precision; in the muon sector the situation is currently more subtle due to hadronic uncertainties and the muon \(g-2\) anomaly (see Sec.\ \ref{sec:gminus2}).

\paragraph{Where the SM is incomplete (empirical).}
Neutrino masses and mixings, dark matter, and the baryon asymmetry require physics beyond the minimal SM. These are not ``aesthetics''; they are direct conflicts between SM predictions and observations.

\paragraph{Where the SM is incomplete (conceptual).}
The Higgs mass parameter and vacuum energy are radiatively sensitive to UV scales. If one regards the SM as an EFT up to \(\Lambda\), then schematically
\begin{equation}
\delta m_H^2 \sim \frac{\kappa}{16\pi^2}\Lambda^2, \qquad
\rho_{\rm vac}\sim \Order(\Lambda^4),
\end{equation}
to be compared with \(m_H\simeq \SI{125}{\GeV}\) and \(\rho_\Lambda\simeq (2.3\ \mathrm{meV})^4\). The degree to which these are ``problems'' depends on assumptions about UV completion; nevertheless they motivate symmetry mechanisms (supersymmetry, compositeness), environmental selection, or new dynamics.

\subsection{General Relativity: EFT view, classical tests, and open regimes}

GR is defined by the Einstein-Hilbert action plus matter:
\begin{equation}
S = \int \dd^4x \sqrt{-g}\left[\frac{\mpl^2}{2}R - \Lambda + \Lag_{\rm matter}\right].
\end{equation}
As an EFT, GR predicts an infinite tower of higher-curvature operators suppressed by some scale \(M\) (often \(M\sim \mpl\) unless new physics enters earlier). In weak fields, deviations are parameterized via parametrized post-Newtonian (PPN) parameters; in the lab and solar system, constraints are extremely tight \cite{Will2014LRR,Turyshev2025PhRvD}.

\paragraph{Precision EP and ISL tests.}
The MICROSCOPE satellite final results constrain the E\"otv\"os parameter at the \(\sim10^{-15}\) level for Ti/Pt test masses \cite{MICROSCOPE2022}. Short-range torsion-balance tests constrain Yukawa deviations from Newtonian gravity down to separations \(\sim 50\,\mu\)m \cite{EotWashISL2020}.

\paragraph{Strong-field and dynamical tests.}
The GW detections by ground-based interferometers probe GR in highly dynamical, nonlinear regimes; their consistency tests constrain deviations in wave propagation, polarizations, and ringdown structure (Sec.\ \ref{sec:GWs}).

\subsection{\LCDM: parameters, tensions, and the role of late- and early-universe probes}
The minimal cosmological model assumes a spatially-flat universe with:
\begin{equation}
H^2(a)=H_0^2\left[\Omega_r a^{-4}+\Omega_m a^{-3}+\Omega_\Lambda\right],
\end{equation}
with initial perturbations characterized by scalar amplitude \(A_s\), tilt \(n_s\), and optical depth \(\tau\). Observations of the CMB and large-scale structure constrain these parameters at percent or sub-percent levels.

As of 2025, two major areas are actively discussed:
\begin{enumerate}[leftmargin=2.2em]
\item \emph{The Hubble constant \(H_0\).}
Early-universe inference in baseline \LCDM\ from the Planck 2018 data gives
\(\Hn = 67.36\pm0.54~\kmsMpc\) \cite{Planck2018Cosmo}.
Late-universe distance-ladder determinations based on Cepheids and SNe~Ia remain higher; a JWST cross-check of Cepheid photometry in calibrator hosts
supports \(H_0 \simeq 72.6\pm 2.0~\kmsMpc\) \cite{Riess2024JWSTValidate}, and a subsequent SH0ES analysis using JWST/HST Cepheids in NGC~4258 reports
\(H_0 = 73.49\pm 0.93~\kmsMpc\) \cite{Riess2025SH0ES}.
Independent late-time approaches based on the TRGB calibration yield values closer to \(\sim 69\text{--}70~\kmsMpc\) with different systematics budgets \cite{Freedman2021H0}.
Time-delay strong-lensing cosmography provides an additional largely independent route; the H0LiCOW analysis reports \(H_0\) near \(73~\kmsMpc\) under standard lens-model assumptions \cite{H0LiCOW13}, while broader lens-model freedom motivates careful degeneracy/systematics accounting \cite{Treu2022TimeDelayReview}.
Standard sirens (beginning with GW170817) provide a calibration-independent approach that is currently statistics-limited but methodologically complementary \cite{Abbott2017StandardSirenH0}.

\item \emph{The clustering amplitude \(S_8\equiv \sigma_8\sqrt{\Omega_m/0.3}\).}
Weak lensing + galaxy clustering constraints continue to test the internal consistency of late-time structure formation.
The KiDS Legacy analysis reports \(S_8 \simeq 0.815^{+0.016}_{-0.020}\) and finds consistency with Planck within current uncertainties \cite{KiDSLegacy2025}.
Interpretation at the percent level requires explicit control of shear calibration, photometric-redshift calibration, intrinsic alignments,
and nonlinear/baryonic modeling, and should be assessed via cross-survey closure tests and shared nuisance-parameter models.
\end{enumerate}
Interpreting these tensions requires careful control of selection effects, calibration, nonlinear baryonic systematics, and internal consistency tests; decisive progress demands multiple \emph{redundant} methods.

% ========================================
\section{Open problems: concrete targets and scales}
\label{sec:open-problems}

A compact list of empirically grounded open problems is given in Table \ref{tab:openproblems}. Each line can be turned into a quantitative science case: it defines an observable, an EFT or model mapping, a target sensitivity, and cross-checks.

\begin{table*}[t!]
\caption{Empirical open problems and representative quantitative targets. Numbers are order-of-magnitude ``decision thresholds'' rather than single-experiment requirements.}
\label{tab:openproblems}
\centering
\renewcommand{\arraystretch}{1.08}
\setlength{\tabcolsep}{2pt}
\begin{tabular}{@{}llll@{}}
\hline\hline
\tcell{0.12\textwidth}{Problem} &
\tcell{0.27\textwidth}{Observable(s)} &
\tcell{0.24\textwidth}{Current state (indicative)} &
\tcell{0.36\textwidth}{Decisive target (indicative)} \\
\hline

\tcell{0.12\textwidth}{Dark matter} &
\tcell{0.27\textwidth}{Direct detection (\(\sigSI(m_\chi)\)), axion couplings \(g_{a\gamma\gamma}\), indirect signals, structure formation} &
\tcell{0.24\textwidth}{\(\sigSI \sim 2.2\times 10^{-48}\ \cm^2\) at \(m_\chi\simeq \SI{40}{GeV}\) (LZ) \cite{LZ2024}; broad axion parameter space partially covered} &
\tcell{0.36\textwidth}{Approach neutrino-floor-limited SI detection for WIMPs; cover QCD axion bands over decades in mass; require multi-channel confirmation} \\

\tcell{0.12\textwidth}{Neutrino masses} &
\tcell{0.27\textwidth}{Absolute mass scale \(m_\beta\), \(\Sigma m_\nu\), \(0\nu\beta\beta\) half-life \(T_{1/2}^{0\nu}\)} &
\tcell{0.24\textwidth}{\(m_\beta < \SI{0.45}{eV}\) (KATRIN, 2025) \cite{KATRIN2025}; \(0\nu\beta\beta\) not observed} &
\tcell{0.36\textwidth}{\(m_{\beta\beta}\sim 10\text{--}20\ \mathrm{meV}\) sensitivity (cover inverted ordering), with controlled NME uncertainties} \\

\tcell{0.12\textwidth}{Baryogenesis} &
\tcell{0.27\textwidth}{EDMs, collider Higgs/phase transition observables, GW backgrounds, flavor/CPV} &
\tcell{0.24\textwidth}{\(|d_e|<4.1\times10^{-30}\ e\cdot\cm\) \cite{Andreev2018ACME,Roussy2023eEDM}; \(|d_n|<1.8\times10^{-26}\ e\cdot\cm\) \cite{Abel2020PRL}} &
\tcell{0.36\textwidth}{1--2 orders improvement in EDMs; detect/limit stochastic GW from first-order PTs; measure Higgs self-coupling to \(\sim 10\%\) with future colliders} \\

\tcell{0.125\textwidth}{Baryon number violation} &
\tcell{0.27\textwidth}{Proton decay \(\tau_p\), \(n\)--\(\bar n\) oscillations \(\tau_{n\bar n}\)} &
\tcell{0.24\textwidth}{\(\tau_p\sim 10^{34}\,\yr\) scale limits; \(\tau_{n\bar n}\) constrained (free and nuclear)} &
\tcell{0.36\textwidth}{Push \(\tau_p\) by \(10\times\)--\(100\times\); push \(\tau_{n\bar n}\) by \(10\times\) with controlled matrix elements} \\

\tcell{0.12\textwidth}{Dark energy / gravity} &
\tcell{0.27\textwidth}{\(w(z)\), growth \(f\sigma_8(z)\), lensing, GW sirens, EP/ISL tests} &
\tcell{0.24\textwidth}{DESI/ACT-era precision with model tensions under debate \cite{ACTDR6Extended,Freedman2024TRGB}} &
\tcell{0.36\textwidth}{Percent-level mapping of expansion + growth with systematics control, consistency across probes; improved EP tests to \(10^{-17}\) in space} \\

\tcell{0.12\textwidth}{Quantum gravity regime} &
\tcell{0.27\textwidth}{GW ringdowns, BH imaging, high-energy astroph propagation, table-top quantum tests of gravity} &
\tcell{0.24\textwidth}{GR passes current tests; quantum regime not directly seen} &
\tcell{0.36\textwidth}{Establish new windows: mid-band GW, precision clocks/AI for tiny GR violations, quantum-to-gravity interface tests} \\

\tcell{0.12\textwidth}{Hierarchy / naturalness} &
\tcell{0.27\textwidth}{Higgs couplings, top/Higgs dynamics, new states, rare processes} &
\tcell{0.24\textwidth}{No confirmed new states below multi-TeV; SMEFT constraints tightening \cite{Cepeda2019HiggsHL}} &
\tcell{0.36\textwidth}{Per-mille Higgs coupling program (lepton colliders) + 100 TeV discovery reach; complement with flavor/EDM/CLFV} \\
\hline
\end{tabular}
\end{table*}

\begin{table*}[h!]
\caption{Platform-independent map from fundamental questions to observables, lever arms, and dominant limitations. The ``best matched'' platform indicates where the \emph{irreducible} lever arm is largest, not where execution is easiest.}
\label{tab:obs_platform_map}
\centering
\renewcommand{\arraystretch}{1.08}
\setlength{\tabcolsep}{2pt}
\begin{tabular}{@{}llll@{}}
\hline\hline
\tcell{0.16\textwidth}{Physics question} &
\tcell{0.24\textwidth}{Observable/parameterization} &
\tcell{0.21\textwidth}{Dominant lever arm} &
\tcell{0.36\textwidth}{Dominant limitation/Best-matched platform} \\
\hline

\tcell{0.16\textwidth}{UFF / EP violation} &
\tcell{0.24\textwidth}{$\eta_{AB}\equiv 2(a_A-a_B)/(a_A+a_B)$; SME coefficients} &
\tcell{0.21\textwidth}{Long free-fall; composition contrast} &
\tcell{0.36\textwidth}{Gravity gradients, spacecraft self-gravity, patch/thermal; torsion balances (short range), drag-free space (long $T$)} \\

\tcell{0.16\textwidth}{Gravitational redshift / LPI} &
\tcell{0.24\textwidth}{$\Delta\nu/\nu=(1+\alpha)\Delta U/c^2$ (clock-dependent $\alpha$)} &
\tcell{0.21\textwidth}{$\Delta U/c^2$ modulation; time transfer} &
\tcell{0.36\textwidth}{Link noise, orbit/potential modeling; eccentric-orbit space + high-accuracy clocks} \\

\tcell{0.16\textwidth}{Time variation of constants} &
\tcell{0.24\textwidth}{$\dot{\alpha}/\alpha$, $\dot{\mu}/\mu$ from clock ratios} &
\tcell{0.21\textwidth}{Long time baselines; high sensitivity coefficients} &
\tcell{0.36\textwidth}{Systematics/BBR, correlations; ground clock networks + fiber} \\

\tcell{0.16\textwidth}{Ultralight scalar DM} &
\tcell{0.24\textwidth}{Oscillatory $\delta X/X$ in clock ratios; transient correlations} &
\tcell{0.21\textwidth}{Coherence time; network baselines} &
\tcell{0.36\textwidth}{Synchronization/time transfer; distributed ground/space clock networks} \\

\tcell{0.16\textwidth}{Short-range fifth forces / ISL} &
\tcell{0.24\textwidth}{Yukawa deviations in $V(r)$ at $\mu$m--m} &
\tcell{0.21\textwidth}{Proximity; shielding} &
\tcell{0.36\textwidth}{EM backgrounds, patch forces; ground lab is dominant} \\

\tcell{0.16\textwidth}{mHz gravitational waves} &
\tcell{0.24\textwidth}{Strain $h(f)$ and propagation tests} &
\tcell{0.21\textwidth}{Access to low-$f$ band} &
\tcell{0.36\textwidth}{Acceleration noise, confusion backgrounds; space interferometry} \\

\tcell{0.16\textwidth}{nHz gravitational waves} &
\tcell{0.24\textwidth}{Timing residual correlations} &
\tcell{0.21\textwidth}{Galaxy-scale baselines} &
\tcell{0.36\textwidth}{Pulsar spin/noise models; PTA (astronomical)} \\

\tcell{0.16\textwidth}{WIMP-like DM scattering} &
\tcell{0.24\textwidth}{Recoil spectrum + target scaling} &
\tcell{0.21\textwidth}{Background suppression} &
\tcell{0.36\textwidth}{Radiogenic + neutrino backgrounds; deep underground ground-based detectors} \\

\tcell{0.16\textwidth}{SEP, PPN, $\dot G/G$, Lorentz/SME, fifth-force/Yukawa} &
\tcell{0.24\textwidth}{Earth--Moon range time series; harmonic signatures; global ephemeris residuals} &
\tcell{0.21\textwidth}{Lunar laser ranging / solar-system laser ranging} &
\tcell{0.36\textwidth}{Geophysical + ephemeris modeling, station time-transfer, atmosphere, reflector modeling} \\
\hline
\end{tabular}
\end{table*}

\subsection{From questions to observables to platform requirements}
\label{sec:obs_to_platform}

The organizing principle of this review is platform-independent: a \emph{fundamental-physics question}
is operationally a question about one or more \emph{observables} (and their likelihood), together with
a parameterization that admits controlled interpretation (EFT Wilson coefficients, symmetry-violation
parameters, or well-defined modifications of propagation/generation laws). Experimental platforms
are then selected by which dimensionless \emph{lever arms} they provide and which noise terms dominate.

For most frontiers, the decisive comparison is not ``ground vs.\ space'' in the abstract, but which of
the following scalings is required by the target observable:
(i) long coherent evolution time ($\propto T$ or $\propto T^2$),
(ii) large potential differences ($\Delta U/c^2$) for redshift/LPI tests,
(iii) long baselines ($L$) for phase accumulation and time/frequency transfer,
(iv) access to frequency bands unavailable on the ground (e.g.\ mHz gravitational waves),
(v) background suppression via shielding/overburden and material control.

Table~\ref{tab:obs_platform_map} summarizes representative cases and the corresponding dominant limitations. The later sections should be read as elaborations of this mapping, not as technology-driven narratives.

% ===========================================
\section{Theory frameworks that organize ``fundamental physics''}
\label{sec:theory-frameworks}

This section emphasizes theoretical directions that (i) are broadly applicable across subfields, and (ii) enable quantitative translation from measurements to fundamental parameters.

\subsection{Effective field theories as the lingua franca}
\label{sec:EFTs-map}

Effective field theories (EFT) is the most important \emph{practical} theoretical framework in 2025 fundamental physics.

\paragraph{SMEFT and electroweak precision.}
At energies below new heavy states, the SM is extended by higher-dimension operators \cite{ColladayKostelecky1997CPT,ColladayKostelecky1998SME,Kostelecky2004GravitySME}:
\begin{equation}
\Lag = \Lag_{\mathrm{SM}} + \sum_i \frac{c_i^{(6)}}{\Lambda^2}\mathcal{O}_i^{(6)} + \sum_j \frac{c_j^{(8)}}{\Lambda^4}\mathcal{O}_j^{(8)}+\cdots.
\end{equation}
A percent-level deviation in a Higgs coupling can correspond to
\begin{equation}
\frac{\delta g}{g}\sim \Order\!\left(\frac{v^2}{\Lambda^2}\right)
\quad \Rightarrow \quad
\Lambda \sim \frac{v}{\sqrt{\delta g/g}}
\approx \SI{2.5}{TeV}\left(\frac{0.01}{\delta g/g}\right)^{1/2},
\end{equation}
modulo operator coefficients and loop factors. This illustrates why a per-mille Higgs program can be a probe of \(\Lambda\sim \Order(10)\ \TeV\) even without new particles produced on-shell.

\paragraph{Chiral EFT, nuclear EFT, and neutrino physics.}
For neutrino scattering and \(0\nu\beta\beta\), hadronic and nuclear matrix elements are often the limiting theory systematics. Chiral EFT organizes nuclear forces and currents with controlled power counting; lattice QCD provides input for low-energy constants and nucleon structure. A long-term frontier is \emph{end-to-end uncertainty propagation}: from QCD to nuclei to detector observables.

\paragraph{Gravitational EFT and cosmological EFTs.}
For late-time cosmology, the EFT of dark energy/modified gravity parameterizes perturbations around FRW via functions of time (e.g.\ \(\alpha_K,\alpha_B,\alpha_M,\alpha_T\)), with stability and causality constraints \cite{Gubitosi2013EFTDE,Bloomfield2013EFTDE,BelliniSawicki2014}. For inflation, the EFT of inflation organizes predictions in terms of the Goldstone mode \(\pi\) of broken time translations and its higher-derivative operators (sound speed, non-Gaussianity templates) \cite{Cheung2008EFTInflation}.

\subsection{Amplitudes, unitarity, analyticity, and positivity}
\label{sec:amplitudesPositivity}

A complementary way to organize beyond-the-Standard-Model inference is to treat the low-energy theory as the infrared limit of a unitary, causal, analytic scattering matrix. Under mild assumptions (crossing symmetry, a mass gap, and sufficient boundedness at high energy), forward-limit dispersion
relations imply \emph{positivity} conditions on low-energy amplitude derivatives. For elastic scattering of identical scalars with amplitude \(\mathcal{A}(s,t)\), one obtains schematically
\begin{equation}
\left.\pdv[2]{s}\mathcal{A}(s,0)\right|_{s=0}
= \frac{2}{\pi}\int_{s_{\rm th}}^\infty \frac{\Im \mathcal{A}(s',0)}{s'^3}\,\dd s' \;>\; 0,
\label{eq:positivityForward}
\end{equation}
so any EFT coefficient controlling the \(s^2\) growth of \(\mathcal{A}\) in the forward limit must be
positive (or satisfy a linear inequality once multiple channels and spins are included).

In SMEFT language, these bounds restrict combinations of higher-dimensional Wilson coefficients that enter multi-boson amplitudes (notably dimension-8 operators in vector-boson scattering). Practically, they can be implemented as theory priors or hard constraints in global fits, reducing degeneracies
among operator directions that are otherwise difficult to separate experimentally. In gravity-coupled EFTs, massless \(t\)-channel exchange (photon/graviton) introduces infrared subtleties; the most robust positivity statements apply to gapped sectors or after explicitly subtracting long-range exchange contributions in a way consistent with the measurement strategy.

The deliverable is therefore not a specific UV model, but a set of \emph{consistency inequalities} that carve out the UV-completable region of Wilson-coefficient space. When an experimental ``best fit'' lands outside this region, the tension is informative: it can indicate underestimated systematics, a breakdown of EFT assumptions for the dataset at hand, or the need to enlarge the effective theory (e.g.\ include additional light degrees of freedom).

\subsection{Statistical inference, global fits, and analysis preservation}
\label{sec:inference}

A technical review of fundamental physics is incomplete without addressing \emph{how} constraints are actually extracted.
Across essentially all frontiers, progress is now limited as much by correlated systematics and theoretical nuisance parameters
as by raw counting statistics.  In this regime, ``sensitivity'' is not a single number: it is a statement about an \emph{inference pipeline} that maps data products to parameters in a model space (typically an EFT), with quantified uncertainties.

\paragraph{Generic structure.}
Let \(d\) denote the data, \(\theta\) the parameters of interest (e.g.\ Wilson coefficients, particle properties, cosmological arameters), and \(\nu\) nuisance parameters encoding instrumental and astrophysical/theory systematics. A consistent treatment requires an explicit likelihood \(\mathcal{L}(d\mid\theta,\nu)\) and priors or constraints on \(\nu\):
\begin{equation} 
p(\theta\mid d)\ \propto\ \int \dd \nu\; \mathcal{L}(d\mid\theta,\nu)\,\pi(\theta)\,\pi(\nu).
\label{eq:posterior_generic}
\end{equation}
Frequentist summaries (profile likelihoods, likelihood-ratio tests, CL$_s$ limits) correspond to optimizing or marginalizing over \(\nu\) with different prescriptions.  In the systematics-dominated regime, the \emph{nuisance model} is often the dominant ``theory'' input.

\paragraph{Global-fit logic.}
A common workhorse is a Gaussian approximation for a vector of summary observables \(O\) with covariance \(C\),
\begin{equation}
\chi^2(\theta)\ =\ \big(O-O_{\rm th}(\theta)\big)^{\!\top} C^{-1}\big(O-O_{\rm th}(\theta)\big),
\label{eq:chi2_generic}
\end{equation}
augmented by nuisance parameters when the experimental covariance is not sufficient to represent known systematics.
This is the backbone of electroweak and SMEFT fits, global neutrino analyses, and many cosmological likelihoods.
For discovery-level claims, it is essential that \(C\) include \emph{correlations} across channels and experiments, and that alternative theory treatments (e.g.\ hadronic models, nonlinear prescriptions) be represented by explicit discrete or continuous nuisance degrees of freedom.

\paragraph{Forecasting and decision thresholds.}
Roadmaps often rely on Fisher forecasts.  For parameters \(\theta_i\),
\begin{equation}
F_{ij}\ =\ -\left\langle\frac{\partial^2 \ln\mathcal{L}}{\partial \theta_i\,\partial \theta_j}\right\rangle,
\qquad
{\rm cov}(\theta)\ \simeq\ F^{-1}.
\label{eq:fisher}
\end{equation}
These provide useful \emph{relative} scaling with exposure, baseline, and noise (e.g.\ \(\propto 1/\sqrt{\Tobs}\) in stationary Gaussian limits), but can be misleading when (i) the likelihood is non-Gaussian, (ii) the dominant uncertainties are model systematics rather than noise, or (iii) parameter degeneracies are broken only by cross-probe combinations. ``Decision-level'' sensitivity therefore should be defined in terms of an end-to-end analysis that includes the intended nuisance model and explicit cross-checks.

\paragraph{Analysis preservation as an enabling technology.}
Because many targets in this review require \emph{combined} constraints (e.g.\ SMEFT across colliders and flavor, dark-energy inference across BAO+SNe+shear, or modified-gravity tests across GWs and cosmology), the community increasingly needs open likelihoods or equivalent sufficiently rich data products (compressed but lossless summaries, public covariances, validated fast surrogates).  Without these, precision gains in one experiment cannot be reliably propagated into improved constraints on fundamental parameters, and apparent ``tensions'' cannot be adjudicated.

\subsection{Nonperturbative and many-body theory}

A growing fraction of ``precision'' fundamental physics is limited not by experimental statistics but by nonperturbative QCD/nuclear inputs and by how theory systematics are propagated into likelihoods. The technical challenge is to deliver \emph{controlled uncertainty budgets} for hadronic and nuclear matrix elements over the kinematic regimes relevant to experiments.

\paragraph{Lattice QCD for precision observables.}
Key targets include hadronic vacuum polarization (HVP) and hadronic light-by-light (HLbL) contributions to \((g-2)_\mu\), nucleon axial/scalar/tensor charges entering \(\beta\) decay and EDM interpretations, and nucleon form factors relevant for DM scattering in the sub-GeV to multi-GeV regime. To fully exploit a \(\mathcal{O}(10^{-10})\) experimental uncertainty in \(a_\mu\), the HVP contribution must be controlled at the \(\lesssim 0.2\%\) level and HLbL at \(\lesssim 10\%\) (so that each contributes
\(\lesssim 10^{-10}\) in absolute uncertainty). Achieving this requires simultaneous control of finite-volume effects, continuum extrapolation, chiral/physical-mass tuning, and renormalization, together with robust validation against dispersive (\(R\)-ratio) determinations where applicable.

\paragraph{Nuclear many-body theory and electroweak responses.}
Many decisive searches (neutrinoless double beta decay, coherent elastic neutrino--nucleus scattering, WIMP/CE\(\nu\)NS discrimination, nuclear Schiff moments in EDM experiments) require nuclear structure and response functions in regimes where correlations and operator renormalization matter.
For \(0\nu\beta\beta\), current nuclear-matrix-element (NME) spreads at the factor-of-\(\sim 2\) level translate
directly into spreads in inferred \(|m_{\beta\beta}|\); a realistic near-term goal is to reduce inter-method discrepancies to \(\lesssim 30\%\) through benchmark nuclei, uncertainty quantification in chiral EFT currents, and cross-isotope consistency checks.

\paragraph{Bootstrap, dispersive, and quantum-simulation approaches.}
The conformal bootstrap and dispersive methods provide nonperturbative constraints on strongly coupled sectors (e.g.\ composite Higgs/hidden valleys) in terms of operator dimensions and OPE coefficients, and can supply priors for EFT coefficients when a conventional weakly coupled UV completion is absent. In parallel, analog/digital quantum simulators for gauge and many-body dynamics are beginning to address regimes with sign problems (real-time dynamics, dense matter) that are difficult for classical Monte Carlo, but the near-term role is best viewed as \emph{validation and toy-model calibration} rather than a replacement for controlled field-theory calculations.

\subsection{Model-building directions most tightly coupled to data}

The space of beyond-SM models is vast; for a review that aims to connect theory to experimental reach, it is more useful to emphasize \emph{parametric structures} that map directly onto observables and that generate correlated signatures across multiple frontiers.

\paragraph{Portals and minimal dark sectors.}
At energies below the electroweak scale, the leading renormalizable couplings between SM fields and a
hidden sector take the form of a small set of ``portal'' operators,
\begin{equation}
\Lag_{\rm portal} \supset \frac{\varepsilon}{2}F'_{\mu\nu}F^{\mu\nu}
+ \lambda_{HS}\,|H|^2 S^2
+ y_N\,\bar L H N
+ \frac{a}{f_a}\,F_{\mu\nu}\tilde F^{\mu\nu}
+\cdots ,
\end{equation}
corresponding (schematically) to kinetic mixing, Higgs portal, neutrino portal, and axion/ALP couplings \cite{Holdom1986TwoU1}. These structures are ``data-coupled'' because they predict both production channels (colliders/beam dumps), precision shifts (EW and flavor), and cosmological/astrophysical signatures (relic abundance, stellar cooling, structure formation). A practical goal is to express experimental results as likelihoods on the portal parameters \((\varepsilon,m_{A'})\), \((\lambda_{HS},m_S)\), \((y_N,M_N)\), \((f_a,g_{a\gamma\gamma},g_{aN})\),
rather than only as detector-specific thresholds.

\paragraph{Neutrino mass, lepton number, and long-lived particles.}
A minimal organizing relation is the seesaw scaling
\begin{equation}
m_\nu \sim \frac{y^2 v^2}{M}
\simeq 0.05~{\rm eV}\left(\frac{y}{10^{-6}}\right)^2\left(\frac{1~{\rm TeV}}{M}\right),
\end{equation}
which makes explicit why TeV-scale neutrino-mass mechanisms often imply very small mixings and hence long-lived sterile states. This motivates a unified treatment of: (i) oscillation + \(\sum m_\nu\) constraints, (ii) \(0\nu\beta\beta\) searches for lepton-number violation, and (iii) displaced-vertex/beam-dump programs
targeting the small-\(y\) regime.

\paragraph{Naturalness, the Higgs sector, and ``no new particles'' possibilities.}
Post-LHC model building increasingly centers on (i) symmetry structures that suppress dangerous operators, (ii) mechanisms that make new states hard to produce (compressed spectra, neutral naturalness, hidden valleys), and (iii) precision programs that treat the SM itself as the EFT to be tested.
Quantitatively, the decisive observables are Higgs couplings at the per-mille to percent level, the Higgs self-coupling \(\lambda_3\) (from double Higgs production), and loop-sensitive electroweak fits; these are the handles that remain even when direct production is absent.

\paragraph{Early-universe microphysics as a model discriminator.}
A key theoretical direction is to treat the early universe as an ``accelerator'' whose outputs are primordial perturbations, relic abundances, and stochastic backgrounds. Here, model-building that is tightly coupled to data focuses on \(\Delta\Neff\), non-Gaussianity, phase-transition GW spectra, and light-field signatures (isocurvature, oscillating constants), with explicit treatment of degeneracies with astrophysical foregrounds and late-time systematics.

\paragraph{Late-time acceleration and modified gravity.}
Any viable modified-gravity or dark-energy model must simultaneously satisfy local tests (screening), cosmological background expansion, and growth/weak-lensing constraints. The technically useful language is the EFT of dark energy / modified gravity, where observables constrain a small set of time-dependent functions (e.g.\ an effective Planck mass evolution, braiding, sound speed) that can be mapped to concrete models.
The practical emphasis should be on model classes that predict \emph{distinct correlated departures} in growth and lensing (rather than only shifting \(H(z)\)), because pure-background modifications are strongly degenerate with systematics and astrophysical nuisance parameters.

\paragraph{Methodological takeaway.}
Across these directions, a recurring theme is that \emph{global inference} requires a consistent mapping from UV parameters \(\rightarrow\) EFT coefficients \(\rightarrow\) observables, with theory uncertainties treated as nuisance parameters (correlated across datasets where appropriate) rather than as informal error bars.

\subsection{Quantum gravity and UV completion: theory directions and quantitative handles}
\label{sec:qgTheory}

Although a complete quantum theory of gravity is not experimentally established, several theory directions supply concrete, quantitatively testable structures and constraints:

\begin{enumerate}[leftmargin=2.2em]
\item \emph{Gravitational EFT and higher-curvature corrections.}
At energies \(E\ll \mpl\), gravity is an EFT \cite{Donoghue1994GRasEFT} with corrections organized as
\(\Lag \supset c_1 R^2/\Lambda_g^2 + c_2 R_{\mu\nu}R^{\mu\nu}/\Lambda_g^2 + \cdots\).
Observable effects scale as powers of \(E/\Lambda_g\) or \(R/\Lambda_g^2\), motivating precision probes of wave propagation (GW dispersion, birefringence), black-hole dynamics (ringdown spectra), and high-curvature environments.

\item \emph{String theory, holography, and ``swampland'' constraints.}
String theory provides UV completions that naturally produce towers of states, axions/ALPs, and extra gauge sectors. Holography relates certain quantum gravities to QFTs, offering calculational control over strongly coupled dynamics and black-hole information flow. Swampland conjectures attempt to delineate which low-energy EFTs admit quantum-gravity UV completions, generating theory priors that can be confronted with cosmology (inflation and dark energy model space).

\item \emph{Asymptotic safety, loop quantum gravity, causal sets, and emergent-gravity approaches.}
These programs target nonperturbative UV behavior or discrete spacetime microstructure. Their phenomenology is often encoded as modified dispersion, deformed symmetries, running couplings, or specific corrections to black-hole entropy and dynamics, which can be tested (or bounded) via precision astrophysics and laboratory interferometry at the quantum-classical interface.
\end{enumerate}

A pragmatic ``state of the art'' viewpoint is that quantum-gravity theory is increasingly constrained not by direct access to \(\mpl\), but by consistency conditions (unitarity/causality/positivity), by cosmological initial-condition data, and by strong-field observations (GWs and BH imaging) that can falsify broad classes of modifications at much lower effective scales.

\subsubsection{Black-hole information: technical progress and remaining gaps}
\label{sec:bh_information}

The black-hole information problem is not presently constrained by laboratory data, but there has been
quantitative theoretical progress in semiclassical gravity that clarifies \emph{what} must be reproduced by any UV completion.
A key object is the generalized entropy
\begin{equation}
S_{\rm gen} = \frac{\mathrm{Area}(\partial I)}{4G_N} + S_{\rm out},
\end{equation}
where $I$ is an ``island'' region and $S_{\rm out}$ is the von Neumann entropy of quantum fields outside.
Extremizing $S_{\rm gen}$ over candidate surfaces (quantum extremal surfaces, QES) produces Page-curve behavior in controlled settings and makes precise the statement that semiclassical gravity must be supplemented by nontrivial entanglement structure to preserve unitarity \cite{Page1993AverageEntropy,EngelhardtWall2015QES,Penington2020EWr,Almheiri2019BulkEntropy,Almheiri2020Islands}.

From a roadmap perspective, the limitation is that these results are most explicit in low-dimensional models and/or holographic settings, and the open problem is to connect them to realistic four-dimensional black holes and to observable discriminants.  The most plausible empirical interfaces remain indirect:
(i) strong-field consistency tests (GW inspiral/ringdown) as constraints on EFT deformations of GR,
(ii) black-hole imaging constraints on near-horizon structure, and
(iii) high-energy propagation effects that test modified dispersion/birefringence.

\subsection{Quantum foundations and the quantum-to-gravity interface: testable parametrizations}
\label{sec:qfoundations}

A non-negligible fraction of what is labeled ``fundamental physics'' concerns whether standard quantum theory (unitary evolution + Born rule + decoherence) is exact, and how it interfaces with gravitation. Unlike many purely interpretational questions, several well-posed deformations yield \emph{quantitative} targets.

\paragraph{Collapse models (e.g., CSL) as falsifiable deformations.}
A common phenomenological parameterization introduces a stochastic, mass-proportional localization with
rate \(\lambda_{\rm CSL}\) and correlation length \(r_{\rm CSL}\) \cite{GRW1986,Pearle1989CSL,Bassi2013CollapseReview}. The implied momentum diffusion produces heating and decoherence rates that scale with system mass and spatial superposition size. In practice, decisive tests require (1) large-mass matter-wave interferometry or optomechanical superpositions, (2) ultra-low environmental decoherence (cryogenic operation, vibration control, patch-potential mitigation), and (3) careful ``null'' channels to bound mundane noise sources that mimic diffusion.

\paragraph{Gravitationally mediated entanglement as a discriminator of the quantum character of gravity (Newtonian regime).}
Proposals in which two masses become entangled through their mutual gravitational interaction would provide a direct empirical discriminator between models in which gravity can transmit quantum information and models in which it cannot.
If gravity can transmit quantum information, two spatially separated masses prepared in spatial superpositions can
become entangled via their mutual Newtonian interaction \cite{Bose2017SpinEntanglement,MarlettoVedral2017GravEntanglement}. A useful order-of-magnitude discriminator is the
gravitational phase accumulated between branches,
\begin{equation}
\phi_g \sim \frac{G m^2}{\hbar r}\,t,
\label{eq:gravphase}
\end{equation}
so achieving \(\phi_g\sim 1\) requires interrogation time
\begin{equation}
t \sim \frac{\hbar r}{G m^2}
\simeq 1.6~{\rm s}\,\Big(\frac{r}{1~\mu{\rm m}}\Big)\Big(\frac{10^{-15}{\rm ~kg}}{m}\Big)^2 .
\label{eq:gravent_time}
\end{equation}
This shows why the target parameter region is ``mesoscopic'': \(m\sim 10^{-15}\)--\(10^{-14}\,{\rm kg}\),
\(r\sim \mu{\rm m}\), and \(t\sim 10^{-2}\)--\(10^{0}\,{\rm s}\), with the dominant challenges being electromagnetic backgrounds (Casimir/patch forces), motional noise, and state preparation/readout. Space platforms can be enabling by providing long free-fall and low-vibration environments, while ground platforms can leverage cryogenics and shielding.

\paragraph{Why this belongs in a 2025 fundamental-physics survey.} Even null results are informative because they bound classes of non-unitary or semiclassical-gravity models that would otherwise remain empirically unconstrained. These tests also connect directly to quantum sensor technology (clocks, interferometers, optomechanics), so they are naturally coupled to the metrology-driven frontier emphasized in Secs.~\ref{sec:exp-frontiers}--\ref{sec:quantum-tech}.

\section{Experimental and observational frontiers}
\label{sec:exp-frontiers}

\subsection{Collider and high-energy frontiers}
\subsubsection{Higgs sector: couplings, width, and self-coupling}
The Higgs is both a portal to new physics and a precision laboratory for symmetry breaking.

\paragraph{Coupling program.}
At the LHC/HL-LHC, Higgs coupling modifiers are often expressed in the \(\kappa\)-framework or via SMEFT global fits. The HL-LHC program aims for percent-level precision on several couplings, with theory systematics increasingly dominant in some channels \cite{Cepeda2019HiggsHL}.

\paragraph{Higgs self-coupling \(\lambda_3\).}
Double-Higgs production probes the trilinear coupling. Projections indicate that HL-LHC precision on \(\kappa_\lambda\equiv \lambda_3/\lambda_3^{\rm SM}\) remains at the \(\sim 50\%\) level  \cite{Monti2025HiggsSelfCoupling}. A decisive test of many electroweak baryogenesis scenarios typically requires substantially better (\(\sim 10\%\)) determination, motivating future colliders.

\paragraph{Energy-scale translation.}
An observed deviation in Higgs couplings can be mapped to EFT scales as in Sec.~\ref{sec:EFTs-map}; however, loop-induced couplings and strongly coupled UV completions can modify this mapping. Therefore \emph{correlated} deviations across multiple processes (including diboson and top/Higgs) are essential.

\subsubsection{Future colliders: discovery reach vs precision}

A technical roadmap requires separating \emph{deliverables} from \emph{machine choices}. The collider landscape is best organized by the distinct physics handles it provides: (i) model-independent Higgs and electroweak measurements at lepton colliders near the Higgsstrahlung and $t\bar t$ thresholds; (ii) direct mass reach, high-energy tails, and multi-boson final states at a $\mathcal{O}(100~\mathrm{TeV})$ hadron collider; and (iii) the option of multi-TeV lepton collisions (linear and/or muon-based) that combine a cleaner initial state with large $\sqrt{s}$.

\paragraph{Lepton ``Higgs factories'' ($e^+e^-$).}
Running near $\sqrt{s}\simeq 240$--$250~\mathrm{GeV}$ targets $e^+e^-\to ZH$ with the recoil-mass technique, enabling
a model-independent extraction of $\sigma(ZH)$ and thus $g_{HZZ}$. A staged program that includes additional energies
(e.g.\ near the $WW$ threshold, $t\bar t$ threshold, and/or above) sharpens the global Higgs-coupling fit, improves width
constraints, and increases sensitivity to top Yukawa effects (directly at higher energy, or indirectly via loop-level
observables), while also delivering high-statistics electroweak precision observables that close flat directions in EFT fits.
The key deliverable is therefore the \emph{global-fit reach} (SMEFT/HEFT) under stated theoretical systematics, not the
particular ring/linac implementation \cite{FCCPhysicsOpportunities2019,CEPC_CDR_PhysDet_2018,BaerILCTDR2013,CLICSummary2018}.

\paragraph{A $\sim 100~\mathrm{TeV}$ $pp$ collider.}
The principal handle is direct production up to multi-TeV masses and a large kinematic lever arm in high-$p_T$ tails
(including vector-boson scattering), which tests energy-growing operator effects and disentangles degeneracies that remain
at the LHC. Even absent discoveries, the program expands the set of rare Higgs and multi-Higgs channels, strengthens constraints on UV completions through both resonance searches and precision tails, and provides the highest-energy environment for stress-testing EFT validity assumptions \cite{FCCPhysicsOpportunities2019}.

\paragraph{Multi-TeV lepton collider options.}
A multi-TeV lepton collider (linear and/or muon-based) offers a cleaner initial state than hadron machines at comparable $\sqrt{s}$, sharpening sensitivity to contact interactions, resonant new states, and energy-growing scattering amplitudes. The decision hinges on whether the required luminosity can be delivered with backgrounds/systematics controlled at the level demanded by precision fits, and how the resulting reach compares (in global analyses) to a staged $e^+e^-$ program plus a $100~\mathrm{TeV}$ $pp$ collider \cite{CLICSummary2018,DelahayeMuonColliders2019}.

The decision metric should therefore be expressed in terms of (i) global-fit reach on well-defined operator sets under stated theory
uncertainties and (ii) direct mass reach for benchmark UV completions, rather than qualitative ``discovery potential'' slogans.

\subsection{Precision electroweak and low-energy probes}
\label{sec:precisionEW}

A complementary discovery strategy to direct production is to measure SM parameters and symmetry-violating observables at momentum transfers \(Q\ll m_Z\). At these scales the relevant new-physics language is an electroweak-chiral EFT or, above \(m_W\), the SMEFT matched onto semileptonic four-fermion operators. 

\subsubsection{The $W$-boson mass and global electroweak consistency}
The $W$ mass is a loop-level precision observable sensitive to radiative corrections from the top/Higgs sector and to BSM contributions that can often be parameterized by oblique corrections. As of the November~2025 PDG update, the quoted world average (excluding the discrepant CDF~II Run~II 2022 result)
is\footnote{The PDG world average currently excludes the CDF~II 2022 result and also does not yet include the updated ATLAS~2024 and first CMS~2025 results.}
\begin{equation}
m_W = 80.3692 \pm 0.0133\ \mathrm{GeV},
\end{equation}
whereas CDF~II reported \(m_W=80.4335\pm0.0094\ \mathrm{GeV}\) \cite{CDFWmass2022}. Newer LHC determinations include ATLAS~2024 \(m_W=80.3665\pm0.0159\ \mathrm{GeV}\) \cite{ATLASWmass2024} and CMS~2025 \(m_W=80.3602\pm0.0099\ \mathrm{GeV}\) \cite{CMSWmass2025}, both close to the quoted world average.  A Standard-Model electroweak fit excluding \(m_W\) implies \(m_W^{\rm SM}=80.356\pm0.006\ \mathrm{GeV}\), so the present tension is dominated by the CDF~II result rather than by a coherent multi-experiment shift \cite{PDGWmass2025}.

\paragraph{Oblique-parameter scale.}
In terms of the Peskin--Takeuchi parameters, a rough scaling (for \(S=U=0\)) is
\begin{equation}
\frac{\delta m_W}{m_W}\simeq \frac{\alpha\,c_W^2}{2(c_W^2-s_W^2)}\,T,
\end{equation}
so that a \(+77\ \mathrm{MeV}\) upward shift relative to the SM fit corresponds to \(T\sim 0.17\) (order-of-magnitude). Any such shift must remain consistent with \(Z\)-pole and other electroweak precision observables, implying correlated scrutiny
of \(\sin^2\theta_{\mathrm{eff}}\), \(m_t\), \(m_h\), and radiative-correction systematics.

\paragraph{What would be decisive next.}
On the experimental side, decisive progress requires percent-level control of recoil modeling, lepton momentum calibration, QED/EW radiative corrections, and PDF systematics (including correlated treatments between experiments), plus additional independent LHC-era measurements with \(\lesssim 10\ \mathrm{MeV}\) total uncertainty. On the theory side, global electroweak fits with transparent correlation models (including theory errors) and explicit BSM interpretations in terms of \((S,T,U)\) and SMEFT coefficients are needed to determine whether the landscape points to a coherent BSM explanation or to an experiment-specific outlier.

A representative parity-violating (PV) basis is
\begin{equation}
\Lag_{\rm PV}=\frac{G_F}{\sqrt{2}}\sum_{q}
\Big[C_{1q}\,(\bar e\gamma_\mu\gamma_5 e)(\bar q\gamma^\mu q)
+ C_{2q}\,(\bar e\gamma_\mu e)(\bar q\gamma^\mu\gamma_5 q)\Big]+\cdots ,
\label{eq:Lpv}
\end{equation}
where the SM predicts the coefficients \(C_{1q},C_{2q}\) including electroweak radiative corrections, and new physics shifts them by \(\delta C\sim (v^2/\Lambda^2)\times\mathcal{O}(c)\) upon matching.

\paragraph{PV electron scattering and the running of \(\sin^2\theta_W\).}
For elastic scattering at four-momentum transfer \(Q^2\), the PV asymmetry scales as
\begin{equation}
A_{\rm PV}\equiv\frac{\sigma_R-\sigma_L}{\sigma_R+\sigma_L}
\simeq \frac{G_F\,Q^2}{4\sqrt{2}\pi\alphaem}\,Q_W^{\rm eff}(Q^2),
\label{eq:Apv}
\end{equation}
so improving \(\delta A_{\rm PV}/A_{\rm PV}\) maps directly into improvements on the effective weak charge and hence the running of \(\sin^2\theta_W(Q)\). Because \(A_{\rm PV}\propto Q^2\), experiments optimize sensitivity by pushing to the highest feasible \(Q^2\) while maintaining control of helicity-correlated systematics (beam polarization, backgrounds, radiative tails).

\paragraph{Atomic parity violation and weak charges.}
In heavy atoms and ions, PV effects probe coherent weak charges \(Q_W\sim -N + Z(1-4\sin^2\theta_W)\) with strong sensitivity enhanced by relativistic \(Z^3\) scaling. The limiting uncertainties are often theoretical: many-body atomic-structure calculations and radiative corrections must be controlled at the \(0.1\%\) level to turn measurements into competitive constraints on semileptonic
operators.

\paragraph{Beta decays and CKM unitarity as EFT tests.}
Superallowed nuclear \(0^+\!\to 0^+\) decays and neutron decay constrain \(|V_{ud}|\) and test first-row unitarity
\(\Delta_{\rm CKM}\equiv 1-(|V_{ud}|^2+|V_{us}|^2+|V_{ub}|^2)\). In SMEFT language, \(\Delta_{\rm CKM}\neq 0\) corresponds to a specific linear combination of semileptonic Wilson coefficients at the scale \(\Lambda\), providing a clean probe of flavor-universal new physics when nuclear and radiative corrections are under control.

\subsection{Flavor, CP, and symmetry tests as ultra-high-scale probes}

\subsubsection{Electric dipole moments (EDMs)}
EDMs probe CP violation with enormous lever arms. For an electron EDM induced by a dimension-6 operator, a crude scaling is
\begin{equation}
d_e \sim \frac{e\, m_e}{16\pi^2 \Lambda^2}\,\Im(c),
\end{equation}
so that improving \(d_e\) bounds by a factor \(10\) probes \(\Lambda\) higher by \(\sqrt{10}\) (for fixed \(\Im(c)\)).  The current best bound is $|d_e| < 4.1\times 10^{-30}\ e\cdot\mathrm{cm}$ (90\% C.L.)~\cite{Roussy2023eEDM}. For comparison, the ACME~II result gave $|d_e|<1.1\times 10^{-29}\ e\cdot\mathrm{cm}$~\cite{Andreev2018ACME}.
The neutron EDM world limit is \(|d_n|<1.8\times10^{-26}\ e\cdot\cm\) \cite{Abel2020PRL}, with next-generation experiments targeting the \(10^{-27}\text{--}10^{-28}\ e\cdot\cm\) regime.\urlfoot{nEDM experiment project page}{https://www.psi.ch/en/nedm}

Electroweak baryogenesis and many TeV-scale CP-violating new physics scenarios generically predict EDMs near current sensitivity unless there are cancellations or alignment mechanisms \cite{Sakharov1967,MorrisseyRamseyMusolf2012EWBG}. Thus EDM improvements test broad classes of models \emph{independently} of collider reach.

\subsubsection{Charged-lepton flavor violation (CLFV)}
CLFV is essentially zero in the SM with neutrino masses (\(\mathrm{BR}\sim 10^{-54}\) for \(\mu\to e\gamma\)), making it a clean BSM null test.  Two efforts are important:  (1) {\(\mu\to e\gamma\).} MEG II reports an upper limit \(\mathrm{BR}(\mu^+\to e^+\gamma)<1.5\times10^{-13}\) (90\% C.L.) based on 2021--2022 data, with further improvements expected \cite{MEGII2025}. (2) {\(\mu N\to e N\) conversion.}
Muon-to-electron (Mu2e) conversion in a nucleus probes different operator combinations and can reach extreme sensitivity. Mu2e targets single-event sensitivity \(\sim 3\times10^{-17}\) in its baseline configuration \cite{Mu2eTDR2015} and probes effective scales up to \(\sim 10^4\ \TeV\) in some scenarios.

\subsubsection{Rare kaon decays: \(K\to \pi\nu\bar\nu\)}
These are ``golden modes'' because SM predictions are unusually clean (CKM-dominated), and new physics can contribute at very high scales. NA62 reports observation and a branching ratio measurement
\begin{equation}
B(K^+\to \pi^+ \nu\bar\nu)=\left(13.0^{+3.3}_{-3.0}\right)\times 10^{-11},
\end{equation}
combining 2016--2022 data \cite{NA622024KpPinunu}. For the neutral mode, KOTO sets \(B(K_L\to \pi^0\nu\bar\nu) < 2.2\times 10^{-9}\) (90\% C.L.) and also constrains invisible boson channels \cite{KOTO2024KLpinunu}.

\subsubsection{Rare $B$ decays and lepton-flavor universality tests}

A prominent theme of the late 2010s were hints of lepton-flavor universality (LFU) violation in $b\to s\ell^+\ell^-$ observables (e.g.\ $R_K$, $R_{K^\ast}$) and in related angular distributions. However, updated LHCb measurements using the full Run~1+2 samples are increasingly consistent with LFU within present uncertainties \cite{LHCbRK2022,LHCbRKst2022}.
A representative recent example is the 2025 high-$q^2$ determination
\begin{equation}
R_K(q^2>15~\mathrm{GeV}^2)=1.08^{+0.11}_{-0.09}\pm0.04,
\end{equation}
which further tightens global fits of semileptonic operators \cite{LHCbRKHighQ22025}. In an EFT language, percent-level measurements of LFU ratios translate into constraints on LFU-violating combinations of Wilson coefficients at the
$\sim \Order(0.1)$ level (weak-scale matching), corresponding to contact scales $\Lambda/\sqrt{c}\sim \Order(10\text{--}30)\ \TeV$ for $\Order(1)$ couplings. Further progress requires global fits across channels, controlled hadronic uncertainties (form factors and charm-loop effects), and cross-validation with
$e^+e^-$ data (Belle~II) and future high-statistics programs.

\subsubsection{Baryon-number violation: proton decay and $n$--$\bar n$ oscillations}

Baryon number is an accidental symmetry of the renormalizable SM. Many UV completions (in particular grand unification) predict baryon-number violation (BNV), making null tests of BNV uniquely sensitive to ultra-high scales.

\paragraph{Proton decay.}
Dimension-6 operators of the schematic form \(qqql/\Lambda^2\) yield proton partial widths
\begin{equation}
\Gamma(p\to e^+\pi^0)\sim \frac{\alpha^2\,m_p^5}{\Lambda^4},
\qquad
\tau_p \sim \Gamma^{-1}\propto \frac{\Lambda^4}{m_p^5},
\label{eq:protondecayScaling}
\end{equation}
where \(\alpha\) encodes hadronic matrix elements and group-theory factors. Current limits at the \(\tau_p\sim 10^{34}\,\yr\) scale already probe \(\Lambda\sim 10^{15\text{--}16}\,\GeV\) for \(\mathcal{O}(1)\) coefficients. A generational improvement in \(\tau_p\) thus
has direct implications for GUT-scale model space.

\paragraph{$n$--$\bar n$ oscillations.}
$\Delta B=2$ oscillations arise from dimension-9 operators \(uddudd/\Lambda^5\). In terms of an effective mixing matrix element \(\delta m\sim \langle \bar n|\mathcal{O}_{\Delta B=2}|n\rangle/\Lambda^5\), the oscillation time is \(\tau_{n\bar n}\sim 1/\delta m\). Using the characteristic hadronic scale \(\langle \bar n|\mathcal{O}|n\rangle\sim \Lambda_{\rm QCD}^6\), one obtains the scaling
\begin{equation}
\tau_{n\bar n}\sim \frac{\Lambda^5}{\Lambda_{\rm QCD}^6},
\label{eq:nnbarScaling}
\end{equation}
so pushing free-neutron searches to \(\tau_{n\bar n}\gtrsim 10^{9\text{--}10}\,\mathrm{s}\) or nuclear-disappearance limits to comparable scales probes \(\Lambda\sim 10^{5\text{--}6}\,\GeV\) (order-of-magnitude, matrix-element dependent). BNV searches therefore access UV physics in a way that is complementary to collider reach and to lepton-number violation tests such as \(0\nu\beta\beta\).

\subsubsection{Lorentz and CPT invariance: SME parameterization and quantitative reach}
\label{sec:lorentzCPT}

Lorentz invariance (and CPT) is a structural pillar of local relativistic QFT. Precision tests are naturally organized by the (Standard-Model Extension) SME, i.e.\ an expansion in observer-covariant operators built from SM and gravitational fields:
\begin{equation}
\Lag_{\rm SME}=\Lag_{\rm SM}+\Lag_{\rm GR}
+\sum_{d\ge 3}\sum_i \frac{k_i^{(d)}}{M^{d-4}}\,\mathcal{O}_i^{(d)} ,
\label{eq:SMEexpansion}
\end{equation}
where \(d\) is operator dimension and \(M\) is a reference scale (often taken as \(M\sim \mpl\) for UV-motivated estimates). Signals are typically \emph{anisotropic} and/or \emph{time-modulated} in the lab frame due to Earth's rotation and orbital motion.

\paragraph{Time-of-flight and dispersion (astronomical baselines).}
A widely used phenomenological form for Lorentz-violating dispersion is a modified group velocity
\(v(E)\simeq c\left[1-\xi \left(E/M_{\rm LV}\right)^n\right]\), giving an energy-dependent delay between photons (or neutrinos) of energies \(E_1,E_2\) over propagation distance \(D\) \cite{JacobPiran2008LIV,Abdo2009GRB090510}:
\begin{equation}
\Delta t \simeq \frac{n+1}{2}\,\xi\,\frac{D}{c}\,
\frac{E_1^n-E_2^n}{M_{\rm LV}^n},
\qquad n=d-4 .
\label{eq:LVToF}
\end{equation}
This shows explicitly why high-energy transients (large \(E\)) and cosmological baselines (large \(D\)) provide extreme leverage, while the limiting systematics are source emission modeling, intrinsic lags, and instrument timing calibration.

\paragraph{Laboratory tests (clocks, resonators, interferometers).}
In the lab, SME coefficients enter as orientation-dependent energy shifts and modified dispersion relations for matter and photons. For nonrelativistic bound systems, the generic scaling is \(\delta E \sim k^{(d)} p^{d-3}\) (up to tensor structure), so higher-\(d\) coefficients are best constrained by either (i) large characteristic momenta \(p\) (relativistic/nuclear systems) or (ii) large phase accumulation time and modulation control (clocks/resonators). The decisive experimental signatures are sidereal/annual modulations, polarization rotation (birefringence), and cross-platform consistency (ground vs space).

\paragraph{What constitutes ``definitive'' progress?}
Because many leading SME coefficients are already strongly bounded, the high-impact frontier is:
(i) higher-dimension operators (\(d\ge 5\)) probed by high-energy messengers via Eq.~\eqref{eq:LVToF};
(ii) gravitationally coupled or environment-dependent Lorentz-violating sectors constrained by clock networks and space redshift tests;
and (iii) redundancy across messengers (photons + neutrinos + GWs) and across platforms (ground + space) to eliminate source-model degeneracies.

\subsection{Muon anomalous magnetic moment and hadronic theory}
\label{sec:gminus2}

The muon anomalous magnetic moment \(a_\mu\equiv (g_\mu-2)/2\) remains a central precision test because (i) the experimental uncertainty has reached the \(\mathcal{O}(10^{-10})\) level, and (ii) the SM prediction is limited by hadronic contributions that are now comparable to (or larger than) the experimental error.

\paragraph{Experimental status.}
The Fermilab Muon \(g-2\) collaboration Final Report (Run\,1--6) quotes
\begin{equation}
a_\mu(\mathrm{FNAL,\ Run\,1\text{--}6}) = 1165920705(148)\times 10^{-12}\,,
\end{equation}
and the combined experimental world average including BNL E821 is \cite{MuonG2Final2025}
\begin{equation}
a_\mu^{\rm exp} = 1165920715(145)\times 10^{-12}
=116592071.5(14.5)\times 10^{-11},
\end{equation}
corresponding to a total experimental fractional uncertainty of \(\simeq 124~\mathrm{ppb}\).

\paragraph{Theory status: the limiting issue is hadronic vacuum polarization (HVP).}
The SM prediction can be written schematically as
\begin{equation}
a_\mu^{\rm SM} = a_\mu^{\rm QED} + a_\mu^{\rm EW} + a_\mu^{\rm had,\ VP} + a_\mu^{\rm had,\ LbL} + \cdots,
\end{equation}
where the hadronic vacuum polarization (HVP) and hadronic light-by-light (HLbL) terms dominate the theoretical uncertainty.
Historically, the Muon \(g-2\) Theory Initiative ``White Paper'' (WP2020) used a data-driven dispersive HVP evaluation and obtained \cite{Aoyama2020MuonG2Theory}
\begin{equation}
a_\mu^{\rm SM}(\mathrm{WP2020}) = 116591810(43)\times 10^{-11},
\end{equation}
which implies a long-discussed discrepancy at the \(\sim 5\text{--}6\sigma\) level when compared to \(a_\mu^{\rm exp}\).

As of 2025, the interpretation is more nuanced: updated theory efforts incorporating lattice-QCD--driven HVP determinations yield a larger SM value with a larger quoted uncertainty. A representative recent update is \cite{MuonG2Theory2025}
\begin{equation}
a_\mu^{\rm SM}(\mathrm{WP2025}) \simeq 116592033(62)\times 10^{-11},
\end{equation}
for which the difference \(\Delta a_\mu \equiv a_\mu^{\rm exp}-a_\mu^{\rm SM}\) is consistent with zero at the \(\lesssim 1\sigma\) level \cite{MuonG2Final2025,MuonG2Theory2025}.
The present ``state of the art'' is therefore not a single anomaly number, but a tension between internally consistent
HVP determinations (dispersive \(e^+e^-\!\to\) hadrons vs.\ lattice-QCD), and between their associated uncertainty models.

\paragraph{What would be decisive next (quantitatively).}
Because the experimental error is now \(\sigma(a_\mu^{\rm exp})\simeq 1.45\times 10^{-10}\),
a decisive resolution requires HVP and HLbL uncertainties to be reduced such that
\(\sigma(a_\mu^{\rm had,\ VP})\lesssim (2\text{--}3)\times 10^{-10}\) and
\(\sigma(a_\mu^{\rm had,\ LbL})\lesssim 1\times 10^{-10}\),
with transparent correlations across methods. Concretely, this means:
(i) new high-precision low-energy \(e^+e^-\) cross-section data and radiative-correction control for dispersive HVP,
(ii) lattice-QCD results with controlled continuum/volume/chiral extrapolations and validated systematics at the sub-percent level for HVP, and (iii) consistent cross-checks between dispersive and lattice determinations, ideally with complementary inputs that reduce shared modeling priors. ``Decisiveness'' should be assessed by convergence of \emph{independent} HVP pipelines rather than by comparing a single preferred SM number to the experiment.

\paragraph{What would be decisive next?}
A decisive resolution requires the following: (i) convergent HVP determinations with quantified correlations, (ii) additional precision observables sensitive to similar operator structures (e.g.\ \(\alphaem\) running, EDMs, collider contact terms), and (iii) explicit UV model discrimination.

\subsection{Neutrino physics frontiers}
\subsubsection{Oscillation program: mass ordering and CP phase}

Long-baseline accelerator experiments target leptonic CP violation and the neutrino mass ordering by combining
$\nu_\mu(\bar\nu_\mu)\!\to\!\nu_e(\bar\nu_e)$ appearance and $\nu_\mu(\bar\nu_\mu)$ disappearance spectra, with
matter effects providing additional ordering sensitivity over $\mathcal{O}(10^3\,\mathrm{km})$ baselines. DUNE will send a high-power beam from Fermilab to SURF\footnote{See details on the Deep Underground Neutrino Experiment (DUNE)  in South Dakota at \url{https://www.dunescience.org/}} ($L\simeq\SI{1300}{km}$) and deploy a staged far detector of order $\sim\SI{40}{kt}$ fiducial liquid-argon time-projection chamber (LAr TPC). The design beam power is $\sim\SI{1.2}{MW}$ with an upgrade path to $\sim\SI{2.4}{MW}$, enabling few-$\sigma$ sensitivity to CP violation for favorable values of $\delta_{\rm CP}$
in nominal projections and robust mass-ordering determination once flux$\times$cross-section and detector-response
systematics are controlled in joint fits \cite{DUNECDR2018,DUNEFDIDR2020,DUNEsite2025}.
Hyper-Kamiokande will pair an upgraded J-PARC beam with a $\sim\SI{260}{kt}$ water-Cherenkov detector
($\sim\SI{190}{kt}$ fiducial) at a $\sim\SI{295}{km}$ baseline, with first physics data expected in the late 2020s
\cite{HyperKDesign2018,HyperKsite2025}.

A decision-level oscillation result is not a ``nonzero best fit'' but a likelihood-level closure test once the dominant systematics are profiled. Concretely, the program becomes decisive when (i) the mass ordering is established at $>5\sigma$ under conservative cross-section and flux nuisance models; (ii) $\delta_{\rm CP}$ constraints reach the $\sim 10^\circ$--$20^\circ$ level in favorable regions \emph{with degeneracy structure validated} by near-detector constraints and energy coverage; and (iii) appearance/disappearance channels and matter-effect systematics are mutually consistent in joint fits, with documented robustness to alternative interaction models.

\subsubsection{Reactor and medium-baseline oscillation program (portfolio complement to long-baseline)}
\label{sec:nu_reactor_medium}

Medium-baseline reactor experiments target precision oscillation spectroscopy in $\bar\nu_e$ disappearance at
$L\sim 50$--$60~\mathrm{km}$, where interference between $\Delta m^2_{31}$ and $\Delta m^2_{21}$ produces
mass-ordering sensitivity and sharp constraints on $(\sin^2\theta_{12},\,\Delta m^2_{21},\,\Delta m^2_{ee})$.
For three-flavor mixing,
\begin{equation}
P_{\bar e\bar e}\simeq 1
-\cos^4\theta_{13}\,\sin^22\theta_{12}\,\sin^2\Delta_{21}
-\sin^22\theta_{13}\left(\cos^2\theta_{12}\sin^2\Delta_{31}+\sin^2\theta_{12}\sin^2\Delta_{32}\right),
\end{equation}
with $\Delta_{ij}\equiv 1.267\,\Delta m^2_{ij}(\mathrm{eV}^2)\,L(\mathrm{km})/E(\mathrm{GeV})$.
Decision-grade ordering sensitivity requires (i) $\sim$percent-level energy-scale linearity, (ii) controlled nonlinear scintillator response, and (iii) validated reactor-spectrum nuisance models. The value of this portfolio is degeneracy breaking when combined with accelerator LBL constraints and atmospheric samples.

\subsubsection{Coherent elastic neutrino--nucleus scattering (CE$\nu$NS): status and why it matters}
\label{sec:cevns}

CE$\nu$NS provides a high-rate, low-threshold channel sensitive to the weak charge and to nonstandard interactions (NSI). For neutrino energy $E_\nu$ and nuclear recoil $E_R$,
\begin{equation}
\frac{\dd\sigma}{\dd E_R} \simeq
\frac{G_F^2\,m_N}{4\pi}\,Q_W^2
\left(1-\frac{m_N E_R}{2E_\nu^2}\right)F^2(E_R),
\qquad
Q_W = N-(1-4\sin^2\theta_W)Z,
\end{equation}
so the rate scales as $\sim N^2$ at low momentum transfer.
Programmatically, CE$\nu$NS now sits at the boundary between neutrino physics and dark-matter systematics:
it is simultaneously (i) a SM test of neutral currents at low $Q^2$, (ii) a calibrator/background for WIMP searches
near the ``neutrino floor,'' and (iii) a probe of BSM NSI parameters $\epsilon_{\alpha\beta}$ that shift $Q_W$ in a
flavor-dependent way. Decisive progress depends on sub-keV thresholds, neutron-background closure, and explicit reactor/SNS flux nuisance models.

\subsubsection{A portfolio map for $0\nu\beta\beta$: isotope complementarity vs nuclear-theory bottlenecks}
\label{sec:0nubb_portfolio}

A roadmap that treats $0\nu\beta\beta$ as a single ``next exposure'' problem is incomplete: the decisive issue is
\emph{operator and nuclear-theory degeneracy}. Even for light-Majorana exchange, uncertainties and method-to-method spreads in $M^{0\nu}$ map directly into inferred $|m_{\beta\beta}|$.  For nonstandard mechanisms (RH currents, short-range operators), the isotope- and topology-dependence
changes, making multi-isotope coverage a physics requirement rather than a sociological preference.

\begin{table*}[t]
\caption{Illustrative $0\nu\beta\beta$ portfolio map: isotope/technology pathways, dominant systematics, and theory bottlenecks. The point is not completeness but to make the operator/isotope/systematics structure explicit.}
\label{tab:0nubb_portfolio}
\centering
\renewcommand{\arraystretch}{1.12}
\setlength{\tabcolsep}{1pt}
\begin{tabular}{@{}llll@{}}
\hline\hline
\tcell{0.07\textwidth}{Isotope} &
\tcell{0.18\textwidth}{Representative technology families} &
\tcell{0.29\textwidth}{Dominant experimental systematics} &
\tcell{0.44\textwidth}{Dominant theory bottlenecks/Decisive cross-check} \\
\hline

\tcell{0.07\textwidth}{$^{76}$Ge} &
\tcell{0.18\textwidth}{HPGe (ultra-low background)} &
\tcell{0.29\textwidth}{Background index in ROI; surface event rejection; energy-scale linearity} &
\tcell{0.44\textwidth}{NME method spread; short-range correlations; decisive cross-check is consistency with other isotopes at comparable $m_{\beta\beta}$ reach} \\

\tcell{0.07\textwidth}{$^{136}$Xe} &
\tcell{0.18\textwidth}{LXe TPC; Xe-loaded scintillator; gaseous Xe TPC} &
\tcell{0.29\textwidth}{Internal radioactivity; $2\nu\beta\beta$ tail; energy resolution/topology} &
\tcell{0.44\textwidth}{NME correlations with $^{76}$Ge; operator discrimination via topology (gas TPC) and isotope comparison} \\

\tcell{0.07\textwidth}{$^{130}$Te} &
\tcell{0.18\textwidth}{Bolometric arrays} &
\tcell{0.29\textwidth}{Surface backgrounds; pileup; energy response stability} &
\tcell{0.44\textwidth}{NME and $g_A$ quenching; decisive cross-check is multi- isotope global inference with correlated NME priors} \\

\tcell{0.07\textwidth}{(others)} &
\tcell{0.18\textwidth}{$^{82}$Se, $^{100}$Mo, \ldots} &
\tcell{0.29\textwidth}{Topology vs background tradeoffs} &
\tcell{0.44\textwidth}{Operator basis matching (hadronic EFT) and NME benchmarking across nuclei} \\
\hline
\end{tabular}
\end{table*}

\begin{comment}
\begin{table*}[t]
\centering
\caption{Illustrative $0\nu\beta\beta$ portfolio map: isotope/technology pathways, dominant systematics, and theory bottlenecks. The point is not completeness but to make the operator/isotope/systematics structure explicit.}
\label{tab:0nubb_portfolio}
\renewcommand{\arraystretch}{1.12}
\begin{tabularx}{\textwidth}{p{0.07\textwidth} p{0.18\textwidth} p{0.29\textwidth} X}
\hline
Isotope &
Representative technology families &
Dominant experimental systematics &
Dominant theory bottlenecks/Decisive cross-check \\
\hline\hline
$^{76}$Ge &
HPGe (ultra-low background) &
Background index in ROI; surface event rejection; energy-scale linearity &
NME method spread; short-range correlations; decisive cross-check is consistency with other isotopes at comparable $m_{\beta\beta}$ reach \\
$^{136}$Xe &
LXe TPC; Xe-loaded scintillator; gaseous Xe TPC &
Internal radioactivity; $2\nu\beta\beta$ tail; energy resolution/topology &
NME correlations with $^{76}$Ge; operator discrimination via topology (gas TPC) and isotope comparison \\
$^{130}$Te &
Bolometric arrays &
Surface backgrounds; pileup; energy response stability &
NME and $g_A$ quenching; decisive cross-check is multi-isotope global inference with correlated NME priors \\
(others) &
$^{82}$Se, $^{100}$Mo, \ldots &
Topology vs background tradeoffs &
Operator basis matching (hadronic EFT) and NME benchmarking across nuclei \\
\hline
\end{tabularx}
\end{table*}
\end{comment}

The near-term \emph{theory deliverable} that upgrades these experiments from ``large exposures'' to ``decision-grade''
constraints is a correlated NME uncertainty model (including inter-isotope correlations) that can be used as a structured
nuisance prior in the global likelihood.

\subsubsection{Neutrino telescopes (MeV to PeV): oscillations, interactions, and BSM reach}
\label{sec:nu_telescopes}

High-energy neutrino telescopes provide (i) neutrino astronomy, (ii) measurements of $\nu N$ interactions at $\sqrt{s}$ beyond accelerators, and (iii) constraints on BSM scenarios that modify propagation or absorption through Earth (secret interactions, Lorentz violation, decay). At lower energies, dense infills and water/ice Cherenkov detectors constrain atmospheric oscillations and can contribute to mass-ordering sensitivity using matter effects. Next-generation programs (e.g.\ IceCube-Gen2-class expansions \cite{IceCubeGen2_2025}) enlarge effective volume and flavor/energy reach, but are systematics-limited by optical properties, atmospheric backgrounds, and neutrino cross-section uncertainties in the small-$x$ PDF regime.

\subsubsection{Absolute mass scale and cosmological mass sum}

KATRIN reports \(m_\beta<\SI{0.45}{eV}\) (90\% C.L., 2025) \cite{KATRIN2025}. Cosmology constrains \(\sum m_\nu\) through CMB lensing and LSS; future surveys aim at \(\sigma(\sum m_\nu)\sim \Order(10)\ \mathrm{meV}\), but require exquisite control of nonlinear and baryonic systematics.

\subsubsection{Neutrinoless double beta decay: reaching the inverted ordering}

For light Majorana exchange,
\begin{equation}
\left(T_{1/2}^{0\nu}\right)^{-1} = G^{0\nu}\,|M^{0\nu}|^2\,\frac{|m_{\beta\beta}|^2}{m_e^2},
\end{equation}
where \(G^{0\nu}\) is phase space, \(M^{0\nu}\) a nuclear matrix element, and \(m_{\beta\beta}\) the effective mass. Covering the inverted ordering typically demands \(m_{\beta\beta}\sim 10\text{--}20\ \mathrm{meV}\), corresponding to \(T_{1/2}^{0\nu}\sim 10^{28}\text{--}10^{29}\ \yr\) depending on isotope and NME \cite{KamLANDZen800_2022}. 

``Definitive'' interpretation requires a portfolio rather than a single exposure number. In practice this means:
multiple isotopes (to break NME degeneracy and probe operator dependence), background indices at or below the
$10^{-5}\ {\rm counts}/(\mathrm{keV\,kg\,yr})$ scale in the ROI with in-situ sideband validation, and a calibration and
topology model that bounds energy-scale nonlinearity and reconstruction biases below the implied $m_{\beta\beta}$ reach.
Equally important is a nuclear-theory uncertainty model that propagates NME correlations across isotopes, so that a claimed
signal (or null) translates into a robust constraint on the underlying lepton-number-violating parameter(s).

\subsection{Dark matter: a multi-decade, multi-platform search problem}

DM searches must span \(\sim 30\) orders of magnitude in mass and wide coupling structures. The field has shifted from a single WIMP paradigm to a portfolio view \cite{LewinSmith1996,GoodmanWitten1985,DrukierFreeseSpergel1986}.

\subsubsection{Direct detection of particle DM}

For elastic nuclear recoils,
\begin{equation}
\frac{\dd R}{\dd E_R} =
\frac{\rho_\chi}{m_\chi}\,
\frac{\sigma_N}{2\mu_{\chi N}^2}\,
F^2(E_R)\,
\eta(v_{\rm min}),
\end{equation}
where \(\rho_\chi\simeq \SI{0.3}{GeV/cm^3}\) (astrophysical prior), \(\mu_{\chi N}\) is the reduced mass, \(F\) the form factor, and \(\eta\) the halo integral.

\paragraph{State of the art (WIMP masses \(\gtrsim\SI{10}{GeV}\)).}
LZ reports a \(\sigSI\) limit of \(2.2\times 10^{-48}\ \cm^2\) at \(m_\chi\simeq \SI{40}{GeV}\) with \(4.2\pm0.1\) tonne-years exposure \cite{LZ2024}. This is close to the regime where coherent neutrino scattering backgrounds (``neutrino floor'') become relevant \cite{Billard2014NeutrinoFloor}, depending on thresholds and target.

\paragraph{What constitutes a ``definitive'' answer?}
A single excess is not definitive. Definitive identification requires the following: (1) target complementarity (different nuclei, scaling with \(A\), spin structure);
(2) spectral consistency (mass inference);
(3) time dependence (annual modulation, directional signatures);
(4) cross-confirmation with collider/astrophysics constraints on the same operator set. Directional detection (gas TPCs, anisotropy) is the clearest route to discriminating DM from neutrino backgrounds in the long run.

\subsubsection{Direct detection landscape (2025 snapshot): xenon, argon, cryogenic, and low-threshold targets}
\label{sec:dm_direct_landscape}

A single anchor limit (e.g.\ LZ) is useful for EFT-level scaling, but it is not a field guide.  In practice,
the discriminating power of direct detection comes from a \emph{portfolio} spanning:
(i) xenon and argon TPCs (multi-tonne, ultra-low backgrounds, WIMP masses $\gtrsim 10~\mathrm{GeV}$),
(ii) cryogenic phonon/ionization calorimeters (sub-GeV nuclear recoils),
(iii) electron-excitation targets (skipper CCDs, semiconductors, superconductors) for $m_\chi\lesssim 1~\mathrm{GeV}$,
and (iv) directional approaches (long-term discriminant against the irreducible CE$\nu$NS background).

\paragraph{Operator-level translation and why multi-target matters.}
Modern comparisons should be made in a nonrelativistic EFT basis for $\chi$--nucleon scattering: different targets weight different linear combinations of isoscalar/isovector couplings and spin responses. Near the neutrino background regime, \emph{target complementarity + at least one additional discriminant} (directionality and/or modulation) is the only robust route to identification.

\begin{table*}[t]
\caption{Representative status of direct-detection program families. Limits are quoted from flagship WIMP-search analyses for each program family and are shown only to provide scale; comparisons require consistent astrophysical priors/signal models.}
\label{tab:dm_landscape_2025}
\centering
\renewcommand{\arraystretch}{1.12}
\setlength{\tabcolsep}{1pt}
\begin{tabular}{@{}lllll@{}}
\hline\hline
\tcell{0.14\textwidth}{Program class} &
\tcell{0.17\textwidth}{Representative experiments (target)} &
\tcell{0.16\textwidth}{Representative exposure (analysis)} &
\tcell{0.16\textwidth}{Representative best limit/Reach} &
\tcell{0.36\textwidth}{Dominant limitations/Decisive next step} \\
\hline

\tcell{0.14\textwidth}{Dual-phase noble TPC (Xe)} &
\tcell{0.17\textwidth}{LZ; XENONnT \cite{XENONnT2023}; PandaX-4T \cite{PandaX4T2024}} &
\tcell{0.16\textwidth}{$\Order(1$--$5)\,$t$\cdot$yr} &
\tcell{0.16\textwidth}{$\sigma_{\rm SI}\sim 10^{-48}$--$10^{-47}\ \mathrm{cm^2}$ at $m_\chi\sim 30$--$50~\mathrm{GeV}$} &
\tcell{0.35\textwidth}{Radiogenic $\beta/\gamma$ + neutrons; detector res- ponse (S1/S2); irreducible CE$\nu$NS; decisive next step is multi-target + discriminant channels near the neutrino background} \\

\tcell{0.14\textwidth}{Dual-phase noble TPC (Ar)} &
\tcell{0.17\textwidth}{DarkSide-50 \cite{DarkSide50LowMass2018}; DarkSide-20k (next-gen)} &
\tcell{0.15\textwidth}{(sub- to multi-t scale, program-dependent)} &
\tcell{0.16\textwidth}{Competitive at high mass only in next-gen Ar; strong low-$^{39}$Ar leverage} &
\tcell{0.35\textwidth}{Pulse-shape discrimination, surface backgrounds, neutron backgrounds; decisive step is multi-t Ar with validated NR calibration and CE$\nu$NS handling} \\

\tcell{0.14\textwidth}{Cryogenic phonon/ioniza- tion (NR)} &
\tcell{0.17\textwidth}{SuperCDMS (HV/HVeV); EDELWEISS; CRESST} &
\tcell{0.16\textwidth}{kg$\cdot$day to kg$\cdot$yr class} &
\tcell{0.16\textwidth}{Sub-GeV NR thresholds; strong reach at $m_\chi\sim 0.1$--$5~\mathrm{GeV}$} &
\tcell{0.35\textwidth}{Luke--Neganov amplification systematics, low-energy backgrounds, near-threshold calibrations; decisive step is background-model closure below $\sim 100$ eV$_{\rm nr}$} \\

\tcell{0.14\textwidth}{Skipper CCD / semiconductor (ER)} &
\tcell{0.17\textwidth}{SENSEI, DAMIC(-M), skipper CCD programs} &
\tcell{0.16\textwidth}{g$\cdot$day to kg$\cdot$day class} &
\tcell{0.16\textwidth}{$e^-$-counting pushes $m_\chi\lesssim 1~\mathrm{GeV}$} &
\tcell{0.35\textwidth}{Dark current and spurious charge, surface events; decisive step is demonstrated multi-site reproducibility of single-$e^-$ background models} \\

\tcell{0.14\textwidth}{Directional/track imaging} &
\tcell{0.17\textwidth}{Gas TPC concepts; negative-ion drift; columnar recombination R\&D} &
\tcell{0.16\textwidth}{technology-limited today} &
\tcell{0.16\textwidth}{Ultimate discriminant vs CE$\nu$NS at very low $\sigma$} &
\tcell{0.35\textwidth}{Angular resolution and head-tail; scalable low-background volume; decisive step is $\Order(10$--$100)$ m$^3$ prototypes with validated direction reconstruction} \\
\hline
\end{tabular}
\end{table*}

\paragraph{``Status anchors'' for the xenon programs.}
The published XENONnT analysis with $1.1$ tonne-year exposure reports a best spin-independent limit $\sigma_{\rm SI}=2.58\times 10^{-47}\ \mathrm{cm^2}$ at $m_\chi=28~\mathrm{GeV}$, while PandaX-4T reports $\sigma_{\rm SI}=1.6\times 10^{-47}\ \mathrm{cm^2}$ at $m_\chi=40~\mathrm{GeV}$ for a $1.54$ tonne-year exposure.  These values are not included to ``rank'' experiments, but to show the scale of present systematics/background regimes across otherwise similar xenon TPC architectures.

\subsubsection{Sub-GeV DM: electron recoils, phonons, magnons, and superconductors}
For \(m_\chi\lesssim\SI{1}{GeV}\), nuclear recoils fall below thresholds and one targets electron excitations (eV scale) or collective modes (meV scale). Key metrics become: (1)
 energy threshold \(E_{\rm th}\) (eV to meV),
(2) background rates at low energies, and 
(3)  control of dark counts and instrumental artifacts.
Technology directions: cryogenic calorimeters, skipper CCDs, superconducting sensors (TES, MKIDs), Dirac materials, and engineered targets with enhanced response.

\subsubsection{Accelerator and fixed-target probes of light dark sectors (portal mediators)}
\label{sec:darkSectorPortalsExp}

A large and well-motivated region of beyond-SM parameter space involves \emph{light mediators} that couple feebly to the SM and may connect to light dark matter. A canonical example is a kinetically mixed dark photon \(A'\) with
\begin{equation}
\Lag \supset -\frac14 F'_{\mu\nu}F'^{\mu\nu}+\frac{\epsilon}{2}F'_{\mu\nu}B^{\mu\nu}+\frac12 m_{A'}^2 A'_\mu A'^{\mu}
+ g_D A'_\mu \bar\chi\gamma^\mu\chi ,
\label{eq:darkPhotonPortal}
\end{equation}
where \(\epsilon\) controls the effective SM coupling and \(\alpha_D\equiv g_D^2/(4\pi)\) controls the dark coupling. The defining feature of intensity-frontier searches is that signal yields scale as powers of \(\epsilon\) and are independent of the astrophysical prior on \(\rho_\chi\).

\paragraph{Connection to direct detection (mediator-mass scaling).}
For DM--electron scattering via a heavy vector mediator (\(m_{A'}\gg q\)), the reference cross section at zero momentum transfer scales as
\begin{equation}
\sigma_e \simeq \frac{16\pi\,\alphaem\,\alpha_D\,\epsilon^2\,\mu_{e\chi}^2}{m_{A'}^4},
\label{eq:sigmaePortalScaling}
\end{equation}
while in the light-mediator regime (\(m_{A'}\ll q\)) the differential rate is enhanced at small momentum transfer,
schematically \(\mathrm{d}\sigma/\mathrm{d}q^2\propto 1/q^4\) up to atomic/solid-state screening effects. This explicit scaling provides a clean translation layer between accelerator constraints on \((\epsilon^2\alpha_D)\) and low-threshold direct-detection programs.

\paragraph{Experimental modalities and decisive systematics.}
Key approaches include (i) visible-decay searches (\(A'\to e^+e^-,\mu^+\mu^-\)) where production \(\propto \epsilon^2\) and backgrounds are dominated by QED tridents and radiative tails; (ii) invisible-decay / missing-momentum searches probing \(A'\to\chi\bar\chi\) where the decisive challenge is control of neutrino-induced and detector veto inefficiencies; and (iii) beam-dump and forward experiments exploiting long baselines and large fluxes. A robust discovery standard is \emph{reproducibility across production modes} (visible vs invisible channels) and consistency with direct-detection scattering expectations via Eq.~\eqref{eq:sigmaePortalScaling}, together with cosmological/astrophysical bounds on additional light degrees of freedom and late decays.

\subsubsection{Axions and ALPs: experimental landscape, program families, and coverage partition}
\label{sec:axion_landscape}

Axion and ALP searches are best organized by \emph{(i) which coupling is targeted} and \emph{(ii) which mass/frequency regime is accessible}, because these determine the signal model, the dominant systematics, and the scan strategy \cite{IrastorzaRedondo2018,DiLuzio2020AxionLandscape}.

\paragraph{Mass--frequency conversion and coherence.}
For nonrelativistic halo DM, the axion oscillation frequency is
\begin{equation}
f_a \simeq \frac{m_a c^2}{h} \simeq 0.2418~\mathrm{GHz}\,\left(\frac{m_a}{\mu\mathrm{eV}}\right),
\end{equation}
and the virial linewidth is $\Delta f/f \sim v^2\sim 10^{-6}$, i.e.\ a quality factor $Q_a\sim 10^6$. This fixes the bandwidth requirements for resonant vs broadband searches.

\paragraph{QCD axion target band.}
For the QCD axion,
\begin{equation}
m_a \simeq 5.7~\mu\mathrm{eV}\left(\frac{10^{12}\ \mathrm{GeV}}{f_a}\right),
\qquad
g_{a\gamma\gamma}=\frac{\alpha}{2\pi f_a}\,C_\gamma,
\end{equation}
with $C_\gamma=(E/N-1.92)$ and typical benchmark magnitudes $|C_\gamma|\sim 0.7$ (DFSZ) or $|C_\gamma|\sim 1.9$ (KSVZ),
implying the familiar near-linearity $g_{a\gamma\gamma}\propto m_a$ for QCD axions.  In contrast, ALPs admit
independent $(m_a,g)$.

\paragraph{Resonant haloscopes (cavity / dielectric / plasma): scan-rate scaling.}
In a resonant haloscope \cite{diCortona2016QCDaxion,Sikivie1983,Sikivie1985}, the expected conversion power scales as
\begin{equation}
P_{a\to\gamma} \sim g_{a\gamma\gamma}^2\,B^2\,V\,C\,Q\,\frac{\rho_a}{m_a},
\end{equation}
and the radiometer equation gives (schematically)
\begin{equation}
\SNR \sim \frac{P_{a\to\gamma}}{k_B T_{\rm sys}}\sqrt{\frac{t_{\rm int}}{\Delta f}},
\qquad \Delta f\sim \frac{f_a}{Q_a}.
\end{equation}
Thus the scan rate for fixed $\SNR$ scales roughly as
$\mathrm{d}f/\mathrm{d}t \propto g_{a\gamma\gamma}^4\,B^4\,V^2\,C^2\,Q^2/T_{\rm sys}^2$, which makes explicit why high-field large-volume magnets, high $Q$, and quantum-limited amplifiers dominate the engineering ``figure of merit''. See Table~\ref{tab:axion_landscape} that shows the current landscape of which approach covers which $(m,g)$ region.

\begin{table*}[t]
\caption{Axion/ALP search families, organized by coupling channel and accessible mass/frequency range. The decisive limitation is listed as the dominant nuisance term that must be closed for a claim.}
\label{tab:axion_landscape}
\centering
\renewcommand{\arraystretch}{1.15}
\setlength{\tabcolsep}{1pt}
\begin{tabular}{@{}llll@{}}
\hline
\tcell{0.16\textwidth}{Coupling/channel} &
\tcell{0.18\textwidth}{Mass regime (typical)} &
\tcell{0.20\textwidth}{Technique class (reso- nant vs broadband)} &
\tcell{0.44\textwidth}{Major limitations/Decisive cross-check} \\
\hline\hline

\tcell{0.16\textwidth}{$g_{a\gamma\gamma}$ (halo DM)} &
\tcell{0.18\textwidth}{$\mu$eV--meV ($\sim$GHz--100\,GHz)} &
\tcell{0.20\textwidth}{Resonant haloscopes (cavities; dielectric haloscopes; plasma haloscopes)} &
\tcell{0.44\textwidth}{Scan coverage vs $Q$ and $T_{\rm sys}$; mode-mapping; spurious narrow lines; decisive cross-check is reproducible, tunable resonance behavior with consistent phase/polarization signatures} \\

\tcell{0.16\textwidth}{$g_{a\gamma\gamma}$ (ultralight)} &
\tcell{0.18\textwidth}{neV--$\mu$eV (kHz--MHz)} &
\tcell{0.20\textwidth}{Lumped-element / LC resonators; broadband pickup loops} &
\tcell{0.44\textwidth}{Magnetic pickup and correlated environmental noise; decisive step is multi-site correlation and magnetic-null channels} \\

\tcell{0.16\textwidth}{Spin couplings ($g_{aN}$, $g_{ae}$)} &
\tcell{0.18\textwidth}{peV--neV (mHz--kHz)} &
\tcell{0.20\textwidth}{NMR / comagnetome- ters / torsion balances; resonant spin precession} &
\tcell{0.44\textwidth}{Magnetic shielding and drifts; spin-relaxation modeling; decisive step is controlled reversals and networked correlation with distinct species} \\

\tcell{0.16\textwidth}{Solar axions (pro- duction in Sun)} &
\tcell{0.18\textwidth}{model-dependent} &
\tcell{0.20\textwidth}{Helioscopes (axion $\to$ X-ray conversion in $B$ field)} &
\tcell{0.44\textwidth}{X-ray backgrounds and pointing/acceptance; decisive step is background-validated off-Sun control + consistent energy spectrum} \\

\tcell{0.16\textwidth}{Laboratory pro- duction (ALPs)} &
\tcell{0.18\textwidth}{broad} &
\tcell{0.20\textwidth}{Light-shining-through-walls} &
\tcell{0.44\textwidth}{Optical systematics; cavity stability; decisive step is multi-configuration null tests and independent regeneration optics} \\
\hline
\end{tabular}
\end{table*}

\paragraph{Astrophysical constraints and how they partition parameter space.}
Stellar cooling (HB stars, red giants, white dwarfs), SN1987A, and black-hole superradiance constrain large regions of ALP parameter space with model-dependent systematics.  In a roadmap context, the correct role of these constraints is to define \emph{priors and targets} for laboratory coverage and to motivate redundancy: a laboratory discovery must be consistent with (or explain tension with) the relevant astrophysical bounds.

\subsubsection{Astronomical DM signals and structure}
Astronomy probes DM self-interactions, free-streaming, and ultra-light wave effects. Examples: halo cores/cusps, cluster mergers, stellar streams, Lyman-\(\alpha\) forest constraints on small-scale power, and black hole superradiance constraints on ultra-light bosons. These probes are systematics-limited by baryonic feedback and astrophysical modeling; progress requires joint inference frameworks.

A complementary (non-particle) DM hypothesis is that some fraction of DM is in compact objects, commonly parameterized as \(f_{\rm PBH}(M)\equiv \Omega_{\rm PBH}(M)/\Omega_{\rm DM}\). Constraints are intrinsically multi-messenger: microlensing and dynamical heating probe gravitational interactions on small scales, CMB anisotropies constrain accretion/energy injection at recombination, and GW merger rates constrain (or motivate) PBH subpopulations depending on the formation scenario. A decisive program requires joint inference of \(f_{\rm PBH}(M)\) across probes with consistent astrophysical priors (halo models, stellar populations, accretion physics), and explicit separation of PBH signals from particle-DM phenomenology in structure data.

\subsection{Cosmology and the early universe}

\subsubsection{Primordial perturbations and inflation}
The tensor-to-scalar ratio \(r\) maps to an inflationary energy scale:
\begin{equation}
V^{1/4} \simeq 1.06\times 10^{16}\ \mathrm{GeV}\left(\frac{r}{0.01}\right)^{1/4}.
\end{equation}
Current CMB $B$-mode constraints give $r_{0.05}<0.036$ (95\% C.L.), corresponding to $V^{1/4}\lesssim 1.5\times10^{16}\ \mathrm{GeV}$ via the above scaling.
Near-term improvements rely on multi-frequency foreground control and delensing; the BICEP/Keck program targets $\sigma(r)\lesssim 0.003$ with extended datasets (analysis/foreground dependent) \cite{BKReview2024}. Thus pushing \(r\) below \(10^{-3}\) constrains high-scale inflation models. A detection at \(r\sim 10^{-3}\) would indicate \(V^{1/4}\sim \mathrm{few}\times 10^{15}\ \mathrm{GeV}\), while a null result at \(r\ll 10^{-4}\) would strongly favor low-scale or nonstandard scenarios (or tuned models).

\subsubsection{Beyond \(r\): primordial non-Gaussianity, small-scale power, and spectral distortions}
While the tensor-to-scalar ratio is a clean inflationary discriminator, it is not the only (or even the most accessible) handle on the microphysics of the primordial era. Three complementary targets are particularly well-motivated:

\paragraph{Local-type primordial non-Gaussianity.}
The squeezed-limit bispectrum amplitude \(f_{\rm NL}^{\rm local}\) is a sharp discriminator of inflationary field content:
single-clock attractor models predict \(|f_{\rm NL}^{\rm local}|\ll 1\), whereas multi-field dynamics can yield \(|f_{\rm NL}^{\rm local}|\gtrsim \mathcal{O}(1)\). A credible \(\sigma(f_{\rm NL}^{\rm local})\sim 1\) measurement from large-scale structure therefore constitutes a decision-level test of whole model classes \cite{Maldacena2003NG,CreminelliZaldarriaga2004Consistency}. In practice, this requires ultra-large-scale control of survey systematics, relativistic projection effects, foregrounds, and multi-tracer strategies.

\paragraph{Small-scale primordial power.}
CMB anisotropies probe \(k\sim 10^{-4}\text{--}0.2\ \mathrm{Mpc^{-1}}\), leaving many decades of smaller scales unconstrained. Small-scale enhancements are tied to primordial black holes, early microhalo formation, and spectral features from non-slow-roll dynamics. Probes include CMB lensing at high \(\ell\), Lyman-\(\alpha\) forests, and (longer-term) 21\,cm fluctuations during the dark ages and cosmic dawn,
where the accessible mode count is enormous but foreground subtraction and calibration are exceptionally challenging.

\paragraph{Spectral distortions of the CMB monopole.}
Dissipation of acoustic modes (Silk damping) at \(10^4\lesssim z\lesssim 10^6\) generates a guaranteed \(\mu\)-distortion signal at the level of \(\mu \sim \mathrm{few}\times 10^{-8}\) in $\Lambda$CDM, with additional contributions from energy injection (decays/annihilations) and small-scale features. A PIXIE-like all-sky spectrometer targets \(\sigma_\mu \sim 10^{-8}\) sensitivity, turning spectral distortions into a quantitative probe of
small-scale primordial power and exotic energy release \cite{PIXIE2024,CabassMu2016}.

\subsubsection{Extra relativistic species and \(\Neff\)}

\(\Neff\) quantifies radiation density beyond photons:
\begin{equation}
\rho_r = \rho_\gamma\left[1+\tfrac78\left(\tfrac{4}{11}\right)^{4/3}\Neff\right].
\end{equation}
A target \(\sigma(\Delta \Neff)\sim 0.06\) is often quoted as ``neutrino-decoupling sensitivity'' that can test many light relic scenarios (motivating CMB-S4-type programs \cite{CMBS4Plan2025}). Achieving this is foreground- and systematics-limited: polarized dust/synchrotron separation, beam calibration, and control of atmospheric and instrumental \(1/f\) noise for ground-based telescopes.

\subsubsection{Late-time expansion and growth: DESI/ACT/weak lensing}

DESI BAO and redshift-space distortions, CMB lensing, and cosmic shear constrain \(w(z)\) and growth. ACT DR6 provides high-precision small-scale CMB power and lensing; extended-model analyses report that some early-dark-energy fixes to \(\Hn\) are not favored by ACT DR6 \cite{ACTDR6Extended}. Weak lensing systematics include shear calibration, intrinsic alignments, photometric redshift calibration, and baryonic feedback. KiDS Legacy reports \(S_8\) compatible with Planck at \(\lesssim 1\sigma\) \cite{KiDSLegacy2025}, illustrating how survey completion and improved calibration can shift global conclusions.

\paragraph{DESI DR2 BAO and dynamical-dark-energy fits (2025).}
DESI DR2 BAO measurements (from $\gtrsim 14$ million galaxies and quasars) now dominate the statistical error budget for many late-time geometric probes. In a flat $\Lambda$CDM fit, DESI reports that BAO-preferred parameters are in mild ($\sim 2.3\sigma$) tension with Planck-era CMB constraints, while remaining consistent with the precisely measured acoustic angular scale $\theta_\ast$. Allowing a time-evolving dark-energy equation of state in the CPL form $w(a)=w_0+w_a(1-a)$ alleviates this tension; DESI finds a favored solution in the quadrant $w_0>-1$ and $w_a<0$, and reports that $w_0w_a$CDM is preferred over $\Lambda$CDM at $3.1\sigma$ for DESI BAO$+$CMB, and at $2.8$--$4.2\sigma$ when also including SNe (depending on the SNe compilation) \cite{DESIDR2BAO2025}.

In a constant-$w$ extension, the DESI$+$CMB$+$DESY5 combination yields
\begin{equation}
w=-0.971\pm0.021
\end{equation}
(68\% credible interval) \cite{DESIDR2BAO2025}.
In the CPL parameterization \cite{ChevallierPolarski2001,Linder2003}, for DESI BAO$+$CMB$+$DESY5 in a $w_0w_a$CDM$+\sum m_\nu$ fit, one finds (68\% credible intervals)
\begin{equation}
w_0 = -0.758\pm 0.058,\qquad
w_a = -0.82^{+0.23}_{-0.21},\qquad
H_0 = 66.75\pm 0.56\ \kmsMpc,
\end{equation}
and $\sum m_\nu<0.129\ \mathrm{eV}$ (95\% upper limit) \cite{DESIDR2BAO2025}.

From DESI$+$CMB alone, DESI quotes 95\% upper limits $\sum m_\nu<0.064\ \mathrm{eV}$ in flat $\Lambda$CDM and $\sum m_\nu<0.16\ \mathrm{eV}$ in $w_0w_a$CDM \cite{DESIDR2BAO2025}. These results illustrate that late-time cosmology is entering a regime where model dependence (parameterization of $w(z)$, treatment of nonlinearities/baryons, and cross-survey covariance) becomes comparable to statistical errors, motivating end-to-end systematics validation and robust cross-checks across probes.

\subsubsection{Closure tests and cross-survey covariance: what ``percent-level'' actually requires}
\label{sec:cosmo_closure_tests}

Late-time cosmology is now limited as much by \emph{shared nuisance structure} as by statistical power. A ``closure test'' is therefore not a rhetorical device but an operational requirement:
a dataset combination is credible only if the inferred parameter shifts are stable under
(1) alternative nuisance parameterizations,
(2) data splits that isolate dominant systematics, and
(3) explicit cross-covariance between overlapping surveys.

\paragraph{Concrete examples of closure tests.}
Representative closure tests that are routinely required for decision-grade inference include:
(i) \emph{geometry vs growth closure}: consistency of $H(z)$ from BAO+SNe with growth constraints from RSD+$f\sigma_8$ and weak lensing under shared $\Omega_m$ priors;
(ii) \emph{shear calibration closure}: image-simulation calibrated multiplicative shear bias $m$ validated by internal nulls (PSF leakage, star-galaxy correlations) and cross-checked by CMB-lensing$\times$galaxy-shear cross-correlations;
(iii) \emph{photo-$z$ closure}: redshift distributions calibrated by spectroscopy, clustering-$z$, and self-organizing-map methods, with residual biases propagated as explicit nuisance modes;
(iv) \emph{foreground closure}: multi-frequency CMB separation validated by jackknives and end-to-end injections.

\paragraph{Cross-survey covariance as a first-class ingredient.}

For overlapping footprints, the correct likelihood is not a product of independent terms. If two surveys provide observables $O_1$ and $O_2$, the combined covariance is block-structured:
\begin{equation}
C=
\begin{pmatrix}
C_{11} & C_{12}\\
C_{21} & C_{22}
\end{pmatrix},
\qquad
\chi^2 = (O-O_{\rm th})^\top C^{-1}(O-O_{\rm th}),
\end{equation}
where $C_{12}$ includes super-sample covariance, shared calibration modes, and shared astrophysical nuisance (e.g.\ baryons). Neglecting $C_{12}$ biases both uncertainties and model selection, especially in $w_0w_a$ fits where the statistical errors are small.

\paragraph{Upcoming space surveys and distinct systematics regimes.}
A comprehensive 2025-era roadmap must distinguish the systematics regimes of the major space datasets:
\emph{Euclid/Roman-like imaging+slitless spectroscopy} deliver stable PSFs and NIR coverage that improves photo-$z$ calibration and lensing, but introduce distinct detector systematics (persistence, nonlinearity, intra-pixel response) and selection functions. Their complementarity with wide-area ground imaging (Rubin/LSST-class) is precisely that the dominant calibration modes differ, enabling cross-survey closure on shear and photo-$z$ nuisance parameters rather than merely reducing statistical error bars.

\subsubsection{21\,cm cosmology: cosmic dawn, reionization, and primordial statistics}
The redshifted 21\,cm hyperfine transition provides a three-dimensional tracer of neutral hydrogen over
\(6\lesssim z \lesssim 30\), with sensitivity to both astrophysics (ionizing sources, heating, radiative feedback) and fundamental microphysics (the small-scale matter power spectrum, baryon--DM interactions, primordial non-Gaussianity, and light relics). A standard approximation for the differential brightness
temperature relative to the CMB is
\begin{equation}
\delta T_b(z) \simeq 27\,x_{\rm HI}(1+\delta_b)\left(\frac{\Omega_b h^2}{0.023}\right)
\left(\frac{0.15}{\Omega_m h^2}\frac{1+z}{10}\right)^{1/2}\left(1-\frac{T_\gamma}{T_S}\right)\ \text{mK},
\end{equation}
where \(x_{\rm HI}\) is the neutral fraction, \(\delta_b\) the baryon overdensity, \(T_S\) the spin temperature,
and \(T_\gamma\) the CMB temperature.

\paragraph{Current state of the art.}
Interferometric EoR experiments have not yet detected the cosmological 21\,cm signal, but have begun to approach astrophysically relevant upper limits. For example, LOFAR analyses report multi-redshift upper limits at \(z\simeq 10.1,\,9.1,\,8.3\) with best limits
\(\Delta^2_{21}<(68.7\,\text{mK})^2\) at \(k=0.076\,h\,\text{cMpc}^{-1}\),
\(\Delta^2_{21}<(54.3\,\text{mK})^2\) at \(k=0.076\,h\,\text{cMpc}^{-1}\),
and \(\Delta^2_{21}<(65.5\,\text{mK})^2\) at \(k=0.083\,h\,\text{cMpc}^{-1}\), respectively
\cite{MertensLOFAR2025}.

\paragraph{Discovery reach: primordial non-Gaussianity.}
A high-leverage target is local-type primordial non-Gaussianity, where the scale-dependent bias implies that sufficiently large-volume surveys can reach \(\sigma(f_{\rm NL}^{\rm local})\sim \Order(1)\). Forecasts for future 21\,cm intensity mapping/interferometric surveys find that, under aggressive foreground mitigation assumptions, one can reach \(\sigma(f_{\rm NL}^{\rm local})\lesssim 1\)
\cite{KaragiannisPNG21cm2021}.

\subsection{Gravitational waves and relativistic astrophysics}
\label{sec:GWs}

Gravitational wave (GW) observations now constitute a direct probe of strong-field gravity and compact-object astrophysics, and are becoming a cosmological tool.

\subsubsection{Frequency bands and detectors}
A schematic mapping (with indicative science reach and dominant noise terms):
\begin{itemize}[leftmargin=2.2em]
\item \emph{Ground interferometers, 2G} (\(\sim 10\text{--}10^3\ \Hz\)): the current Advanced LIGO/Virgo/KAGRA network.  
Sources: stellar-mass binaries, nearby core-collapse supernovae, stochastic backgrounds at \(\gtrsim 10\ \Hz\).
Dominant limitations: seismic and Newtonian gravity-gradient noise at low \(f\), thermal noise at intermediate \(f\), quantum shot noise at high \(f\).

\item \emph{Ground interferometers, 3G} (\(\sim 1\text{--}10^3\ \Hz\)): Einstein Telescope (ET) and Cosmic Explorer (CE) concepts. These target \(\mathcal{O}(10)\) strain-amplitude improvement over 2G across much of the band, which (for compact binaries) translates to \(\mathcal{O}(10^3)\) increase in sampled spacetime volume.  Representative science consequences include BNS detections out to \(z\gtrsim 2\), BBH detections to \(z\gtrsim 10\) (population and mass-dependent), high-SNR ringdowns for precision GR tests, and large standard-siren samples for sub-percent cosmology \cite{ETDesign2020,EvansCE2021,BranchesiETCE2023}.

\item \emph{Space interferometers, mHz} (\(\sim 10^{-4}\text{--}10^{-1}\ \Hz\)): massive black hole binaries, extreme mass-ratio inspirals, Galactic binaries, and potential cosmological backgrounds.  LISA (arm length \(\sim 2.5\times 10^6\ \mathrm{km}\), planned for the mid-2030s) is the anchor mission; other mHz concepts such as Taiji and TianQin target related bands with different constellations/baselines and may provide network-level improvements for sky localization and polarization disentangling \cite{CaiNetwork2024}.

\item \emph{Decihertz (``mid-band'') detectors} (\(\sim 10^{-2}\text{--}10\ \Hz\)): proposed DECIGO/BBO-class laser interferometers and atom-interferometer concepts (e.g.\ AEDGE/SAGE-like) fill the gap between LISA and ground-based detectors.  This band enables multiband tracking of stellar binaries (months--years before merger), early-warning for multimessenger campaigns, and strong leverage on certain primordial backgrounds and cosmic strings, provided foreground subtraction and correlated-noise control are adequate \cite{KawamuraDECIGO2021,YagiDECIGO2011,AbouElNeajAEDGE2020,TinoSAGE2019}.

\item \emph{Solar-system binaries / precision ranging, $\mu$Hz ($\sim 10^{-7}$--$10^{-4}\,$Hz):}  Long-baseline laser ranging (LLR; also tracking of interplanetary spacecraft) can probe an ultra-low-frequency stochastic gravitational-wave background (SGWB) via \emph{binary resonance}: GW power at discrete harmonics $f_n = n f_{\rm orb}$ drives a random walk in the fitted osculating elements and hence produces sidebands in post-fit range residuals.  In the diffusion regime,
\(
\mathrm{Var}\!\left[\Delta X(T)\right]\simeq D_X(f_n)\,T\) and \(
D_X(f_n)\propto S_h(f_n),
\)
with
\(
\Omega_{\rm GW}(f)=({2\pi^2}/{3H_0^2})\,f^{3} S_h(f)\) \cite{BlasJenkinsPRL2022,BlasJenkinsPRD2022,HuiMcWilliamsYang2013}. Dominant nuisances are orbit/ephemeris modeling, tides and geophysics, station coordinates/time transfer, troposphere, and reflector/CCR response. Improvements hinge on the new-gen CCRs and high-power CW LLR facilities  \cite{Turyshev2025HighPowerLLR,Turyshev2025HighPowerLLR-ErBu,Turyshev2025CCR}.

\item \emph{Pulsar timing arrays (PTAs), nHz} (\(\sim 10^{-9}\text{--}10^{-7}\ \Hz\)): supermassive black hole binary background and potentially early-universe contributions. Systematics: pulsar spin noise, interstellar medium effects, clock errors, and Solar-System ephemeris uncertainties.
\end{itemize}

\paragraph{Current observational status (late 2025).}
In the \(\sim 10\text{--}10^3\ \Hz\) band, the LVK fourth observing run (O4; May~24~2023--Nov.~18~2025) produced \(\sim 250\) low-latency GW candidates, with 128 significant events in the first analyzed segment (GWTC-4.0), and more detections expected as the remaining O4 data are analyzed \cite{LVKO4Complete2025}. In the nHz band, multiple PTAs report evidence for a common low-frequency stochastic process and (in several analyses) angular correlations consistent with the Hellings--Downs quadrupolar signature expected for an isotropic stochastic GW background \cite{HellingsDowns1983}. A common parameterization is a power-law characteristic strain $h_c(f)=A(f/f_{\rm yr})^{\alpha}$ with $f_{\rm yr}=1\,\mathrm{yr}^{-1}$; for the inspiralling-supermassive-black-hole expectation $\alpha=-2/3$, the IPTA DR2 analysis reported $A=2.8^{+1.2}_{-0.8}\times 10^{-15}$ \cite{IPTADR2GWB2022}. More recently, NANOGrav's 15-year dataset reports evidence for Hellings--Downs correlations and infers $A=2.4^{+0.7}_{-0.6}\times10^{-15}$ (90\% credible interval, $\alpha=-2/3$) \cite{NANOGrav15yr2023}, while EPTA DR2 finds $A=2.5\pm0.7\times10^{-15}$ for fixed $\alpha=-2/3$ \cite{EPTADR2GWB2023}.
This corresponds to an energy-density scale $\Omega_{\rm GW}(f_{\rm yr})\sim 10^{-8}$ at $f_{\rm yr}$ via
\begin{equation}
\Omega_{\rm GW}(f)=\frac{2\pi^2}{3H_0^2}\,f^2\,h_c^2(f).
\end{equation}
Going from ``detection`` to ``physics`` now hinges on: (i) cross-consistency between PTA datasets and noise/ephemeris modeling choices, (ii) spectral-shape discrimination (SMBHBs vs cosmic strings/phase transitions), and (iii) identification of individually resolvable nHz sources in the brightest-tail regime.

\subsubsection{Quantitative GW strain sensitivity and ``decisive'' tests}

Detectors measure strain \(h\sim \Delta L/L\). For a monochromatic source, \(\SNR\) scales as
\begin{equation}
\SNR^2 \sim 4\int \frac{| \tilde h(f)|^2}{S_n(f)}\,\dd f,
\end{equation}
so improvements target lowering noise power spectral density \(S_n\), increasing observation time \(\Tobs\), and accessing more cycles. The relevant fundamental-physics deliverables include:
\begin{itemize}[leftmargin=2.2em]
\item \emph{Propagation}: constrain deviations in GW speed, dispersion (massive graviton), and extra polarizations.
\item \emph{Ringdown}: test Kerr structure; look for echoes or additional modes.
\item \emph{Cosmology}: standard sirens yield \(\Hn\) with different systematics than distance ladders (needs redshifts from EM counterparts or statistical associations).
\item \emph{Phase transitions}: stochastic GW backgrounds probe beyond-SM first-order transitions at \(\Order(0.1\text{--}10)\ \TeV\) for LISA-band signals (model dependent).
\end{itemize}

\subsubsection{Dense matter, the nuclear equation of state, and strong-interaction physics}

Compact objects provide a unique laboratory for QCD at baryon densities inaccessible on Earth. Key observables include the neutron-star mass--radius relation \(M(R)\), tidal deformability, cooling, and post-merger GW spectra.

\paragraph{Tidal deformability as an EOS observable.}
For a neutron star of mass \(M\), radius \(R\), and Love number \(k_2\), the dimensionless tidal deformability is
\begin{equation}
\Lambda \equiv \frac{2}{3}k_2\Big(\frac{c^2 R}{G M}\Big)^5 .
\label{eq:tidalLambda}
\end{equation}
Binary-inspiral waveforms constrain combinations such as \(\tilde\Lambda\), providing EOS information that can be cross-validated with x-ray pulse-profile measurements and radio timing. ``Decisive'' progress requires joint inference that propagates waveform systematics, spin priors, and population selection effects, and that interfaces consistently to nuclear-theory priors (chiral EFT at low densities and phenomenological extensions at higher densities).

\section{Ground-based experiments,  astronomy, and in-situ space tests}
\label{sec:ground-space}

This section separates three observational modes that are often conflated.
(i) Ground-based controlled experiments: active control of geometry, fields, materials, and calibration,
typically enabling the strongest rejection of instrumental systematics.
(ii) Passive astronomical observations: the ``source'' is provided by nature, enabling extreme baselines
and strong-field environments at the cost of astrophysical nuisance modeling.
(iii) In-situ space experiments: controlled probes in long free-fall and large potential differences,
often changing the relevant scaling with interrogation time or baseline.
For each mode we state the accessible observables, the dominant noise/systematic terms, the key
scaling laws, and the realistic ceiling set by irreducible limitations.

\subsection{Measurement classes and relevant error budgets}
\label{sec:platform-taxonomy}

To keep the discussion physics-driven rather than technology-driven, it is useful to separate three measurement
classes by (i) what is actually observed, (ii) what dominates the nuisance model, and (iii) which sensitivity scalings
can be exploited without introducing comparably large systematics.

\begin{table*}[t]
\caption{Platform classes for fundamental-physics measurements. ``Reach'' is indicative and should be interpreted as decision-level only when the stated dominant systematics are controlled at commensurate levels.}
\label{tab:platform-taxonomy}
\centering
\renewcommand{\arraystretch}{1.12}
\setlength{\tabcolsep}{1pt}
\begin{tabular}{@{}llll@{}}
\hline
\tcell{0.12\textwidth}{Class} &
\tcell{0.255\textwidth}{Accessible observables} &
\tcell{0.28\textwidth}{Dominant limitations (typical)} &
\tcell{0.32\textwidth}{Realistic decision leverage} \\
\hline\hline

\tcell{0.13\textwidth}{Ground-based controlled experiments} &
\tcell{0.255\textwidth}{Clock ratios $y(t)$; differential ac- celerations $\Delta a(t)$; null asymme- tries (EDM/CLFV/PV); short-range gravity} &
\tcell{0.28\textwidth}{Environmental couplings (thermal, EM, vibration); gravity gradients/ self-gravity; calibration drifts; the- ory matching (hadronic/nuclear)} &
\tcell{0.30\textwidth}{Best for symmetry null tests and con- trolled reversals; decisive when config- uration changes separate signal from dominant backgrounds} \\

\tcell{0.13\textwidth}{Passive astrono- my (photons, neutrinos, GWs)} &
\tcell{0.255\textwidth}{Propagation (dispersion/birefri- ngence); spectra and line shifts; waveform consistency; populati- on statistics; CMB/LSS fields} &
\tcell{0.28\textwidth}{Source modeling and intrinsic lags; selection functions; cosmic variance; foregrounds; waveform and population systematics} &
\tcell{0.31\textwidth}{Unique access to strong fields, early universe, ultra-high energies; decisive when multi-messenger \& multi-band cross-checks break source degeneracies} \\

\tcell{0.13\textwidth}{In-situ space experiments (active probes)} &
\tcell{0.255\textwidth}{Redshift and potential modulation; EP tests in free fall; long-baseline interferometry/ time transfer; precision tracking} &
\tcell{0.28\textwidth}{Spacecraft systematics (thermal gradients, charging); drag-free/ attitude control couplings; orbit determination; link calibration} &
\tcell{0.32\textwidth}{Decisive when the terrestrial limitati- on is irreducible (seismic/Newtonian noise, gravity sag), a closed end-to-end error budget preserves the scaling gain} \\
\hline
\end{tabular}
\end{table*}

In the remainder of this section, ``ground vs space'' is therefore treated as an error-budget question:
space is enabling only when the dominant terrestrial floor is fundamental or extremely hard to engineer away,
and when the mission architecture provides built-in redundancy (differential channels, modulation strategies, and independent links).

\subsection{Space as a sensitivity lever: when scalings improve and when they do not}
\label{sec:space-levers}

Space is not intrinsically ``better'' than the ground; it changes a small set of control parameters that enter precision-sensitivity scalings and, crucially, it reshapes which systematics dominate the error budget. The scientific case for orbit is strongest when (i) the leading terrestrial limitation is effectively irreducible
(seismic/Newtonian noise, gravity sag, atmospheric fluctuations, gravity-induced gradients), or (ii) the desired observable requires large and well-calibrated modulations in free-fall time, baseline, or gravitational potential \cite{Turyshev2007IJMPD}.

\paragraph{The quantitative levers.}
To first order, the unique ``knobs'' enabled by space are:
(i) long coherent interrogation time \(T\) in near-free fall;
(ii) large baselines \(L\) (including inter-spacecraft links) and well-characterized potential differences \(\Delta U\);
and (iii) access to frequency bands where terrestrial environments impose hard floors (notably \(\lesssim 10~\mathrm{Hz}\) for gravitational and inertial measurements). None of these levers is useful unless the platform introduces
systematics that remain below the corresponding signal levels; therefore, the relevant comparison is always \emph{end-to-end} (payload + spacecraft + environment), not component-level performance.

\paragraph{Free fall and inertial sensors (atom interferometers and accelerometers).}
For a light-pulse atom interferometer (AI), the acceleration-induced phase is, schematically,
\begin{equation}
\Delta\phi_a \simeq k_{\rm eff}\, a\, T^2 ,
\label{eq:space_AI_phase}
\end{equation}
and the shot-noise-limited acceleration sensitivity scales as (cf.\ Sec.~\ref{sec:quantum-tech}, Eq.~\eqref{eq:AIaccel})
\begin{equation}
\delta a_{\rm SN} \sim \frac{1}{C\,k_{\rm eff}\,T^2\,\sqrt{N}} ,
\label{eq:space_AI_accel}
\end{equation}
where \(C\) is fringe contrast and \(N\) the detected atom number per shot. The central ``space gain'' is the \(T^2\) scaling: moving from \(T\sim 0.3\text{--}1~\mathrm{s}\) (typical on Earth once
vibrations and gravity sag are controlled) to \(T\sim 10~\mathrm{s}\) yields a raw gain of \(\sim 10^2\) in \(\delta a\);
pushing to \(T\sim 30\text{--}100~\mathrm{s}\) corresponds to a raw gain of \(\sim 10^3\text{--}10^4\). In practice, the realized gain is limited by effects that scale similarly to (or faster than) the signal: platform acceleration noise and rotation (Coriolis/Sagnac phases), wavefront aberrations and pointing jitter, gravity gradients and self-gravity (spacecraft mass distribution), magnetic and electric field gradients, and finite-temperature/expansion of the atomic ensemble. Thus, exploiting long \(T\) typically requires either (a) a drag-free reference and a low-noise control loop in the relevant band, or (b) a differential geometry (gradiometer / multi-cloud AI) that suppresses common-mode platform motion while preserving sensitivity to the targeted signal (e.g., EP-violating differential accelerations, ultra-light-field forces).

\paragraph{Clocks: gravitational redshift, potential modulation, and networks.}
For clocks, the primary space lever is the gravitational redshift,
\begin{equation}
\frac{\Delta f}{f} = \frac{\Delta U}{c^2},
\label{eq:redshift_basic}
\end{equation}
and the ability to realize large, well-modeled \(\Delta U\) while maintaining stable time/frequency transfer. Representative Earth-orbit potential differences relative to the geoid are
\(\Delta U/c^2 \simeq 4.1\times 10^{-11}\) for a \(h\simeq 400~\mathrm{km}\) LEO orbit and
\(\Delta U/c^2 \simeq (5\text{--}6)\times 10^{-10}\) for MEO/GEO-class altitudes; the asymptotic limit
to infinity is \(\Delta U/c^2 \to GM_\oplus/(R_\oplus c^2)\simeq 6.9\times 10^{-10}\).
A fractional clock uncertainty \(\delta y\) corresponds to a resolvable potential difference
\begin{equation}
\delta U \simeq c^2\,\delta y \;\;\Rightarrow\;\;
\delta h \simeq \frac{\delta U}{g} \simeq 0.9~\mathrm{cm}\left(\frac{\delta y}{10^{-18}}\right)
\left(\frac{9.8~\mathrm{m\,s^{-2}}}{g}\right),
\label{eq:clock_height_equiv}
\end{equation}
so \(10^{-18}\) comparisons are, in principle, sensitive to centimeter-equivalent potential changes. However, \emph{redshift tests and new-physics searches are transfer-limited unless} the time/frequency link (ground--space or space--space) reaches comparable stability and calibration (typically \(\lesssim 10^{-18}\) over the relevant averaging time) and unless relativistic orbit determination and propagation corrections are controlled at the same level. Space also enables deliberate potential modulation (e.g., eccentric orbits) and long-baseline clock networks, which are valuable for searches for oscillating/transient signals in fundamental constants (ultra-light fields) because common-mode instrumental noise can be rejected and coherence times can be exploited \cite{Turyshev2024PhRvD-tetra}.

\paragraph{Low-frequency observables and long baselines (including GWs).}
Below \(\sim 10~\mathrm{Hz}\), ground experiments face hard environmental floors (seismic motion and Newtonian
gravity-gradient noise), motivating space platforms for both GW and non-GW inertial measurements. For laser-interferometric GW detectors, the basic scaling \(\delta L \sim h\,L\) makes the advantage of large baselines explicit: increasing \(L\) by orders of magnitude relaxes the displacement-noise requirement for a fixed strain target \(h\), while shifting the instrument into bands inaccessible from the ground (Sec.~\ref{sec:GWs}). For precision force measurements and very-low-frequency inertial signals, orbit removes atmospheric loading and many near-field couplings, enabling long integration times and clean modulation strategies, but it does \emph{not} remove the need for careful gravity-gradient and self-gravity control (spacecraft mass distribution and thermal drifts can dominate).

\paragraph{Space-specific systematics and the ``credible space case'' criterion.}
Space introduces distinct and often mission-critical systematics: radiation-induced degradation and charging, spacecraft charging/plasma interactions, thermal gradients and thermo-elastic drift, limited maintenance/debugging, tight mass/power/thermal margins, and coupling between spacecraft attitude/drag-free control loops and the payload. Accordingly, a credible space case for a given fundamental-physics parameter is not ``larger \(T\), \(L\), or \(\Delta U\)'' in isolation, but a demonstrated net gain in the \emph{total} uncertainty after propagating:
(i) platform noise into the science channel,
(ii) calibration and orbit-determination uncertainties into relativistic corrections, and (iii) realistic rejection of correlated disturbances.
The highest-return space opportunities are therefore those where the targeted observable is fundamentally band-limited or modulation-limited on Earth, and where the mission architecture provides a transparent, closed error budget with built-in cross-checks (e.g., differential channels, redundant links, or multi-platform consistency).

\subsection{Astronomical messengers as fundamental physics probes}

Astronomical facilities are decisive when they (i) open an observable window (new frequency band, new messenger,
new redshift regime) or (ii) reduce calibration/foreground/systematic floors so that inference is limited by cosmic variance or by a controlled nuisance model rather than by uncontrolled systematics. Decision-grade progress therefore requires explicit ``closure tests'': 
(1) redundant observables with shared nuisance parameterizations;
(2) end-to-end injection/recovery on simulations spanning astrophysical and instrumental uncertainties; and
(3) cross-calibration between surveys/instruments with independent systematics.

Concrete near- to mid-term targets include:
(i) CMB polarization measurements pushing $r \lesssim 10^{-3}$ and $\sigma(\Delta N_{\rm eff})\sim 0.03$--$0.06$
with validated foreground separation;
(ii) large-scale structure and weak lensing analyses achieving percent-level cross-probe closure on growth and geometry
under explicit nonlinear+baryonic nuisance models;
(iii) standard-siren cosmology reaching sub-percent $H_0$ only if redshift association and selection functions
are jointly modeled and cross-validated;
(iv) multi-messenger constraints (EM+GW+neutrinos) that become decisive only when population priors and selection
effects are propagated consistently through the likelihood.

\subsubsection{Photons}
Electromagnetic observations constrain fundamental physics primarily through propagation effects (dispersion and birefringence), high-resolution spectroscopy (searches for variations of constants), and the CMB as a calibrated snapshot of the primordial perturbations.

\paragraph{Spectroscopy and variation of constants.}
Quasar absorption systems and molecular lines constrain \(\Delta\alpha/\alpha\) and \(\Delta\mu/\mu\) over gigaparsec baselines. Current bounds at the \(\sim 10^{-6}\) level are typically dominated by wavelength-calibration and astrophysical systematics; a decisive improvement requires stabilized spectrographs, frequency-comb calibration, and control of kinematic segregation in absorbers. Laboratory optical clocks provide complementary local tests with much cleaner systematics and well-defined coupling maps (see Sec.~\ref{sec:quantum-tech}).

\paragraph{High-energy photon propagation.}
Time-of-flight dispersion tests and polarization-based birefringence tests constrain Lorentz/CPT-violating operators and quantum-gravity-inspired dispersion. The cleanest leverage comes from short-duration transients (GRBs, flares) with broad energy coverage and well-characterized intrinsic emission lags.
For many operators, the limiting factor is no longer photon counting but source modelling and instrument timing calibration.

\paragraph{CMB temperature and polarization.}
The CMB remains the most controlled arena for primordial physics: tensors (\(r\)), light relics (\(\Delta \Neff\)),
isocurvature, and parity-violating birefringence.  The decision thresholds and limiting systematics (foreground separation, beam calibration, polarization angle calibration, and \(1/f\) noise) are discussed in Sec.~\ref{sec:sensitivity} and in the early-universe subsection of Sec.~\ref{sec:exp-frontiers}.

\subsubsection{High-energy astrophysical probes of dark sectors (beyond GW/cosmology)}
\label{sec:he_astro_darksectors}

High-energy photons, cosmic rays, and neutrinos constrain dark sectors in ways that are complementary to direct detection: they probe \emph{annihilation/decay} and \emph{portal-mediated production} rather than elastic scattering. The fundamental-physics content is encoded in an emission model, an astrophysical ``$J$-factor'' (or $D$-factor), and a structured nuisance model for backgrounds.

\paragraph{Canonical flux mapping (annihilation/decay).}
For annihilation into a spectrum $\dd N/\dd E$, the differential photon flux is
\begin{equation}
\frac{\dd\Phi}{\dd E} =
\frac{\langle\sigma v\rangle}{8\pi m_\chi^2}\,
J\,\frac{\dd N}{\dd E},
\qquad
J\equiv \int_{\Delta\Omega}\dd\Omega\int_{\rm l.o.s.}\rho_\chi^2(s)\,\dd s,
\end{equation}
while for decay one has
$\dd\Phi/\dd E = (1/4\pi m_\chi\tau_\chi)\,D\,\dd N/\dd E$ with $D=\int \rho_\chi\,\dd s$.
Thus the decisive nuisance parameters are: the halo profile (dominant for the GC), substructure boosts (cluster-scale),
and the instrument+astrophysical foreground model.

\paragraph{Why this belongs in a ``fundamental physics'' roadmap.}
These probes are uniquely sensitive to:
(i) heavy DM masses well above collider reach,
(ii) velocity-dependent annihilation (Sommerfeld/enhanced scenarios),
(iii) long-lived particles and dark cascades that produce distinctive spectra,
(iv) Lorentz-violation / quantum-gravity dispersion constraints from transients. They become decision-grade only when multiple targets (dwarfs vs GC vs clusters) and multiple messengers
($\gamma$+$\nu$+CR) produce consistent inferences under shared nuisance models.

\subsubsection{Neutrinos}
Astrophysical neutrinos probe both particle physics and dense-matter astrophysics in regimes that cannot be reproduced on Earth. The core strength is the \emph{lever arm}: long baselines and extreme energies/densities, with complementary systematics compared to laboratory beams.

\paragraph{Low-energy (solar, atmospheric, accelerator-connected) neutrinos.}
These constrain mixing and matter effects and provide consistency checks across production environments.
At present, the limiting uncertainties are often cross sections and flux modelling rather than detector statistics.

\paragraph{MeV neutrinos from core-collapse supernovae.}
A Galactic supernova would provide \(\mathcal{O}(10^3\text{--}10^5)\) events depending on detector class (water Cherenkov, liquid scintillator, LAr), enabling time- and energy-resolved tests of neutrino transport, collective oscillations, and exotic cooling channels. Here the ``sensitivity'' is dominated by \emph{readiness}: uptime, low backgrounds, and calibrated response when a rare event occurs.

\paragraph{TeV--PeV neutrinos.}
High-energy astrophysical neutrinos probe neutrino--nucleon interactions at center-of-mass energies beyond accelerators and can constrain BSM interactions that modify attenuation through the Earth or flavor composition.  Robust interpretation requires joint modelling of source populations, detector acceptance, and Standard-Model uncertainties in parton distributions at small \(x\).

\subsubsection{Gravitational waves}
Gravitational waves (GWs) are both a new messenger and a precision tool for strong-field gravity. They constrain the propagation sector (speed, dispersion, extra polarizations), the generation sector (inspiral/ringdown consistency with Kerr),
and cosmology (standard sirens)  \cite{Schutz1986StandardSirens}. The ``astronomical signal'' aspect is crucial: interpretation depends on source populations and astrophysical priors. A technically robust program therefore requires multi-band and multi-messenger datasets (ground mHz--kHz, PTAs nHz, and EM counterparts where available) analyzed with shared waveform systematics and population models, so that putative deviations from GR can be tested against astrophysical alternatives rather than fit in isolation. As in Sec.\ \ref{sec:GWs}, GWs provide access to strong gravity and early-universe phase transitions; multi-band detection (space + ground) adds parameter leverage.

\subsection{In-situ space experiments: EP, redshift, time/frequency transfer}

\subsubsection{Equivalence principle and inverse-square law}
The weak equivalence principle (WEP) is commonly quantified by the E\"otv\"os parameter for two test bodies,
\begin{equation}
\eta \equiv 2\frac{a_1-a_2}{a_1+a_2} \simeq \frac{\Delta a}{g},
\label{eq:eotvos}
\end{equation}
while short-range deviations from Newtonian gravity are often parameterized by a Yukawa correction
\(V(r)=-(Gm_1m_2/r)\left[1+\alpha\,e^{-r/\lambda}\right]\).
MICROSCOPE reached \(|\eta|\sim 10^{-15}\) for composition dipoles in orbit, which already constrains a wide class of light scalar/vector ``fifth-force'' models when mapped onto composition-dependent charges.

A comprehensive ``state of the art'' assessment should also track complementary regimes: lunar laser ranging (LLR)
tests of the strong equivalence principle via Earth--Moon free fall toward the Sun (sensitive to gravitational self-energy), and solar-system radio science constrains post-Newtonian parameters (e.g.\ the Shapiro time delay constraint on \(\gamma-1\) at the \(\sim 10^{-5}\) level \cite{Bertotti2003Cassini}). Binary pulsars and pulsar--white-dwarf systems provide independent strong-field constraints on dipolar radiation and time variation of \(G\), and are therefore a key cross-check on any modified-gravity interpretation of cosmological data.

\paragraph{Lunar laser ranging (LLR) and strong-EP tests.}
LLR measures the Earth--Moon distance by timing laser pulses returned from lunar retroreflector arrays, enabling precision reconstruction of the lunar orbit and rotation \cite{Merkowitz2010LLR}. This provides sensitive tests of relativistic gravity, in particular strong-equivalence-principle (Nordtvedt-effect) tests of differential free fall of the Earth and Moon toward the Sun (including sensitivity to gravitational self-energy) \cite{Williams2004LLRPRL,Williams2009LLREarthMoon}, along with constraints on possible time variation of $G$ and bounds on post-Newtonian phenomenology through detailed orbital modeling. Modern analyses and dedicated facilities (e.g., APOLLO) have pushed LLR into the millimeter-precision regime \cite{Murphy2012APOLLO,Mueller2012LLRRelativity,WilliamsTuryshevBoggs2012CQG}, and next-generation concepts using higher-return laser links (including high-power CW approaches \cite{Turyshev2025HighPowerLLR,Turyshev2025HighPowerLLR-ErBu}) aim to further improve signal statistics and reduce systematics, strengthening the complementarity of LLR with laboratory and in-orbit EP tests.

Beyond its classic sensitivity to the strong equivalence principle (SEP), the multi-decade LLR data set functions as a global-fit experiment for relativistic gravity in the solar system: it constrains post-Newtonian dynamics (e.g. combinations of PPN parameters), searches for a possible time variation $\dot G/G$, tests relativistic orbital precessions (including geodetic/de Sitter terms) \cite{Williams2004LLRPRL,Williams2009LLREarthMoon,BiskupekMuellerTorre2020LLR}, and provides
competitive bounds on preferred-frame and Lorentz-symmetry-violating effects in gravity (e.g. in the
SME framework) \cite{BattatChandlerStubbs2007PRL,BaileyEverettOverduin2013LLR_Lorentz}. In contrast to single-parameter laboratory null tests, LLR is typically limited by a coupled modeling  framework (ephemerides, station time-transfer, atmosphere, and lunar geophysics), so decisiveness comes from multi-station redundancy and independent analysis pipelines.

\paragraph{What is realistically needed next?}
A credible next step beyond \(|\eta|\sim10^{-15}\) requires an explicit systematics budget showing simultaneous control of:
(i) differential accelerometer drifts and thermal gradients, (ii) electrostatic charging and patch potentials,
(iii) gravity-gradient couplings and spacecraft mass-motion, and (iv) composition dependence across multiple test pairs.
On the ISL side, improvements at \(\lambda\sim 10\text{--}100~\mu{\rm m}\) remain dominated by control of electrostatic, Casimir, and seismic backgrounds, motivating continued short-range torsion-balance and micro-cantilever programs in parallel with any space concept.

\subsubsection{Antimatter gravity: free fall of antihydrogen}
Testing whether neutral antimatter falls with the same acceleration as matter probes the universality of free fall in a qualitatively new regime and constrains exotic long-range forces (e.g.\ vector forces coupled to baryon/lepton number) that could masquerade as ``antigravity'' in some effective descriptions. The ALPHA-g experiment reported a direct measurement of the vertical gravitational acceleration of antihydrogen,
with best-fit
\begin{equation}
g_{\bar H} = \left(0.75 \pm 0.13_{\rm stat+sys} \pm 0.16_{\rm sim}\right) g,
\end{equation}
consistent with attractive gravity but at \(\sim 25\%\) precision \cite{ALPHAg2023Nature}. Reaching the \(\lesssim 1\%\) regime requires order-of-magnitude reductions in initial-condition and magnetic-field-gradient systematics, as well as complementary approaches (AEgIS, GBAR) with different systematics.

\subsubsection{Gravitational redshift with clocks}
The gravitational redshift between two potentials differs by
\begin{equation}
\frac{\Delta f}{f} \simeq \frac{\Delta U}{c^2}.
\end{equation}
Thus a clock comparison at \(10^{-18}\) fractional frequency corresponds to sensitivity to potential differences \(\Delta U/c^2\sim 10^{-18}\), i.e.\ height differences of order centimeters on Earth (\(\Delta U\sim g\Delta h\)). Space missions enable much larger \(\Delta U\) modulation across eccentric orbits, turning clock precision into tests of GR and searches for ultra-light fields coupled to SM parameters.

\subsubsection{Time/frequency transfer: fiber, free-space, and satellite links}

A global clock network requires transfer stability at or below clock instability. State-of-the-art fiber frequency transfer reaches fractional instabilities at or below \(10^{-19}\) in favorable conditions (demonstrations include \(\sim 3\times10^{-19}\) in deployed fiber contexts \cite{Hoghooghi2024MCF} and \(\sim 3\times10^{-19}\) on long-haul links \cite{Calonico2014Fiber}). Satellite links remain a bottleneck for global comparisons; ACES (installed on the ISS in April 2025 for a $\sim30$-month mission) combines PHARAO (cold-atom Cs fountain) with a space hydrogen maser and microwave+laser links, enabling $\sim10^{-17}$ comparisons within a few days when stations are simultaneously in view, and $\sim$week-scale intercontinental comparisons---about 1--2 orders of magnitude beyond GNSS-based timing links \cite{ACESESA}.

What is needed for a decisive ``clock cosmology / DM'' program? The following elements are useful: (1) optical time/frequency transfer with stability \(\lesssim 10^{-18}\) over intercontinental baselines (space or hybrid fiber+space),
(2)  multi-clock comparisons with different sensitivity coefficients to \(\alphaem\), \(\mu\), nuclear parameters,
(3)  synchronized networks to search for transient signals (topological defects) and coherent oscillations (ultralight scalars).

\section{Coherent sensors (clocks, atom interferometers, and quantum-limited readout) as parameter probes}
\label{sec:quantum-tech}

We treat ``quantum technology'' instruments only insofar as they deliver new or improved constraints on well-defined fundamental-physics parameters. The relevant distinction is operational: these devices use long-lived quantum coherence (or quantum-limited readout) to convert small Hamiltonian perturbations into measurable phase or frequency shifts. The resulting reach is set by (i) the mapping from a model parameter to an effective phase/frequency perturbation, (ii) how the signal scales with interrogation time, baseline, or potential difference, and (iii) the dominant systematic floor and time-transfer noise.

We use the phrase \emph{quantum sensor} only when the demonstrated sensitivity is traceable to a physics-based advantage (e.g.\ coherence time, squeezed readout reducing projection noise, quantum-limited amplifiers), and we state explicitly whether the measurement is statistics-limited or systematics-limited in the regime of interest.

\subsection{Terminology and criteria: when does ``quantum'' provide a physics advantage?}
\label{sec:quantum-terminology}

The term ``quantum sensor'' is used in this review only when a demonstrable, physics-based advantage follows from
(i) long phase coherence time, (ii) quantum-noise-limited readout (projection noise, shot noise, or their squeezed/entangled variants), or (iii) access to a coupling channel that is intrinsically quantum (e.g., spin-dependent interactions). In particular, the label is \emph{not} used as a synonym for ``new technology'': many mature instruments (e.g., SQUIDs, masers, lasers) are quantum at the microscopic level but provide fundamental-physics leverage only when their noise model and systematics
enable a cleaner constraint on a well-defined parameter (EFT Wilson coefficient, equivalence-violation parameter, or coupling constant).

Two examples make the point quantitatively:

\paragraph{Redshift/EEP tests with clocks.}
A common parameterization of local position invariance violation writes
\begin{equation}
\frac{\Delta f}{f} = (1+\alpha_{\rm LPI})\frac{\Delta U}{c^2}.
\end{equation}
A measurement with fractional uncertainty $\delta y$ constrains $\delta\alpha_{\rm LPI}\sim \delta y / (\Delta U/c^2)$.
Thus, even with $\delta y\sim 10^{-18}$, a decisive improvement requires a \emph{transfer- and orbit-determination}
error budget at or below $10^{-18}$ on the relevant averaging time; otherwise the gain in clock accuracy does not translate
into a gain in $\alpha_{\rm LPI}$.

\paragraph{EP tests with atom interferometers.}
For light-pulse interferometers, $\delta a\sim 1/(k_{\rm eff}T^2\sqrt{N})$ is only the starting point: reaching $\eta\sim 10^{-17}$ implies differential phases at the $10^{-7}$ rad level for representative $(k_{\rm eff},T)$, so the decisive question is whether wavefront, Coriolis, gravity-gradient, and platform-noise systematics can be bounded below that level by reversals, auxiliary metrology, and independent cross-check configurations.

\subsection{Precision clocks}
\label{sec:clocks}

\paragraph{Parameter map.}
Clock comparisons constrain (a) gravitational redshift / local position invariance through $\Delta\nu/\nu=(1+\alpha)\Delta U/c^2$ (with possible clock-dependent $\alpha$), (b) time variation of dimensionless constants through the sensitivity coefficients of the relevant transitions, and (c) couplings of ultralight fields that induce oscillatory or transient shifts of those constants. The interpretation therefore depends on the clock species and transition (which fixes the coefficients), and on the fidelity of time/frequency transfer (which sets the correlation structure across sites).

\paragraph{Scaling and limiting noise.}
For averaging times where the instability is dominated by uncorrelated noise, the statistical sensitivity improves as $\tau^{-1/2}$; however, for fundamental-physics applications the limiting floor is often a systematic offset (BBR shifts, Stark/Zeeman shifts, collisional shifts) and the link noise in long-distance comparisons. For redshift tests, the physically relevant lever arm is $\Delta U/c^2$, so the comparison between ground and space is set by achievable potential differences and the transfer noise budget, not by the clock fractional stability alone.

\paragraph{Incremental vs.\ transformative.}
Optical clocks are transformative when they enable a qualitatively new observable class (e.g.\ correlated networks for ultralight-field transients, or redshift tests with large $\Delta U/c^2$ modulation). They are incremental when they primarily refine an already over-constrained parameter with the same degeneracy and
systematic structure.

\subsubsection{Optical clocks: Performance metrics}
Two key metrics: (1)  \emph{Systematic uncertainty} (accuracy): fractional bias control, now at or below \(10^{-18}\) in leading systems. (2)  \emph{Stability} (Allan deviation): \(\sigma_y(\tau)\sim \sigma_0/\sqrt{\tau}\) at short times, until limited by oscillator noise, environment, or systematics.
Quantum projection noise (QPN) gives
\begin{equation}
\sigma_y^{\rm QPN}(\tau)\sim \frac{1}{Q}\sqrt{\frac{T_c}{N\,\tau}},
\end{equation}
where \(Q\) is the quality factor, \(T_c\) the cycle time, and \(N\) atom number (or effective number with entanglement). Squeezing can improve scaling beyond standard quantum limit.

\subsubsection{Systematics budgets (representative)}
Dominant shifts include blackbody radiation (BBR), lattice Stark shifts (for lattice clocks), Zeeman shifts, density shifts, probe light shifts, and relativistic Doppler/gravity potentials. Achieving \(10^{-19}\) class accuracy generally requires:
(1)  sub-Kelvin effective BBR environment knowledge (or cryogenic operation),
(2)  high-fidelity field control and in-vacuum thermometry,
(3)  improved atomic data and shift coefficients.

State-of-the-art optical lattice clocks are now reporting systematic uncertainty below \(10^{-18}\) (e.g.\ \(\sim 8\times 10^{-19}\) in a strontium system), with demonstrated \(10^{-19}\)-class stability at \(\tau\sim 10^4\ \mathrm{s}\) under favorable conditions \cite{SrClock2024PRL}.

\subsubsection{Physics enabled}
Precision clocks allow for a series of important experiments, including  (1) tests of GR redshift at unprecedented precision;
(2)  constraints on time variation of constants (laboratory limits have reached \(\lesssim 10^{-18}/\yr\) in some analyses \cite{DzubaAlphaVar2023});
(3) searches for ultralight DM via oscillating constants and transient events \cite{Turyshev2025PhRvD};
(4)  detection of tiny forces via chronometric geodesy (potential mapping).

\subsubsection{Ultralight fields with clocks: oscillations and transients}

A particularly quantitative mapping exists between clock performance and couplings of ultralight bosonic fields that may constitute some or all of the dark matter \cite{Turyshev2025PhRvD}. For a coherently oscillating scalar field \(\phi\) with mass \(m_\phi\),
\begin{equation}
\phi(t)\simeq \phi_0 \cos(m_\phi t), \qquad 
\phi_0 \simeq \frac{\sqrt{2\rho_{\rm DM}}}{m_\phi},
\label{eq:phiDM}
\end{equation}
where \(\rho_{\rm DM}\) is the local dark-matter density. Linear couplings to SM parameters can be parameterized as
\begin{equation}
\frac{\delta \alphaem}{\alphaem}= d_e\,\frac{\phi}{\Lambda_\phi},\qquad
\frac{\delta m_e}{m_e}= d_{m_e}\,\frac{\phi}{\Lambda_\phi},\qquad
\frac{\delta m_q}{m_q}= d_{m_q}\,\frac{\phi}{\Lambda_\phi},\ \ldots
\label{eq:scalarCouplings}
\end{equation}
which induce oscillations of clock transition frequencies. For a ratio of two clock frequencies,
\begin{equation}
\frac{\delta(\nu_1/\nu_2)}{\nu_1/\nu_2}
\simeq \Delta K_\alpha\,\frac{\delta\alphaem}{\alphaem}
+ \Delta K_\mu\,\frac{\delta\mu}{\mu}
+ \Delta K_q\,\frac{\delta X_q}{X_q}+\cdots ,
\label{eq:clockDM}
\end{equation}
where \(\Delta K\) are known sensitivity-coefficient differences and \(X_q\) denotes relevant nuclear parameters. This makes a global network of heterogeneous clocks (different \(\Delta K\)) with transfer stability \(\lesssim 10^{-18}\) a directly interpretable probe of \((d/\Lambda_\phi)\) over a wide mass range (from periods shorter than the clock cycle time up to coherence
times of years).

\subsection{Atom interferometers (AIs)}
\subsubsection{Basic phase scaling}
For a light-pulse Mach-Zehnder AI measuring acceleration \(a\),
\begin{equation}
\Delta \phi \simeq \keti\, a\, T^2,
\label{eq:AIphase}
\end{equation}
so shot-noise-limited acceleration sensitivity is
\begin{equation}
\delta a \sim \frac{1}{\keti T^2 \sqrt{N}},
\label{eq:AIaccel}
\end{equation}
with \(N\) detected atoms and interrogation time \(T\). Large momentum transfer (LMT) increases \(\keti\). Microgravity enables larger \(T\) without large apparatus, potentially giving orders-of-magnitude improvement.

\paragraph{E\"otv\"os-parameter reach and required phase sensitivity.}
The EP figure of merit is the E\"otv\"os parameter $\eta$ defined in Eq.~\eqref{eq:eotvos}. For $\eta\ll 1$ one has
$\eta\simeq \Delta a/g$, so a target $\eta\sim 10^{-17}$ corresponds to $\Delta a\sim 10^{-16}\,\mathrm{m\,s^{-2}}$.
For representative parameters $k_{\rm eff}\sim 10^{7}\,\mathrm{m^{-1}}$ (LMT-enhanced) and $T\sim 10\,\mathrm{s}$
(microgravity), the implied differential phase requirement is $\delta\phi\sim k_{\rm eff}\Delta a\,T^2\sim 10^{-7}\ \mathrm{rad}$.
Achieving this level requires extreme common-mode vibration rejection and in-situ closure on dominant systematics
(wavefront/pointing, Coriolis terms, magnetic gradients, and gravity-gradient/self-gravity effects) through configuration
reversals and auxiliary metrology.

\paragraph{Representative implementations and realistic scale-up paths.}
On the ground, current and near-term instruments provide both physics reach and technology validation. Examples include \emph{MAGIS-100} (a 100\,m-scale atom interferometer under construction) and \emph{AION-10} (a 10\,m class strontium AI designed as a pathfinder for larger baselines). These facilities directly test key subsystems that dominate real performance---large-momentum-transfer optics, phase noise, wavefront control, common-mode vibration rejection, and long-term stability---and therefore de-risk extrapolations to \(\mathcal{O}(100\text{--}1000\,\mathrm{m})\) baselines for mid-band GW/ULDM searches.  Space concepts such as \emph{AEDGE} and \emph{SAGE} propose to exploit microgravity to extend \(T\) and to access cleaner low-frequency environments, but remain contingent on drag-free and thermal-control performance at levels comparable to the targeted AI noise budgets
\cite{AbeMAGIS2021,AION10TDR2025,AbouElNeajAEDGE2020,TinoSAGE2019}.

\subsubsection{Dominant systematics and noise sources}
For a three-pulse ($0,T,2T$) light-pulse interferometer, inertial signals enter through the sensitivity function
$g(t)$, giving a phase response to platform acceleration $a(t)$ along $\mathbf{k}_{\rm eff}$,
\begin{equation}
\Delta\phi_a = k_{\rm eff}\!\int_0^{2T}\!dt\, g(t)\, a(t),\qquad
g(t)=
\begin{cases}
t & 0<t<T,\\
2T-t & T<t<2T,
\end{cases}
\end{equation}
and an acceleration-noise variance
\begin{equation}
\sigma_\phi^2 \simeq k_{\rm eff}^2 \int_0^\infty df\; S_a(f)\,
\frac{16\,\sin^4(\pi f T)}{(2\pi f)^4}.
\end{equation}
This makes explicit why increasing $T$ boosts signal $\propto T^2$ while simultaneously tightening low-frequency
requirements on residual acceleration noise.

\paragraph{Platform acceleration and vibration.}
Residual mirror/platform motion couples through the transfer function above. Ground-based instruments rely on isolation and feed-forward; space instruments translate this requirement into drag-free performance and attitude control at the frequencies that dominate the sensitivity integral.

\paragraph{Rotation and Coriolis/Sagnac terms.}
A residual rotation $\boldsymbol{\Omega}$ and transverse velocity $\mathbf{v}_\perp$ generate a Coriolis phase of order
$\Delta\phi_\Omega \sim 2\,k_{\rm eff}(\mathbf{v}_\perp\times\boldsymbol{\Omega})\!\cdot\!\hat{\mathbf{k}}\,T^2$.
Controlling (or modulating) $\boldsymbol{\Omega}$ and matching $\mathbf{v}_\perp$ across species is therefore essential in differential measurements.

\paragraph{Wavefront aberrations and beam-pointing jitter.}
Spatially varying laser phase (wavefront curvature/aberrations) combined with finite-temperature atomic clouds produces both biases and excess phase noise. The scaling is set by the cloud size and velocity spread together with optical quality and pointing stability; mitigation typically requires large beam waists, high-quality optics, active pointing control, and in-sit characterization (e.g.\ $k$-reversal, auxiliary wavefront diagnostics, or multi-configuration null tests).

\paragraph{Gravity gradients and self-gravity.}
Tidal fields $\Gamma_{ij}=\partial_i\partial_j U$ couple to finite arm separation and to species-dependent trajectory differences, leading to apparent differential accelerations $\delta a\sim \Gamma\,\delta z$ and contrast loss. In addition to the terrestrial gradient ($\Gamma_\oplus\sim 3\times 10^{-6}\,\mathrm{s^{-2}}$ vertically), local mass distributions (vacuum hardware, nearby source masses, spacecraft) must be modeled and controlled at a level commensurate with the target $\eta$.

\paragraph{Internal-state shifts and magnetic/electric backgrounds.}
State-dependent Zeeman/Stark shifts and field gradients can mimic inertial phases if correlated with the pulse sequence.
Dual-species operation suppresses common-mode effects, but imperfect matching of internal states and trajectories limits rejection, so stability/monitoring of fields and deliberate modulation strategies are usually part of the systematics budget.

\subsubsection{Physics enabled}
Long-baseline atom interferometers function as differential accelerometers and gradiometers. The same phase response used for inertial sensing can be mapped onto new-physics observables through controlled changes of species, internal states, geometry, and source-mass configurations.

\paragraph{Equivalence-principle and composition tests.}
Using two atomic species (or two internal states), one targets the E\"otv\"os parameter via $\Delta\eta \simeq \Delta\phi/(k_{\rm eff} g T^2)$. The discriminant for new physics is robustness under null tests such as $k$-reversal, deliberate variation of trajectories, and comparisons across multiple species choices, rather than a single headline sensitivity number.

\paragraph{Fifth forces and screened interactions.}
With a nearby source mass or modulated ambient potential, AIs probe non-Newtonian forces (e.g.\ Yukawa-type modifications) by measuring induced differential accelerations at the modulation frequency. The reach is set by the smallest resolvable $\delta a$ together with geometric modeling of the source mass and control of self-gravity and gradients.

\paragraph{Mid-band gravitational waves.}
In a gradiometer geometry with baseline $L$, a GW strain $h$ induces a tidal acceleration $a_{\rm GW}\sim (L/2)\,\ddot h$, leading to a differential phase scaling $\Delta\phi_{\rm GW}\sim k_{\rm eff}(L/2)(2\pi f)^2 h\,T^2$. This highlights the complementarity to optical laser interferometers: sensitivity improves with $L$, $k_{\rm eff}$, and $T$, but requires extremely low residual platform noise in the relevant band.

\paragraph{Geodesy and inertial navigation as enabling capabilities.}
Even when the primary target is fundamental physics, precision gravity/gradient/rotation sensing is often needed to calibrate self-gravity models and to maintain long-term stability in EP and force-search configurations.

\subsection{Quantum-limited electrometry and magnetometry}

SQUIDs, SERF magnetometers, NV-center ensembles, trapped ions, and superconducting microwave/LC resonators enable searches for axion/ALP couplings, ultra-light dark sectors, and exotic spin-dependent interactions. The relevant ``instrument performance'' variables are typically a field (or flux) noise spectral density, an effective sensing volume, bandwidth, and a validated model for magnetic/electric backgrounds.

\paragraph{Noise and scaling.}
For a broadband magnetometer with (one-sided) field-noise spectral density \(S_B(f)\) (in \(\mathrm{T^2/Hz}\)),
the minimum detectable coherent near-monochromatic field amplitude after integration time \(T_{\rm int}\) scales as
\begin{equation}
B_{\rm min}(f) \sim \sqrt{\frac{S_B(f)}{T_{\rm int}}}\,,
\end{equation}
up to \(\mathcal{O}(1)\) factors from filtering and signal coherence. Typical state-of-the-art sensitivities are
in the fT/\(\sqrt{\rm Hz}\) range for SERF and SQUID magnetometers in their optimal bands, and in the pT/\(\sqrt{\rm Hz}\) range for NV ensembles near dc; the frequency dependence matters as much as the headline number.

\paragraph{Ultra-light fields and coherence.}
Ultra-light bosonic DM behaves as a classical oscillatory field with quality factor \(Q_a\sim 10^6\), so the intrinsic
coherence time is \(\tau_{\rm coh}\sim Q_a/\omega\). This motivates two complementary strategies:
(i) broadband searches for transient or wideband signals (topological defects, bursts, broadband couplings), and
(ii) resonant enhancement (high-\(Q\) circuits or spin precession near resonance) for narrowband coherent signals, with
scanning strategies that properly account for trials factors and astrophysical priors on the DM velocity distribution.

\paragraph{Physics channels that are realistically discriminating.}
Examples include: axion/ALP couplings to spins that appear as an oscillating pseudo-magnetic field in comagnetometers; oscillating nuclear EDMs and spin-precession signals (NMR-based concepts); and kinetically mixed dark photons that can drive currents/fields in resonant circuits. In all cases, a ``definitive'' claim requires demonstrated scaling with control parameters (e.g.\ applied static fields, shielding configuration, resonance tuning), reproducible phase relations across separated sensors, and null channels that bound magnetic pickup and environmental correlations.

\paragraph{Enabling technologies and systematics.}
The dominant technical risks are environmental magnetic-field noise, Johnson noise in shields, vibration-to-field coupling, and calibration drifts. Progress is therefore coupled to improved magnetic shielding (including active compensation), cryogenic low-loss resonators, low-noise optical readout, and cross-site correlation networks. Space deployment is potentially useful primarily when it enables a cleaner low-frequency environment and long coherent integration, but it must close a full error budget including spacecraft magnetic cleanliness and charging.

\section{Sensitivity targets for ``definitive'' answers}
\label{sec:sensitivity}

Table \ref{tab:targets} expands the logic: it states \emph{what would constitute a decisive measurement} and the approximate sensitivity scale needed, with cross-check requirements.

% assumes this is already defined (once in preamble):
% \newcommand{\tcell}[2]{\parbox[t]{#1}{\raggedright\strut #2\strut}}

\renewcommand{\arraystretch}{1.10}
\setlength{\tabcolsep}{1pt}

\begin{longtable}{@{}llll@{}}
\caption{Selected decision-level targets (non-exhaustive).}
\label{tab:targets}\\
\hline
\tcell{0.17\linewidth}{Question} &
\tcell{0.17\linewidth}{Observable} &
\tcell{0.24\linewidth}{Sensitivity scale} &
\tcell{0.35\linewidth}{Notes on decisiveness / cross-checks}\\
\hline\hline
\endfirsthead

\hline
\tcell{0.17\linewidth}{Question} &
\tcell{0.17\linewidth}{Observable} &
\tcell{0.24\linewidth}{Sensitivity scale} &
\tcell{0.35\linewidth}{Notes on decisiveness / cross-checks}\\
\hline\hline
\endhead

\tcell{0.17\linewidth}{Is DM a WIMP?} &
\tcell{0.17\linewidth}{\(\sigSI(m_\chi)\) + modula- tion / direction} &
\tcell{0.24\linewidth}{Approach neutrino floor; directional anisotropy} &
\tcell{0.35\linewidth}{Require multi-target confirmation, spectral + time/dir consistency, astrophysical priors}\\

\tcell{0.17\linewidth}{Is DM an axion?} &
\tcell{0.17\linewidth}{\(g_{a\gamma\gamma}\) vs \(m_a\) scanning} &
\tcell{0.24\linewidth}{Cover QCD axion band across decades in \(m_a\)} &
\tcell{0.35\linewidth}{Insist on signal reproducibility with tunable parameters, multiple detectors}\\

\tcell{0.17\linewidth}{Majorana neutrinos?} &
\tcell{0.17\linewidth}{\(0\nu\beta\beta\): \(T_{1/2}^{0\nu}\)} &
\tcell{0.24\linewidth}{\(10^{28}\text{--}10^{29}\ \yr\) class} &
\tcell{0.35\linewidth}{Multi-isotope, validate backgrounds, nuclear theory uncertainty control}\\

\tcell{0.17\linewidth}{Leptonic CPV?} &
\tcell{0.17\linewidth}{\(\delta_{\rm CP}\) in oscillations} &
\tcell{0.24\linewidth}{\(\sim 10^\circ\text{--}20^\circ\) (region-dependent)} &
\tcell{0.35\linewidth}{Robustness to cross-section and matter-effect systematics}\\

\tcell{0.17\linewidth}{Is baryon number violated?} &
\tcell{0.17\linewidth}{\(\tau_p\), \(\tau_{n\bar n}\)} &
\tcell{0.24\linewidth}{\(\tau_p \gtrsim 10^{35\text{--}36}\,\yr\); \(\tau_{n\bar n}\gtrsim 10^{9\text{--}10}\,\mathrm{s}\)} &
\tcell{0.35\linewidth}{Cross-check multiple modes/isotopes; robust hadronic matrix elements; complementarity with \(0\nu\beta\beta\) and EDMs}\\

\tcell{0.17\linewidth}{New CP phases?} &
\tcell{0.17\linewidth}{EDMs \(d_e,d_n\)} &
\tcell{0.24\linewidth}{10--100$\times$ improvements beyond current} &
\tcell{0.35\linewidth}{Combine e/n/atomic EDMs to resolve operator structure}\\

\tcell{0.17\linewidth}{EW phase transition?} &
\tcell{0.17\linewidth}{Higgs self-coupling + GW background} &
\tcell{0.24\linewidth}{\(\sim 10\%\) on \(\lambda_3\) + stochastic GW sensitivity} &
\tcell{0.35\linewidth}{Collider + GW synergy; model-dependent mapping}\\

\tcell{0.17\linewidth}{Extra light relics?} &
\tcell{0.17\linewidth}{\(\Delta \Neff\)} &
\tcell{0.24\linewidth}{\(\sigma(\Delta \Neff)\sim 0.03\text{--}0.06\)} &
\tcell{0.35\linewidth}{Foreground and calibration dominated; cross-check with BBN}\\

\tcell{0.17\linewidth}{GR violations?} &
\tcell{0.17\linewidth}{EP parameter \(\eta\), GW propagation, PPN} &
\tcell{0.24\linewidth}{\(\eta\sim 10^{-17}\) (space) + multi-band GW tests} &
\tcell{0.35\linewidth}{Insist on multiple channels: EP + GW + clocks}\\

\tcell{0.17\linewidth}{Is the $H_0$ tension physical?} &
\tcell{0.17\linewidth}{$H_0$ from standard sirens, strong lenses, TRGB/JAGB, BAO+BBN} &
\tcell{0.24\linewidth}{$\sigma(H_0)/H_0 \lesssim 0.5\%$ with mutually independent calibration chains} &
\tcell{0.35\linewidth}{Require internal-consistency closures (distance ladder vs sirens vs BAO), explicit systematics budgets, and agreement of multiple late-time methods}\\

\tcell{0.17\linewidth}{Is growth consistent with GR+$\Lambda$CDM?} &
\tcell{0.17\linewidth}{$S_8$, $f\sigma_8(z)$, CMB lensing vs shear vs RSD consistency} &
\tcell{0.24\linewidth}{Percent-level cross-probe clo- sure with baryonic/nonlinear systematics controlled} &
\tcell{0.35\linewidth}{Focus on joint inference with shared nuisa- nce models and validation on simulations/ data splits}\\
\hline
\end{longtable}

In  Table~\ref{tab:targets}, the quoted targets are intended as \emph{decision-level} sensitivities: levels at which a null result or a discovery would substantially eliminate broad classes of models (or force qualitatively new theoretical structure), \emph{assuming} that the dominant experimental and theory systematics are controlled at commensurate levels. They should therefore not be read as simple ``instrument requirements''. In several cases (e.g.,\ DM direct detection near the neutrino floor, precision cosmology in the presence of nonlinear baryonic effects, or SMEFT fits limited by hadronic/nuclear matrix elements), improvements in raw exposure or detector noise alone are insufficient: progress requires joint investment in calibration, end-to-end simulations, and theory inputs that reduce nuisance-parameter degeneracies.

\section{Roadmap: principles, decision points, and timelines}
\label{sec:roadmap}

This roadmap is written in terms of \emph{physics decision points} rather than projects. In the current regime, many flagship constraints are limited by structured nuisance models (instrument calibration, astrophysical foregrounds, hadronic and nuclear matrix elements, waveform systematics) rather than by counting statistics.  A roadmap is therefore required for a technical reason: to ensure that new measurements actually move constraints in \emph{model space} (EFT coefficients / symmetry-violation parameters), and that they do so with redundancy sufficient to distinguish new physics from underestimated systematics. Accordingly, each staged step below is defined by (i) a target observable and scaling lever arm, (ii) a closed error budget for the dominant systematics, and (iii) at least one independent cross-check channel that constrains the same parameter combination.

The preceding sections imply that a roadmap should be  formulated as a sequence of \emph{decision problems} (model discrimination and parameter inference tasks), not as a facility list. In the regimes that dominate current opportunities---systematics-limited precision tests, background-limited rare-event searches, and theory-limited interpretation---progress is controlled by \emph{closure} rather than by marginal improvements in a single noise metric.

A useful operational definition is the following. For a set of datasets $\{D_k\}$ with experiment-specific nuisance parameters $\nu_k$ and shared physics parameters $\theta$ (e.g.\ EFT coefficients, cosmological parameters), each channel provides a likelihood $\mathcal{L}_k(D_k\!\mid\!\theta,\nu_k)$. A result becomes \emph{actionable} only when the inferred constraints on $\theta$ are stable under (i) substantially different nuisance models within a channel and (ii) at least one conceptually independent channel with different dominant couplings and different systematics. In other words, the goal is not $\Delta$(single-sensor sensitivity), but demonstrated stability of the
inference under controlled model and nuisance deformations.

\subsection{Principle I: redundancy and closure tests}

Many frontier measurements now sit at or near a systematics floor. Decisive progress therefore requires two or more methods that probe the same underlying parameter(s) with  \emph{orthogonal} dominant error terms. Concretely: late-time cosmology demands closure between distance-ladder calibrations, strong-lensing time delays, standard sirens, and BAO+BBN; equivalence-principle and fifth-force searches demand probes with different coupling structures (torsion balances, atom interferometers, space accelerometers, clock/redshift comparisons); and any dark-matter claim requires target complementarity (nuclear/electron/collective excitations), at least one distinctive discriminant (directionality and/or temporal modulation), and consistency with accelerator/astrophysical bounds in the same operator basis.

Several enabling efforts are common across essentially all decision points:
(i) nonperturbative QCD/nuclear inputs with quantified uncertainties and correlations (EDMs, $0\nu\beta\beta$,
rare decays, neutrino scattering);
(ii) open likelihoods or validated compressed surrogates to enable combined inference under shared nuisance models;
(iii) end-to-end calibration and simulation pipelines that propagate instrument systematics into the final likelihood; and (iv) explicit theory-consistency constraints (unitarity/analyticity/positivity; stability/causality in cosmological EFTs) implemented as quantitative priors when appropriate. These are not managerial preferences; they are necessary conditions for improved measurements to translate into falsifiable constraints on UV structure.

\subsection{Principle II: inference pipeline as a first-class deliverable}

From the start, each program should commit to three publishable deliverables. First, a public mapping from instrumental parameters (noise spectra, thresholds, duty cycle, scan strategy) to reach in a standardized signal model or operator basis. Second, a complete uncertainty budget that separates statistical noise, calibration systematics, backgrounds, and theoretical/nuclear/hadronic inputs, and propagates them into the final likelihood. Third, an explicit cross-validation plan (redundant observables, blinding protocols, calibration lines, auxiliary sensors, and planned null tests) that can falsify the leading false-positive scenarios.

Every major program should specify, \emph{before} hardware freezes, a translation chain (see Figure~\ref{fig:translation_chain})
\[
\text{instrument model} \;\rightarrow\; \text{forward model for observables} \;\rightarrow\; \text{likelihood}
\;\rightarrow\; \text{EFT/UV interpretation}.
\]
Technically, this means: (i) an end-to-end response model  (calibration, nonidealities, environmental couplings) that is validated in situ; (ii) a likelihood (or a documented approximate likelihood) that propagates correlated systematics and nuisance parameters; (iii) versioned simulation + reconstruction with reproducible pipelines; and (iv) an explicit matching layer (operator basis choices, running/mixing where relevant, and theory-uncertainty models) so that results can be combined consistently across laboratory, astrophysical, and cosmological data.

This principle is especially important in areas where the limiting uncertainties are \emph{theory dominated} (hadronic/nuclear matrix elements, waveform systematics, astrophysical population priors) or where discovery claims require multi-experiment confirmation (DM, EP/ISL, subtle EFT deviations).

\subsection{Principle III: stage by go/no-go thresholds}

The staging below is  organized by \emph{physics decision points} rather than by platform. Each decision point lists the measurement modes that constrain the same underlying parameter combinations.

\subsubsection*{2025--2030: establish cross-probe baselines and close dominant systematics}

\paragraph*{a. Is there evidence for new weak-scale structure in the Higgs/electroweak sector?}
Near-term priority is to convert HL-LHC-era measurements into global SMEFT constraints with transparent theory uncertainties and validated EFT truncation assumptions, and to cross-check with low-energy constraints (EDMs, CLFV, rare decays) that probe complementary operator directions at much higher effective scales.  The decisive output is not a channel-by-channel coupling modifier but a closed global fit with stated nuisance-model robustness.

\paragraph*{b. What dark-matter hypotheses remain viable after next-generation null results?}
Push spin-independent WIMP searches into the neutrino-background regime with multi-target portfolios, while simultaneously expanding sub-GeV searches (electron recoils and collective excitations) and axion/ALP coverage with both resonant and broadband techniques.  In parallel, strengthen operator-level connections to accelerator portal searches and to structure probes that test free streaming and self-interactions. The decision metric is the reduction of viable coupling--mass volume after applying cross-consistency requirements, not the improvement of any single limit curve.

\paragraph*{c. Neutrino mass and lepton number: can the field reach the inverted-ordering frontier with controlled theory?}
Advance the long-baseline oscillation program toward robust mass-ordering and $\delta_{\rm CP}$ constraints under conservative interaction-model nuisance frameworks, while tightening the absolute mass scale via kinematic measurements and cosmological mass-sum inference.  For $0\nu\beta\beta$, the near-term technical bottleneck is correlated nuclear-theory uncertainty: reduce NME spreads and quantify inter-isotope correlations so that upcoming exposures translate into decision-grade constraints on the underlying lepton-number-violating
parameter(s).

\paragraph*{d. Are late-time cosmology tensions robust to improved modeling?}
Use DESI/ACT-era data and early Euclid/Rubin analyses to test whether inferred departures from $\Lambda$CDM persist under explicit nonlinear/baryonic and selection nuisance models.  The decisive output is a cross-survey, cross-probe closure test in which the same nuisance parameterization is carried consistently through geometric and growth observables.

\paragraph*{e. Ultimate EP/GR tests, improved ISL tests.}
On the ground/near-space side, upgrades to LLR (including high-power laser links and next-generation ranging architectures \cite{Turyshev2025HighPowerLLR,Turyshev2025HighPowerLLR-ErBu}) provide a practical path to strengthen solar-system tests of relativistic gravity and long-range deviations in parallel with dedicated space EP missions.

\paragraph*{f. Strong-field gravity and multi-messenger baselines.}
Exploit the growing GW and PTA datasets to stress-test GR in propagation and ringdown sectors with population-systematics explicitly modeled.  The decision metric is whether any deviation is stable under alternative waveform systematics and selection functions, and whether it is consistent with independent weak-field constraints (EP/redshift/PPN) in the same deformation basis.

\subsubsection*{2030--2040: open new observable windows and build decisive redundancy}

\paragraph*{a. mHz gravitational waves and multi-band inference.}
Realize LISA-class mHz interferometry and integrate with ground-based networks for multi-band parameter inference and propagation tests.  The key fundamental-physics outputs are (i) strong-field consistency tests with high-SNR ringdowns and (ii) stochastic-background discrimination (astrophysical vs early-universe) under foreground and instrument-noise correlation models.

\paragraph*{b. Next-generation ground-based GW detectors and high-statistics standard sirens.}
3G ground detectors (ET/CE-class) are uniquely enabling for large standard-siren samples and precision tests of compact-object dynamics.  The coupling to cosmology is decision-grade only if redshift association and selection effects are controlled and if the same inference framework is cross-validated against independent distance indicators.

\paragraph*{c. Long-baseline neutrino program maturity.}
DUNE/Hyper-K-class exposures should enable robust CP-phase constraints and precision tests of the three-flavor framework; the decisive criterion is stability under realistic flux/cross-section nuisance models validated by near-detector data and external hadron-production inputs.

\paragraph*{d. Precision CP and rare-process portfolio.}
Order-of-magnitude improvements in EDMs, CLFV, and rare kaon/meson channels remain among the most efficient
routes to ultra-high-scale sensitivity.  The physics deliverable is operator-structure discrimination via a \emph{portfolio} (electron, neutron, diamagnetic atoms/molecules; multiple CLFV channels) rather than a single headline limit.

\paragraph*{e. Controlled redshift and EP tests with in-situ missions.}
In-situ experiments and missions are decisive where they change the dominant scaling (free-fall time, potential modulation, low-frequency environment) and where the spacecraft+payload error budget closes end-to-end. The decision metric is net reduction of the total propagated uncertainty in EP/redshift parameters after including orbit determination, time/frequency transfer, self-gravity, charging, and thermal gradients.

\subsubsection*{2040+: extend direct reach and precision to qualitatively new regimes}

The long-term frontier depends on what the 2025--2040 era reveals. If no new states appear but coherent EFT deviations accumulate, the highest priority becomes facilities that resolve EFT flat directions and test energy-growing amplitudes (lepton-collider precision programs and/or $\sim 100~\mathrm{TeV}$ hadron-collider direct reach, judged by global-fit impact rather than by slogans). If anomalies persist in cosmology/gravitation, the priority becomes multi-messenger closure with improved foreground control and with complementary weak-field tests that bound screening and environment dependence. If credible dark-sector hints emerge, the priority becomes cross-confirmation across laboratory, collider, and astronomical channels with operator-level consistency.

\subsection{A short watchlist: what specific closure tests would settle them?}
\label{sec:contentious_claims}

Several high-visibility claims in 2025-era cosmology are not settled by raw statistics but by whether the \emph{same inference} survives end-to-end systematics closure across independent probes.

\paragraph{$H_0$ discrepancy: what would settle it?}
Decision-grade settlement requires (i) $\sigma(H_0)/H_0\lesssim 0.5\%$ from at least two calibration chains with demonstrably different dominant systematics (distance ladder vs sirens vs BAO+BBN vs lenses), and (ii) demonstrated stability to selection-function and population priors (for sirens and lenses).

\paragraph{Dynamical-dark-energy preferences in BAO+SNe+shear combinations.}
For any reported preference for evolving $w(z)$ (e.g.\ CPL $w_0$--$w_a$), the decisive closure tests are: (i) consistent BAO distances across tracers/redshift bins under a shared reconstruction nuisance model, (ii) supernova photometric-calibration and population-drift control verified by data splits, (iii) explicit cross-survey covariance included for overlapping footprints, and (iv) growth-vs-geometry closure under a shared baryonic/nonlinear nuisance model.

\paragraph{$S_8$ / growth tensions.}
The decisive tests are cross-survey shear calibration closure (independent pipelines), cross-correlation with CMB lensing, and explicit baryonic feedback marginalization validated on simulations spanning plausible feedback models.

\section{Conclusions}
\label{sec:concl}

A central point of this review is that \emph{physics is not platform-defined}. Whether a result is obtained on the ground, in space, or via astronomy matters only insofar as the platform changes the accessible observables and the dominant error terms. Accordingly, we separated three measurement archetypes and evaluate each by its observable content, scaling laws, and realistic noise floors:
(i) \emph{ground-based controlled experiments}, where null tests and calibration hierarchies can be made explicit but environmental noise (seismic, Newtonian, thermal, electromagnetic) often sets the floor;
(ii) \emph{passive astronomical observations}, where long baselines and strong-field/early-universe conditions are available but inference is frequently limited by astrophysical modeling and selection systematics; and
(iii) \emph{in-situ space experiments}, where free-fall times, gravitational-potential modulation, and low-frequency force noise can improve particular scalings, but only if time/frequency transfer, drag-free performance, and spacecraft systematics are subdominant.

As of 2025, ``fundamental physics'' is best viewed as a quantitatively constrained program with a clear baseline
(SM+GR+\LCDM) and a finite set of empirically motivated failure modes: missing particle content (dark matter, neutrino masses), missing dynamical ingredients (baryogenesis, dark energy), and possible deformations of spacetime structure and quantum theory. The practical implication is that the field is no longer organized by a single energy frontier.  Instead it is organized by \emph{decision problems}: which measurements, taken together, can exclude broad classes of models once correlated systematics and theory uncertainties are propagated transparently.

A central message of this review is that the relevant performance metric is increasingly a \emph{decision threshold}, not a marginal sensitivity improvement.  Tables~\ref{tab:openproblems} and \ref{tab:targets} make this explicit by tying each open question to (i) an observable (or set of observables), (ii) an approximate sensitivity scale that would qualitatively change the
inference, and (iii) the cross-checks required for robustness.  Examples include reaching \(T^{0\nu}_{1/2}\sim 10^{28}\)--\(10^{29}\,\mathrm{yr}\) with multiple isotopes for Majorana neutrinos, pushing spin-independent WIMP searches into the neutrino-floor neighborhood together with modulation/directionality discriminants, \(\sigma(\Delta N_{\rm eff})\sim 0.03\)--0.06 for light relics with foreground- and calibration-dominated error budgets, order-of-magnitude improvements in EDM and CLFV sensitivities with operator-level disentangling, and space EP tests targeting \(\eta\sim 10^{-17}\) with composition diversity and independent gravity probes.

Achieving ``definitive'' answers in these regimes requires cross-cutting infrastructure that is often underemphasized relative to hardware.  The same limiting factors recur across otherwise distinct subfields: (i) hadronic and nuclear uncertainties (\(0\nu\beta\beta\) matrix elements, EDM and rare-process matching), (ii) nonperturbative QCD inputs in precision tests (e.g.\ hadronic contributions to \((g-2)_\mu\)), (iii) nonlinear and selection systematics in cosmological inference (shear and photometric calibration, intrinsic alignments, baryonic feedback), and (iv) waveform and population systematics for GW-based
tests of GR and cosmography.  A technically credible roadmap therefore requires end-to-end likelihoods, shared simulation and systematics models, and global fits that translate instrument performance into EFT-level constraints that can be compared across laboratory, astrophysical, and cosmological data.

On the instrumentation side, the most credible near-term gains arise when new technology delivers a new observable channel or a qualitatively different systematics regime. Quantum sensors are exemplary: optical clocks plus coherent time-transfer networks enable $10^{-18}$--$10^{-19}$ fractional frequency comparisons; atom interferometers and optomechanical inertial sensors extend precision acceleration/gradient measurements into low-frequency regimes relevant to fifth-force and quantum--gravity interface tests; and quantum-limited EM sensors enable broadband ultralight-field searches via coherent phase accumulation and correlated networks. Space platforms are enabling when they suppress the dominant noise terms by orders of magnitude (extended free-fall, quieter low-frequency environment, longer baselines, orbital modulation), but their value depends on closure of the end-to-end nuisance budget through in-flight calibration and independent redundancy. Comparable ``new-window'' leverage also comes from multi-band GW observations, next-generation colliders that lift EFT flat directions with differential data in global fits, and cosmology programs that enforce closure under shared nuisance models via cross-survey null/injection tests.

The near-term priority is to translate improved raw sensitivity into \emph{identified} constraints by fixing (i) the EFT
target space (operator basis/parameter combinations), (ii) the likelihood with an explicit nuisance sector, and (iii) the
closure tests that separate signal from systematics; only then can forthcoming collider, intensity-frontier,
multi-messenger, and quantum-precision data be compressed into a small set of viable UV structures beyond
SM$+$GR$+\Lambda$CDM.

A useful synthesis is to separate realistic discovery potential by time horizon into (i) \emph{new observables} and (ii) \emph{incremental precision}. In the near term (2025--2030), the most likely decisive outcomes are those already limited by systematics closure: (i) long-baseline neutrinos with joint oscillation+cross-section inference; (ii) direct DM searches approaching the neutrino background, where confirmation requires multi-target complementarity and at least one discriminant (modulation and/or directionality); (iii) EDM/CLFV advances that exclude broad CPV/flavor operator classes once hadronic/nuclear matching uncertainties are propagated; and (iv) late-time cosmology cross-checks where the dominant error is shared nuisance modeling rather than counting statistics. In the mid term (2030--2040), the clearest genuinely new windows are multi-band GW astronomy and precision gravitational-potential modulation with space time/frequency transfer, contingent on end-to-end error budgets staying below the targeted redshift/EP channels. In the long term (2040+), step changes come from new energy or baseline regimes: next-generation colliders (direct reach plus global SMEFT closure) and ultimate low-frequency/long-baseline precision experiments once terrestrial floors become irreducible. Across all horizons, the technical criterion is invariant: constraints stated at the operator level with an explicit nuisance sector and redundancy across independent measurement classes.

Finally, no single platform is privileged by the physics: ground-based control, astronomical baselines, and in-situ space
conditions are complementary. The only meaningful comparison is at the level of \emph{observables, scalings, and error
budgets}. Space is advantageous only when it changes the dominant noise term or unlocks an otherwise inaccessible
scaling (extended free fall, gravitational-potential modulation, reduced low-frequency force noise) beyond terrestrial
limits; conversely, ground laboratories often dominate when calibration fidelity and systematic closure set the floor.
Accordingly, we organized the landscape by the mapping \emph{measurement $\rightarrow$ likelihood (with an explicit
nuisance sector) $\rightarrow$ parameter inference}, and treated redundancy as the operational definition of a decisive
test.

With that criterion in hand, the path forward is a disciplined portfolio optimized for \emph{identifiable inference}: invest where a new observable is opened or where the leading noise term is \emph{replaced} (not merely reduced), and require closure---explicit nuisance models/priors, preserved likelihood products, and pre-specified validation tests---before translating sensitivity gains into statements about fundamental parameters. Concretely, decisive targets should be posed as operator-level reach under stated nuisance assumptions, with an end-to-end research program that is validated by null tests, data splits, and orthogonal observables.  The payoff is commensurate: a coherent set of complementary results that does not merely tighten bounds, but
either decisively excises broad regions of theory space, or produces anomalies that survive rigorous cross-checks and
full uncertainty propagation from instrument and environment to the inferred parameters---thereby rising to the level of
credible evidence for new physics.
\vskip -20pt
\section*{Acknowledgments} 
\vskip -5pt
The work described here was carried out at the Jet Propulsion Laboratory, California Institute of Technology, Pasadena, California, under a contract with the National Aeronautics and Space Administration.
% \textcopyright 2025. California Institute of Technology. Government sponsorship acknowledged.

\vskip -15pt

%\bibliography{fundamental-physics}


\vskip -18pt
\appendix
\section{Selected experiment classes, dominant systematics, and parameter combinations}
\label{app:landscapes}

\vskip -15pt
This appendix collects compact landscape tables that connect broad measurement classes to:
(i) the data products entering the likelihood (primary observables),
(ii) the dominant correlated nuisance terms that set robustness once statistical errors are subdominant, and
(iii) the combinations of phenomenological/EFT parameters constrained after marginalization. The tables are schematic by design; their purpose is cross-program comparability---to make operator/parameter complementarity and the leading bottlenecks explicit at a glance. In addition to the dark-matter, neutrino, and precision/symmetry portfolios in Tables~\ref{tab:landscape_dm}--\ref{tab:landscape_precision}, Table~\ref{tab:landscape_gravity}
summarizes gravity/spacetime tests (UFF/EP, ISL/fifth forces, solar-system ranging including LLR, and clocks/time transfer).
Table~\ref{tab:landscape_gw} provides an analogous summary for strong-field and propagation tests from
binary-pulsar timing, GWs, and PTAs.

%\subsection{Dark matter landscape table}

\begin{table*}[h!]
\vskip -15pt
\caption{Dark-matter landscape: experiment classes and the parameter combinations they constrain.}
\label{tab:landscape_dm}
\centering
\renewcommand{\arraystretch}{1.10}
\setlength{\tabcolsep}{0pt}
\begin{tabular}{@{}llll@{}}
\hline
\tcell{0.15\textwidth}{Class} &
\tcell{0.22\textwidth}{Observable(s) / mapping} &
\tcell{0.26\textwidth}{Dominant systematics} &
\tcell{0.32\textwidth}{Parameters constrained} \\
\hline\hline

\tcell{0.15\textwidth}{Noble TPC (NR)} &
\tcell{0.22\textwidth}{Recoil spectrum $\dd R/\dd E_R$} &
\tcell{0.26\textwidth}{Radiogenics, neutrons, CE$\nu$NS, response model} &
\tcell{0.32\textwidth}{NR EFT couplings $c_i^{p,n}$ (SI/SD + momentum-dependent operators)} \\

\tcell{0.15\textwidth}{Cryogenic NR} &
\tcell{0.22\textwidth}{Threshold recoil counts; phonon/ionization partition} &
\tcell{0.26\textwidth}{Sub-keV calibration, surface backgrounds} &
\tcell{0.32\textwidth}{Low-mass NR EFT; light mediators via $q$-dependence} \\

\tcell{0.15\textwidth}{Semiconductor / skipper (ER)} &
\tcell{0.22\textwidth}{$e^-$-counting; ER spectra} &
\tcell{0.26\textwidth}{Dark current, spurious charge} &
\tcell{0.32\textwidth}{$\chi$--$e$ scattering $\sigma_e(q)$; dark photon portals $(\epsilon,m_{A'})$} \\

\tcell{0.15\textwidth}{Axion haloscopes} &
\tcell{0.22\textwidth}{Narrowband excess at $f_a$} &
\tcell{0.30\textwidth}{Cavity systematics, magnetic pickup} &
\tcell{0.31\textwidth}{$g_{a\gamma\gamma}$ vs $m_a$ (and ALP generalizations)} \\

\tcell{0.15\textwidth}{Indirect $\gamma/\nu$/CR} &
\tcell{0.22\textwidth}{Spectra + morphology} &
\tcell{0.26\textwidth}{$J$-factors, foregrounds, instrument response} &
\tcell{0.31\textwidth}{$\langle\sigma v\rangle$, $\tau_\chi$, cascade models} \\
\hline
\end{tabular}
%\end{table*}

\begin{comment}
\begin{table*}[h!]
\vskip -10pt
\centering
\caption{Dark-matter landscape: experiment classes and the parameter combinations they constrain.}
\label{tab:landscape_dm}
\renewcommand{\arraystretch}{1.10}
\begin{tabularx}{\textwidth}{p{0.20\textwidth} p{0.22\textwidth} p{0.24\textwidth} X}
\hline
Class & Observable(s)  / mapping & Dominant systematics & Parameters constrained \\
\hline\hline
Noble TPC (NR) & Recoil spectrum $\dd R/\dd E_R$ & Radiogenics, neutrons, CE$\nu$NS, response model & NR EFT couplings $c_i^{p,n}$ (SI/SD + momentum-dependent operators) \\

Cryogenic NR & Threshold recoil counts; phonon/ionization partition & Sub-keV calibration, surface backgrounds & Low-mass NR EFT; light mediators via $q$-dependence \\

Semiconductor / skipper (ER) & $e^-$-counting; ER spectra & Dark current, spurious charge & $\chi$--$e$ scattering $\sigma_e(q)$; Dark photon portals $(\epsilon,m_{A'})$ \\

Axion haloscopes & Narrowband excess at $f_a$ & Cavity systematics, magnetic pickup & $g_{a\gamma\gamma}$ vs $m_a$ (and ALP generalizations) \\

Indirect $\gamma/\nu$/CR & Spectra + morphology & $J$-factors, foregrounds, instrument response & $\langle\sigma v\rangle$, $\tau_\chi$, cascade models \\
\hline
\end{tabularx}
\end{table*}
\end{comment}

%\subsection{Neutrino landscape table}

%\begin{table*}[h!]
\vskip -5pt
\caption{Neutrino landscape: major program classes beyond a single long-baseline narrative.}
\label{tab:landscape_nu}
\centering
\renewcommand{\arraystretch}{1.10}
\setlength{\tabcolsep}{1pt}
\begin{tabular}{@{}llll@{}}
\hline\hline
\tcell{0.14\textwidth}{Class} &
\tcell{0.21\textwidth}{Observable(s) / mapping} &
\tcell{0.28\textwidth}{Dominant systematics} &
\tcell{0.33\textwidth}{Parameters constrained} \\
\hline

\tcell{0.14\textwidth}{Long-baseline acceleration} &
\tcell{0.21\textwidth}{Appearance/disappearance spectra} &
\tcell{0.28\textwidth}{Flux$\times$XS nuisance, detector response} &
\tcell{0.33\textwidth}{$(\delta_{\rm CP},\theta_{23},\Delta m^2_{31})$, ordering (w/ matter effects)} \\

\tcell{0.14\textwidth}{Reactor medium-baseline} &
\tcell{0.21\textwidth}{$\bar\nu_e$ disappearance fine structure} &
\tcell{0.28\textwidth}{Energy nonlinearity, spectrum nuisance} &
\tcell{0.33\textwidth}{Ordering-sensitive interference; precision $\Delta m^2$ and $\theta_{12}$} \\

\tcell{0.14\textwidth}{CE$\nu$NS} &
\tcell{0.21\textwidth}{Low-$E_R$ spectra} &
\tcell{0.28\textwidth}{Neutron backgs., threshold model} &
\tcell{0.33\textwidth}{SM weak charge; NSI $\epsilon_{\alpha\beta}$} \\

\tcell{0.14\textwidth}{Telescopes (TeV-- PeV)} &
\tcell{0.21\textwidth}{Flavor/energy distribu- tions; Earth absorption} &
\tcell{0.28\textwidth}{Optical medium, atm. backgrounds, PDFs} &
\tcell{0.33\textwidth}{$\nu N$ cross sections; secret interactions; LV dispersion} \\

\tcell{0.14\textwidth}{$0\nu\beta\beta$ portfolio} &
\tcell{0.21\textwidth}{$T^{0\nu}_{1/2}$ across isotopes} &
\tcell{0.28\textwidth}{ROI backgrnds.; NME correlations} &
\tcell{0.33\textwidth}{$m_{\beta\beta}$ (light) + short-range LNV operators} \\
\hline
\end{tabular}
\end{table*}

\begin{comment}
\begin{table*}[h!]
\vskip 5pt
\centering
\caption{Neutrino landscape: major program classes beyond a single long-baseline narrative.}
\label{tab:landscape_nu}
\renewcommand{\arraystretch}{1.10}
\begin{tabularx}{\textwidth}{p{0.14\textwidth} p{0.21\textwidth} p{0.28\textwidth} X}
\hline
Class & Observable(s) / mapping & Dominant systematics & Parameters constrained \\
\hline\hline
Long-baseline acceleration & Appearance/disappearance spectra & Flux$\times$XS nuisance, detector response & $(\delta_{\rm CP},\theta_{23},\Delta m^2_{31})$, ordering (w/ matter effects) \\

Reactor medium-baseline & $\bar\nu_e$ disappearance fine structure & Energy nonlinearity, spectrum nuisance & Ordering-sensitive interference; precision $\Delta m^2$ and $\theta_{12}$ \\

CE$\nu$NS & Low-$E_R$ spectra & Neutron backgs., threshold model & SM weak charge; NSI $\epsilon_{\alpha\beta}$ \\

Telescopes (TeV--PeV) & Flavor/energy distributions; Earth absorption & Optical medium, atm. backgrounds, PDFs & $\nu N$ cross sections; secret interactions; LV dispersion \\

$0\nu\beta\beta$ portfolio & $T^{0\nu}_{1/2}$ across isotopes & ROI backgrnds.; NME correlations & $m_{\beta\beta}$ (light) + short-range LNV operators \\
\hline
\end{tabularx}
\end{table*}
\end{comment}

%\subsection{Precision/EDM landscape table}

\begin{table*}[h!]
\vskip 5pt
\caption{Precision symmetry tests: dominant systematics and EFT parameter combinations.}
\label{tab:landscape_precision}
\centering
\renewcommand{\arraystretch}{1.10}
\setlength{\tabcolsep}{1pt}
\begin{tabular}{@{}llll@{}}
\hline
\tcell{0.20\textwidth}{Class} &
\tcell{0.20\textwidth}{Observable(s) / mapping} &
\tcell{0.24\textwidth}{Dominant systematics} &
\tcell{0.34\textwidth}{Parameters constrained} \\
\hline\hline

\tcell{0.20\textwidth}{EDMs (e, n, atoms)} &
\tcell{0.20\textwidth}{$d_e$, $d_n$, diamagnetic EDMs} &
\tcell{0.24\textwidth}{Fields/leakage, many-body/nuclear theory} &
\tcell{0.34\textwidth}{CPV dipoles, Weinberg operator, semilep- tonic CPV} \\

\tcell{0.20\textwidth}{CLFV} &
\tcell{0.20\textwidth}{$\mu\to e\gamma$, $\mu N\to e N$} &
\tcell{0.24\textwidth}{Accidentals, beam-related bkg} &
\tcell{0.33\textwidth}{Dipole + 4-fermion CLFV operators} \\

\tcell{0.20\textwidth}{PV (e scattering/atoms)} &
\tcell{0.20\textwidth}{$A_{\rm PV}(Q^2)$, weak charges} &
\tcell{0.24\textwidth}{Radiative tails, atomic theory} &
\tcell{0.34\textwidth}{Semileptonic $C_{1q},C_{2q}$ (SMEFT matching)} \\

\tcell{0.20\textwidth}{Rare meson decays} &
\tcell{0.20\textwidth}{$K\to\pi\nu\bar\nu$, $B$ anomalies} &
\tcell{0.24\textwidth}{Hadronic form factors, charm loops} &
\tcell{0.34\textwidth}{Semileptonic EFT operators at $\sim 10$--$100$ TeV scales} \\
\hline
\end{tabular}
%\end{table*}

\begin{comment}
\begin{table*}[h!] \vskip -15pt
\vskip  5pt
\centering
\caption{Precision symmetry tests: dominant systematics and EFT parameter combinations.}
\label{tab:landscape_precision}
\renewcommand{\arraystretch}{1.10}
\begin{tabularx}{\textwidth}{p{0.20\textwidth} p{0.20\textwidth} p{0.24\textwidth} X}
\hline
Class & Observable(s) / mapping & Dominant systematics & Parameters constrained \\
\hline\hline

EDMs (e, n, atoms) & $d_e$, $d_n$, diamagnetic EDMs & Fields/leakage, many-body/nuclear theory & CPV dipoles, Weinberg operator, semileptonic CPV \\

CLFV & $\mu\to e\gamma$, $\mu N\to e N$ & Accidentals, beam-related bkg & dipole + 4-fermion CLFV operators \\

PV (e scattering / atoms) & $A_{\rm PV}(Q^2)$, weak charges & Radiative tails, atomic theory & Semileptonic $C_{1q},C_{2q}$ (SMEFT matching) \\

Rare meson decays & $K\to\pi\nu\bar\nu$, $B$ anomalies & Hadronic form factors, charm loops & Semileptonic EFT operators at $\sim 10$--$100$ TeV scales \\
\hline
\end{tabularx}
\end{table*}
\end{comment}

%\subsection{Gravity/spacetime landscape table}

%\begin{table*}[h!]
\vskip 10pt
\caption{Gravity/spacetime tests: measurement classes, dominant systematics, and parameter combinations.}
\label{tab:landscape_gravity}
\centering
\renewcommand{\arraystretch}{1.10}
\setlength{\tabcolsep}{1pt}
\begin{tabular}{@{}llll@{}}
\hline
\tcell{0.15\textwidth}{Class} &
\tcell{0.24\textwidth}{Observable(s) / mapping} &
\tcell{0.26\textwidth}{Dominant systematics} &
\tcell{0.35\textwidth}{Parameters constrained} \\
\hline\hline

\tcell{0.15\textwidth}{Torsion balances (UFF/ISL)} &
\tcell{0.24\textwidth}{$\Delta\tau(\varphi)$ or $\Delta a(t)$; Yukawa $\alpha(\lambda)$} &
\tcell{0.26\textwidth}{Patch/electrostatics; alignment/ metrology; thermal drifts; gravity gradients} &
\tcell{0.35\textwidth}{$\eta_{AB}$; $\alpha(\lambda)$ (typically $\lambda\sim 10^{-4}$--$10^{1}\,$m); screened scalar charges} \\

\tcell{0.15\textwidth}{Space EP (drag-free)} &
\tcell{0.24\textwidth}{$\Delta a(t)$ in orbit; $\eta_{AB}\simeq \Delta a/g$} &
\tcell{0.26\textwidth}{Thermal gradients; charging/ patch; self-gravity; drag-free/ attitude coupling} &
\tcell{0.35\textwidth}{$\eta_{AB}$ (demonstrated $\sim 10^{-15}$; targets $\sim 10^{-17}$); long-range composition-dependent forces} \\

\tcell{0.15\textwidth}{Atom-interferometer EP/ISL} &
\tcell{0.24\textwidth}{$\Delta\phi\simeq k_{\rm eff}\Delta a\,T^2$; gradiometry $\Delta a\sim \Gamma L$} &
\tcell{0.26\textwidth}{Wavefront/pointing; Coriolis/ rotation; gradients; platform ~~$a(t)$ noise} &
\tcell{0.35\textwidth}{$\eta_{AB}$; fifth-force couplings $\alpha(\lambda)$ (geometry-dependent); ULDM force couplings} \\

\tcell{0.15\textwidth}{LLR / ILR} &
\tcell{0.24\textwidth}{Two-way range normal points $\rho(t)$; global orbit/rotation fit; diffusion/sidebands in fitted $(a,e,\ldots)$ at $f\simeq n\,\Omega_{\rm orb}$} &
\tcell{0.26\textwidth}{Station timing/coords; troposphere; reflector/array response; ephemeris/tides} &
\tcell{0.35\textwidth}{$\eta_{\rm SEP}$ (Nordtvedt); $\dot G/G$; PPN combos ($\beta,\gamma$); SME gravity $\bar s^{\mu\nu}$; $\alpha(\lambda)$ at AU--EM scales; also $\Omega_{\rm GW}(f\simeq n\,\Omega_{\rm orb})$ via binary resonance} \\

\tcell{0.15\textwidth}{Planetary ranging / Shapiro} &
\tcell{0.24\textwidth}{Range/Doppler; $\Delta t_{\rm Shapiro}\propto (1+\gamma)$} &
\tcell{0.26\textwidth}{Solar plasma; transponder/clock biases; non-gravitational \ acceleratoons; ephemerides} &
\tcell{0.35\textwidth}{PPN $\gamma$ (time delay), $\beta$ (ephemeris/perihelia); $J_2$; $\dot G/G$; preferred-frame params} \\

\tcell{0.15\textwidth}{Clocks + time transfer} &
\tcell{0.24\textwidth}{Redshift $(1+\alpha_{\rm LPI})\Delta U/c^2$; clock ratios for $\dot X/X$} &
\tcell{0.26\textwidth}{Link stability/calibration; $U(\mathbf{x},t)$ knowledge; environment  shifts (BBR/Zeeman)} &
\tcell{0.35\textwidth}{$\alpha_{\rm LPI}$; $\dot\alpha/\alpha$, $\dot\mu/\mu$; ULDM couplings $d_i/\Lambda_\phi$; SME matter-sector coefficients} \\
\hline
\end{tabular}
\vskip 10pt
\caption{Astrophysical tests (binary pulsars, GWs, PTAs): dominant systematics and constrained deformation sectors.}
\label{tab:landscape_gw}
\centering
\renewcommand{\arraystretch}{1.10}
\setlength{\tabcolsep}{1pt}
\begin{tabular}{@{}llll@{}}
\hline
\tcell{0.13\textwidth}{Class} &
\tcell{0.22\textwidth}{Observable(s) / mapping} &
\tcell{0.26\textwidth}{Dominant systematics} &
\tcell{0.36\textwidth}{Parameters constrained} \\
\hline\hline

\tcell{0.13\textwidth}{Binary pulsars (timing)} &
\tcell{0.22\textwidth}{Post-Keplerian params; $\dot P_b$; Shapiro $(r,s)$} &
\tcell{0.26\textwidth}{Timing red noise; DM/ISM; ephemeris; geometry/mass priors} &
\tcell{0.36\textwidth}{Dipole radiation; scalar-tensor couplings; strong-field SEP; $\dot G/G$} \\

\tcell{0.13\textwidth}{Ground GWs (10--$10^3$ Hz)} &
\tcell{0.22\textwidth}{Inspiral phasing + ring- down QNMs; residual tests} &
\tcell{0.26\textwidth}{Waveform systematics; calibra- tion; selection/population priors} &
\tcell{0.36\textwidth}{Generation-sector PN deformations; extra pols.; GW dispersion ($m_g$); parity/LV tests} \\

\tcell{0.13\textwidth}{Space GWs ($10^{-4}$--$10^{-1}$ Hz)} &
\tcell{0.22\textwidth}{MBHB/EMRI waveforms; propag.\ over Gpc baselines} &
\tcell{0.26\textwidth}{Acceleration noise; confusion foregrounds; orbit/clock errors} &
\tcell{0.36\textwidth}{Low-$f$ propagation tests; extra pols.; dispersion; strong-field consistency (Kerr)} \\

\tcell{0.13\textwidth}{PTAs ($10^{-9}$--$10^{-7}$ Hz)} &
\tcell{0.22\textwidth}{Cross-correlated timing residuals; spectral shape} &
\tcell{0.26\textwidth}{Solar-system ephemeris; pulsar noise; ISM corrections; clock errors} &
\tcell{0.36\textwidth}{Stochastic backgrounds (SMBHB/strings/PT); non-GR pols.; propagation-sector LV} \\

\tcell{0.13\textwidth}{Multi-messenger propagation} &
\tcell{0.22\textwidth}{$\Delta t_{\rm GW-EM}$; polarization/ dispersion constraints} &
\tcell{0.26\textwidth}{Intrinsic emission lags; localization; selection effects} &
\tcell{0.36\textwidth}{$c_g/c$, dispersion, birefringence/parity violation (model-dependent)} \\
\hline
\end{tabular}
\end{table*}

\end{document}